\documentclass[sigconf]{acmart}

\AtBeginDocument{%
  }

\setcopyright{acmlicensed}
\copyrightyear{2026}
\acmYear{2026}
\acmDOI{XXXXXXX.XXXXXXX}
\acmConference[CHI '26]{2026 ACM CHI Conference on Human Factors in Computing Systems}{Apr 13--17, 2026}{Barcelona, Spain}

\acmISBN{978-1-4503-XXXX-X/2018/06}

\usepackage{graphicx}
\usepackage{booktabs}
\usepackage{color}
\usepackage{subfigure}
\usepackage{subcaption}

\copyrightyear{2026}
\acmYear{2026}
\setcopyright{cc}
\setcctype{by-nc-nd}
\acmConference[CHI '26]{Proceedings of the 2026 CHI Conference on Human Factors in Computing Systems}{April 13--17, 2026}{Barcelona, Spain}
\acmBooktitle{Proceedings of the 2026 CHI Conference on Human Factors in Computing Systems (CHI '26), April 13--17, 2026, Barcelona, Spain}
\acmPrice{}
\acmDOI{10.1145/3772318.3791845}
\acmISBN{979-8-4007-2278-3/2026/04}

\begin{document}

\title{Are We Automating the Joy Out of Work?  Designing AI to Augment Work, Not Meaning}
%


\author{Jaspreet Ranjit}
\orcid{0009-0000-5843-2701}
\affiliation{%
  \institution{University of Southern California}
  \city{Los Angeles}
  \state{CA}
  \country{United States}
}
\email{jranjit@usc.edu}

\author{Ke Zhou}
\orcid{0000-0001-7177-9152}
\affiliation{
\institution{Nokia Bell Labs}
\city{Cambridge}
\country{United Kingdom}}
\affiliation{\institution{University of Nottingham}
\city{Nottingham}
\country{United Kingdom}}
\email{ke.zhou@nokia-bell-labs.com}

\author{Swabha Swayamdipta}
\orcid{0000-0002-5851-8254}
\affiliation{%
  \institution{University of Southern California}
  \city{Los Angeles}
  \state{CA}
  \country{United States}
}
\email{swabhas@usc.edu}

\author{Daniele Quercia}
\orcid{0000-0001-9461-5804}
\affiliation{\institution{Nokia Bell Labs} 
\city{Cambridge}
\country{United Kingdom}}
\affiliation{\institution{Politecnico di Torino}
\city{Turin}
\country{Italy}}
\email{quercia@cantab.net}



\begin{abstract}

Prior work has mapped which workplace tasks are exposed to AI, but less is known about whether workers perceive these tasks as meaningful or as busywork.
We examined: (1) which dimensions of meaningful work do workers associate with tasks exposed to AI; and (2) how do the traits of existing AI systems compare to the traits workers want.
We surveyed workers and developers on a representative sample of 171 tasks and use language models (LMs) to scale ratings to 10{,}131 computer-assisted tasks across all U.S. occupations. 
Worryingly, we find that tasks that workers associate with a sense of agency or happiness may be disproportionately exposed to AI.
We also document design gaps: developers report emphasizing politeness, strictness, and imagination in system design; by contrast, workers prefer systems that are straightforward, tolerant, and practical.
To address these gaps, we call for AI whose design explicitly focuses on meaningful work and worker needs, proposing a five-part research agenda. 
\begin{figure*}[h!]
  \centering
  \includegraphics[width=\linewidth]{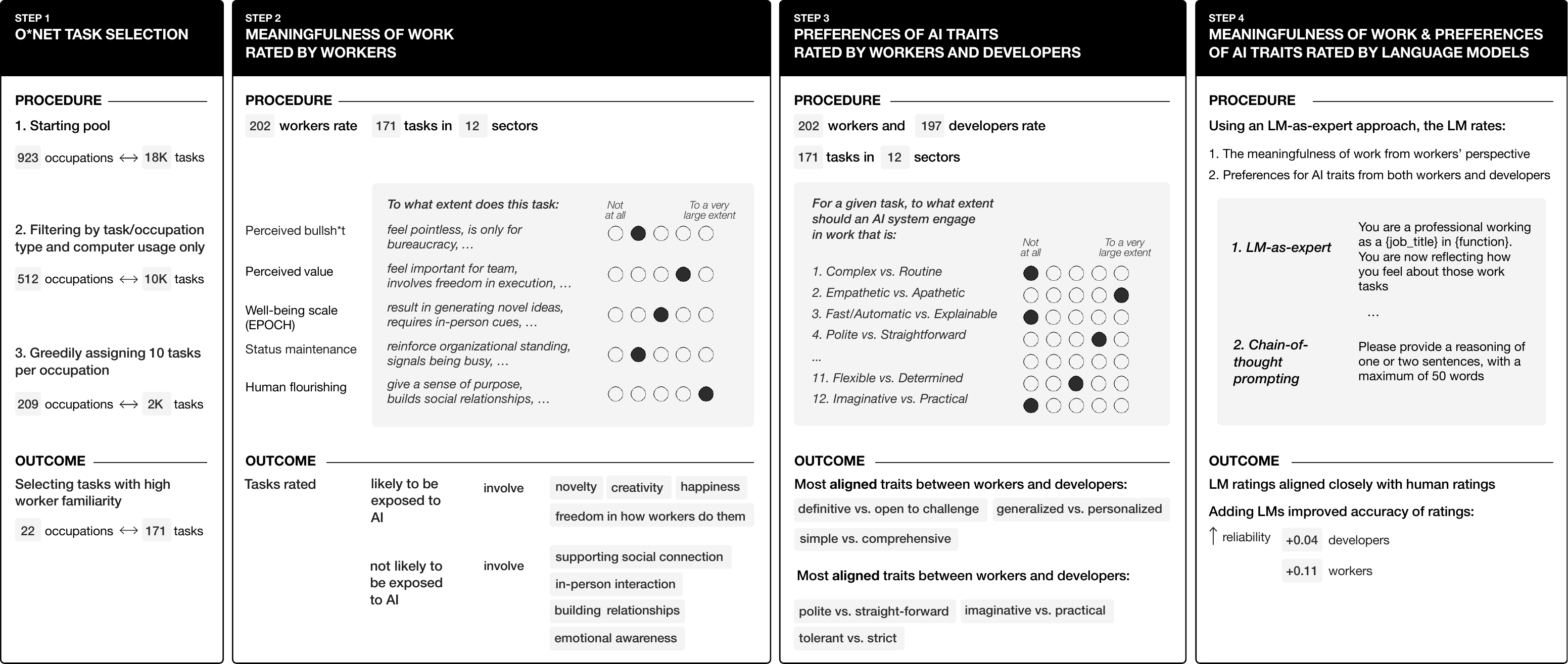}
  \caption{\textbf{Overview of Study Design}: \emph{(Step 1)} Workplace tasks were restricted to those primarily completed on a computer and performed daily or weekly, then filtered by Prolific availability, AI Impact Index, and worker familiarity.
\emph{(Step 2)} Workers rated tasks across five dimensions: perceived bullsh*t, perceived value, well-being scale, status maintenance, and human flourishing. Tasks more likely exposed to AI scored higher on novelty, creativity, happiness, and freedom, while those less likely emphasized emotional awareness, in-person interaction, relationships, and social connection.
\emph{(Step 3)} Workers and developers rated which psychological traits an AI system should possess when augmenting tasks. When designing AI augmented tasks, developers emphasized polite, strict and imaginative systems whereas, workers preferred straightforward, tolerant, and practical systems.
\emph{(Step 4)} LMs were prompted as experts to simulate worker and developer ratings, leading to moderate to high intra-class correlation with human responses.
} 
\Description{Four-panel figure showing study design steps. Step 1: task selection, narrowing to 171 tasks across 22 occupations. Step 2: workers rating meaningfulness of work across five dimensions, distinguishing tasks more vs. less likely to be augmented. Step 3: workers and developers rating desired AI traits, with aligned and misaligned traits highlighted. Step 4: language models simulating ratings with reliability gains (+0.09 for developers, +0.11 for workers).}

\label{fig:pitchfig}
\end{figure*}

\end{abstract}





\maketitle

\section{Introduction}
\label{sec:introduction}

Public debate about AI and occupations often focuses on job loss versus job growth. Some studies have predicted broad job displacement~\cite{Webb2019, Felten2023Abilities, Frank2019}, while others have anticipated growth, with AI complementing workers in ways that are associated with higher productivity and the emergence of new roles~\cite{Autor2015augmentation, Brynjolfsson2018SML, NoyZhang2023, Peng2023, Cui2024, Brynjolfsson2023Support, septiandri2024potential}. These outcomes are not fixed, however, and are widely argued to vary with how teams design AI systems. In this paper, \emph{AI} refers to software systems (e.g., LM tools, agents) that automate or augment computer-based tasks by generating, transforming, or routing information. By \emph{teams}, we mean the broader set of actors involved in AI system development, including developers, UI/UX designers, product managers, and others who contribute to how these systems are built and used. Because the very possibility of drawing a clear distinction between automation and augmentation is debated in the literature~\cite{Acemoglu2022AIJobs, Autor2024ApplyingAI, autor2022newfrontiers, felten2021occupational},
with different works adopting different operationalizations (e.g., based on time saved by AI to do a task~\cite{eloundou2024gpts}), we clarify our terminology as follows: throughout, we use \emph{AI exposure}~\cite{felten2021occupational}, our main construct of interest, to refer to tasks that current or near-term AI systems could plausibly perform or substantially speed up, operationalized as those above the $75^{th}$ percentile of the distribution of the patent-based AI Impact Index~\cite{septiandri2024potential}; \emph{AI automation} for cases where AI can perform a task end-to-end with minimal or no human involvement; and \emph{AI augmentation} for cases where AI supports or enhances human work, while humans retain primary responsibility and decision-making authority, with human involvement measured, for example, using the scale in~\cite{shao2025future}.

To inform AI design, we examine: (1) Which tasks are likely to be exposed to AI? (2) How do workers evaluate these tasks in their daily work? (3) Do AI teams design systems that meet workers' needs? When these three aspects align, growth-oriented scenarios become more plausible; when any one fails, organizations may face greater waste and resistance, in the workplace and beyond.

The bulk of existing research lies in economics and has concentrated on the first aspect: identifying which tasks are exposed to AI, and measuring the resulting labor-market effects. This body of work forms the backbone of the literature: it maps technologies (e.g., via patents~\cite{Autor2015augmentation,septiandri2024potential}) to tasks, and quantifies how AI is associated with changes in occupations~\cite{loaiza2024epoch,shao2025future,septiandri2024potential, hazra2025ai}. A growing line of research has used LMs to annotate tasks~\cite{eloundou2024gpts}; for example, labeling a task as `exposed’ when AI is defined as enabling at least a 50\% reduction in reported time to complete the task at equal or higher quality. By this definition, one study estimated that LMs are relevant for the majority of tasks in just 1.8\% of U.S. occupations, but this share is estimated to reach over 46\% when AI is considered together with productivity software~\cite{eloundou2024gpts}. Yet, the same study also estimated that only about 1.86\% of tasks are fully automatable without human oversight~\cite{eloundou2024gpts}. Therefore, while complete automation is uncommon, augmentation is expected to play the larger role~\cite{ma2025towards}. In customer support, for example, the use of generative AI systems is associated with roughly 14\% higher productivity, on average, with increases of about 35\% in the number of issues resolved per hour for less-experienced workers~\cite{shao2025future}.

The second and third aspects---(2) how workers evaluate tasks exposed to AI, and (3) whether teams design AI to meet workers' needs---remain underexplored. These gaps present a promising research agenda for HCI. Recent work has begun to map where workers prefer human intervention, suggesting that workers report being comfortable with AI handling information-centric tasks, while preferring to focus on interpersonal and organizational work~\cite{shao2025future}. Yet, despite rich literature on meaningful work in management science~\cite{tomlinson_souto_otero2025, rostain2025meaningful, lepisto2017meaningful, lipswiersma2020worthy, rosso2010meaning, bailey2019review, hackman1976} and on human-centered systems in HCI~\cite{norman2019human, shneiderman2020human, amershi2019guidelines, muller2020participatory}, we still know little about which dimensions of meaningful work workers associate with tasks potentially exposed to AI, and whether AI teams design AI tools with the traits workers want. To address this gap, we asked two research questions:
\begin{enumerate}
  \item[] (RQ1) Which dimensions of meaningful work do workers associate with tasks exposed to AI in their daily work?
  \item[] (RQ2) Do teams design AI systems with traits that align with the traits workers want?
\end{enumerate}

In so doing, we made the following main contributions:
\begin{enumerate}
\item We identified Prolific workers who reported high familiarity with a representative set of 171 tasks spanning 22 occupations drawn from the U.S. Department of Labor's Occupational Information Network (O*NET) (methodology in Section~\ref{sect:selecting-tasks}; task representativeness in Figure~\ref{fig:task_dist}).

\item We conducted a scoping review of research on meaningful work (Section~\ref{sect:scoping}), which informed the construction of survey items for workers (Section~\ref{sect:q1-33}). Building on recent work specifying desirable AI traits across occupations~\cite{dong2024fears}, we then developed parallel survey modules: one for workers, eliciting the traits they want AI to possess (Section~\ref{sect:q1-33}), and one for developers, eliciting the traits they intend to design in AI systems (Section~\ref{sect:q34-45}).

\item We administered the surveys to 202 previously identified workers and to a new sample of 197 developers across 171 tasks in 12 sectors (Section~\ref{sect:administering}). We then scaled up their responses by measuring LM agreement with human ratings and, under reasonable agreement, generated task-level annotations for 10{,}131 tasks across 512 occupations and 19 sectors (Section~\ref{sec:scaling}). We make these two human-generated datasets publicly available\footnote{Project website: \textbf{\url{http://social-dynamics.net/ai-impact/automating-joy/}}}.

\item Our first main contribution, addressing RQ1 (Section~\ref{sec:answer-rq1}; Figure~\ref{fig:rq1_bar}), shows that tasks identified as likely to be exposed to AI are more strongly associated by workers with novelty/creativity, positive affect, and autonomy (Figure~\ref{fig:rq1_sector_likely}). This challenges the usual narrative that automation mainly targets routine tasks~\cite{septiandri2024potential}. By contrast, tasks rated as unlikely to be exposed to AI are more often linked by workers to emotional awareness, in-person interaction, relationship building, and social connection (Figure~\ref{fig:rq1_sector_not_likely}).

\item Our second main contribution, in addressing RQ2 (Section~\ref{sec:answer-rq2}; Figure~\ref{fig:rq2_overall}), reveals gaps between workers' preferences and developers' design intentions: workers prefer straightforward systems, whereas developers intend to design polite systems (Figure~\ref{fig:rq2_overall}), consistent with reports of LM sycophancy~\cite{Danry2025Deceptive}, which is a systematic bias of LMs toward agreeing with users' views irrespective of correctness. More generally, developers favor polite, strict, and imaginative systems, whereas workers often describe these design choices as unnecessary friction or rigidity rather than genuine support. By contrast, both groups converge on the need of personalized systems.
\end{enumerate}

Our contributions and findings motivate design principles that aim to preserve meaningful work and better align with worker needs (Section~\ref{sec:discussion}), culminating in a five-part research agenda for HCI researchers (Section~\ref{sec:conclusion} and \autoref{tab:agenda_blueprint}).

\section{Related Work}

To mirror our research questions in Section~\ref{sec:introduction}, we review two strands of literature: (1) which tasks can be exposed to AI (Section~\ref{sec:ai-augmented}), and (2) how teams design AI to meet workers’ needs (Section~\ref{sec:how-teams}; RQ2). We then add a third strand on the meaningfulness of work by conducting a formal scoping review (Section~\ref{sect:scoping}; Appendix~\autoref{tab:foundational_citations}).

\subsection{Tasks exposed to AI} 
\label{sec:ai-augmented}

Across field deployments, controlled trials, and exposure analyses, three properties are often associated with AI exposure being likely: (1) the task can be decomposed into explicit steps; (2) the work can be partitioned into sub-tasks under human direction; and (3) outputs can be checked against clear rules, tests, or ground truth. These properties are common in computer-based workflows \citep{eloundou2024gpts,Felten2023Abilities}. In such settings, LMs can accelerate routine components while people supply context and judgment. For example, laboratory studies report faster completion and higher quality in settings where humans set goals and verify results~\citep{NoyZhang2023,Peng2023,Cui2024,Brynjolfsson2023Support,ChoiSchwarcz2023Law}.

However, studies also report that few tasks can be automated end-to-end, whereas many can be augmented, often within existing human workflows \citep{eloundou2024gpts,Felten2023Abilities}. Patent-to-task analyses document substantial AI exposure in skilled, non-routine domains (e.g., clinical image review, routing, programming) where professionals nonetheless retain final responsibility \citep{Webb2019,septiandri2024potential}. AI exposure is already evident across domains: in software development, AI supports debugging with error traces, proposes refactoring, generates tests, analyzes logs, and prepares code reviews~\citep{Peng2023}; in professional writing, workers use AI to outline, draft, adjust tone, convert formats, and check citations and style~\citep{NoyZhang2023, hwang2025cowriting}; and in specialist review, clinicians use AI to prepare imaging pre-reads, and draft reports for subsequent human review and sign-off~\citep{septiandri2024potential}.

Even when many tasks could be exposed to AI, empirical studies find that workers often report preferring to retain activities involving judgment, interpersonal interaction, and coordination~\citep{Autor2015augmentation,Brynjolfsson2018SML,Felten2023Abilities,eloundou2024gpts}, while accepting greater AI exposure for repetitive digital work~\citep{Autor2015augmentation,Brynjolfsson2018SML,Felten2023Abilities,eloundou2024gpts}.

\subsection{How teams design AI to meet workers' needs}
\label{sec:how-teams}
Prior work suggests that, when tasks feel more meaningful, workers often report preferring to maintain ownership; when they feel less meaningful, workers report being more willing to offload work to AI.
Workers judge AI systems typically along two dimensions: warmth (benevolent intent), and competence (ability) \citep{Fiske2007}. 
People are more willing to delegate work when they believe the AI system is competent at it \citep{erlei2024choice}.
Design choices are reported to shape these beliefs: anthropomorphic features or friendly conversational styles can make AI systems seem warmer and more capable in experimental settings~\citep{liu2024perceptions,ladak2024care}. 
When people further attribute a `mind' to AI in such studies, they tend to collaborate with it more, and also blame it more when it makes errors, with these effects amplified by human-like cues~\citep{Gray2007,Waytz2014}. HCI studies show that cues about system expertise, humanness, and fit to the setting influence perceived warmth and competence, with expertise cues most strongly predicting reported trust and use~\citep{krop2024expertise, kim2025fostering}. To complicate matters, system developers often evaluate the trustworthiness of a system differently than system users do~\citep{vereschak2024trust, li2025confidence, li2025text}.

While the warmth–competence framework advances theory, workers rarely judge AI in simple binary terms. Recent evidence suggests people judge whether AI is suitable for a job by the traits the job requires~\citep{dong2024fears}. \citet{dong2024fears} finds that traits such as fairness, sincerity, warmth, competence, determination, intelligence, tolerance, and imagination were treated as distinct dimensions in assessing whether AI is suitable for a job.
Additional studies similarly find that expert cues and context-task fit increase acceptance in workplace settings~\citep{krop2024expertise} where examples include interface features that calibrate reliance (e.g., uncertainty cues, targeted explanations), reduce over-reliance, and help people make better decisions~\citep{vasconcelos2023explanations,erlei2024choice, liang2024survey}. 
In cooperative tasks, people rely on AI more when it seems warm and competent than when it simply performs well~\citep{mckee2024cooperation}.
This suggests that AI system design should distinguish relevant traits based on roles (e.g., fairness for managers, sincerity for clinicians)~\citep{mckee2024cooperation}.
In our work, we examine which traits workers want in tasks exposed to AI, and which traits developers design for, revealing role-specific priorities (e.g., fairness for managerial tasks; sincerity for clinical tasks).

\mbox{} \\
\noindent
\textbf{Research Gap.} Prior work has identified which tasks are exposed to AI and has explored design choices that promote trust and reliability~\cite{kim2024m}. 
Yet two important questions remain underexplored. 
First, we lack understanding of how AI exposure reshapes the meaningfulness of work, and whether tasks exposed to AI feel purposeful to workers or feel merely like bureaucratic busywork. 
Second, we do not know whether teams design AI systems with the traits that workers actually want.
Related work in high-stakes domains (e.g., judiciary) has begun to surface user expectations and requirements for AI tools~\cite{solovey2025interacting}, but little is known about how these expectations translate to everyday work practices. 
In the pursuit of productivity and speed, with the introduction of AI, teams may risk changes to autonomy, care for others, excellence, and fairness~\cite{hazra2025ai}, qualities that prior work links to experiences of meaningful work.

\section{Methods}
\label{sec:methods}

To close this gap, we addressed two research questions (RQs): 
\begin{enumerate}
  \item[] (RQ1) Which dimensions of meaningful work do workers associate with tasks exposed to AI in their daily work?
  \item[] (RQ2) Do teams design AI systems with traits that align with the traits workers want?
\end{enumerate}

To answer these two questions, our methodology followed four steps (\autoref{fig:pitchfig}). First, we selected representative tasks from the O*NET 29.3 database\footnote{\url{https://www.onetcenter.org/database.html} accessed July 2025}, and recruited samples of workers and AI developers on the crowd-sourcing platform of Prolific. We initially identified 171 tasks spanning 22 representative occupations (Section~\ref{sect:selecting-tasks}). 
By ``representative'', we mean that the task sample was stratified to approximate the distribution of occupational sectors reported by the U.S. Bureau of Labor Statistics, ending up with modest deviations (Figure~\ref{fig:task_dist}). Second, we  conducted a scoping review synthesizing research on the meaningfulness of work (Section~\ref{sect:scoping}). Guided by this review, we measured workers' experiences using items that capture the meaningfulness of work (Q1--Q33; Section~\ref{sect:q1-33}).
We then measured workers' and developers' views on the design of AI systems using items capturing design traits (Q34--Q45; Section~\ref{sect:q34-45}). 
Third, we administered the survey to workers and developers under consistent protocols on Prolific (Section~\ref{sect:administering}). 
Fourth, to scale the analysis, we used LMs to simulate workers and developers, generated task-level annotations for 10{,}131 tasks across 512 occupations and 19 sectors, and validated the model-derived annotations against our human data (Section~\ref{sec:scaling}).

\subsection{Selecting Tasks and Recruiting Workers and Developers}
\label{sect:selecting-tasks}

\noindent
\textbf{O*NET 29.3 Database.} The O*NET database provides standardized information on U.S. sectors\footnote{Sectors refers to O*NET major groups \url{https://www.onetcenter.org/taxonomy/2019/structure.html}.}, occupations and their associated tasks. 
It organizes 923 occupations into sectors where each occupation is broken down into task statements\footnote{\url{https://www.onetcenter.org/dictionary/29.3/excel/task_statements.html}} describing work activities. 
In total, O*NET contains 18{,}796 tasks, where each task is classified as \textit{core} or \textit{supplementary}, and annotated with how frequently it's performed (e.g., yearly or less, monthly, weekly, daily, or hourly)\footnote{\url{https://www.onetcenter.org/dictionary/29.3/excel/task_ratings.html}}.

\mbox{ } \\
\noindent
\textbf{Selecting Tasks Exposed to AI.} Our study primarily focuses on workplace tasks likely to be exposed to AI. 
We follow prior work~\cite{shao2025future}, and apply a multi-step filtering pipeline (Step 1; \autoref{fig:pitchfig}) to identify a representative set of tasks for our study. 
First, we ensured each task is classified as either \textit{core} or \textit{supplementary} to ensure its relevance to the occupation. 
Then, we characterized and filtered occupations and tasks by two criteria as determined by GPT-4o annotations in line with \cite{shao2025future}: (1) the \textit{occupation} primarily involves computer use and, (2) the \textit{task} can be completed on a computer. 
Upon manual inspection, we found that GPT-4o occasionally excluded occupations (e.g., nursing, education professionals) that are widely recognized as exposed to AI~\cite{rony2024advancing, robert2019artificial, rahm2023education}.
To ensure these occupations were represented in our dataset, we manually included a list of 427 occupations that were exempted from these filters (examples in Appendix \autoref{tab:occupations_not_filtered}). 
Following these filtering steps, our dataset contained 10{,}131 tasks spanning 512 occupations.

\mbox{ } \\
\noindent
\textbf{Recruiting Workers and Selecting Tasks Familiar to Them.} We used Prolific to recruit U.S. workers and developers (Appendix Table~\ref{tab:prolific-demographics}). We applied the ``work function'' screener to identify participants likely to be familiar with the tasks they evaluated. Both workers and developers were compensated at a rate of \$11 per hour. Each O*NET occupation was mapped to one of 21 work functions (Appendix \autoref{tab:prolific_jobs_mapping}), yielding 4,473 tasks across 209 occupations. To keep surveys tractable and reduce participant fatigue, we downsampled to at most 10 tasks per occupation using a greedy selection criterion based on task frequency annotations (e.g., daily) and estimated AI exposure~\cite{septiandri2024potential}, resulting in 2{,}078 tasks across 209 occupations.  Our study required participants who were experts in their domains, which made recruitment especially challenging: if participants are not familiar with the tasks they are asked to evaluate, their responses risk being speculative rather than grounded in real-world practice.
To target U.S. workers who are experts in their domains, we first ran a preliminary survey which identified workers that: (1) belonged to one of 21 professional work functions on Prolific, (2) passed attention checks, and (3) reported being highly familiar with at least one O*NET task in their occupation. 
Following this rigorous pre-screening, we recruited 202 workers, excluding those who failed our attention checks.  From these responses, we retained tasks rated as ``Very familiar'' or ``Extremely familiar'' by at least three participants, producing a final set of 171 tasks across 22 occupations and 12 sectors. This subset provided broad coverage (12 of 22 total sectors in O*NET), while focusing on tasks that were central to the occupation (i.e., core or frequently performed), likely to be exposed to AI, and validated as highly familiar to workers.

\mbox{ } \\
\noindent
\textbf{Recruiting Developers.}  For developers, we focused on U.S.-based  AI practitioners who met the following screening criteria: weekly AI use (ranging from once a week to multiple times daily), individual contributor or non-manager role, employment in coding, technical writing, or systems administration, and primary function in engineering (e.g., software) or research. 
We recruited 197 developers who each rated 10 tasks, drawn from the same pool of tasks the workers rated. Each task received ratings from at least three distinct developers.
Given the heterogeneity of ``AI developers'' on Prolific and the ambiguity of O*NET task descriptions that allows for interpretive variation, we detail the distribution of developers’ technical roles, AI usage patterns, and work functions in Appendix Figure~\ref{fig:developers_representation}. We found that the majority of developers in our sample hold software, data, IT infrastructure, or ML/AI engineering roles. Developers also reported extensive use of AI tools in their workflow, including LM assistants, code-generation tools, and data-analysis systems, indicating active engagement with contemporary AI technologies. Work-function distributions further show representation across engineering, research, analytics, and operations roles.
\vspace{-5pt}
\subsection{Scoping Review on Meaningful Work} 
\label{sect:scoping}

After selecting tasks and recruiting workers, we next determined which questions would best capture the extent to which workers perceive their tasks as ``meaningful''. 
To ground these questions in the literature, we conducted a scoping review following the five-stage framework in \cite{arksey2005}:

\mbox{ } \\
\noindent
\textbf{Step 1. Identifying the research question.} The main research question was: \emph{What are the documented, theorized, or studied dimensions of task meaningfulness, symbolic work, impression management, and status signaling in today's workplaces?} This question covers both personal views of meaningfulness, and the social or symbolic factors that can make work performative, strategic, or status-driven.

\mbox{ } \\
\noindent
\textbf{Step 2. Identifying relevant literature.} To ensure coverage across disciplines, we went on Google Scholar and JSTOR, and we used the Boolean search string: {\small\texttt{(``meaningful work'' OR ``task significance'' OR ``work motivation''  OR ``impression management'' OR ``status threat'' OR ``symbolic work'' OR ``performative work'') AND (work OR job OR employee OR organization OR labor)}}. 
The review included only English-language, peer-reviewed work, with no date restrictions. 


\mbox{ } \\
\noindent
\textbf{Step 3. Selecting the articles.} We used the following inclusion rules: articles must discuss task meaningfulness, symbolic or performed work, status signaling, or impression management in a work or organizational setting; both empirical and theoretical work was eligible; and full-text access had to be available. 
Articles outside of work or organizational settings, and those limited to consumer behavior or marketing without reference to employees or task meaning, were excluded, resulting in 56 articles.
After removing duplicates and screening titles and abstracts, 42 remained for full-text review, and we included 21 that met all criteria.

\mbox{ } \\
\noindent
\textbf{Step 4. Charting the data.} We coded each article with a structured form. Key fields included: main constructs (e.g., task significance, performative work);  theory used (e.g., Job Characteristics Model, Institutional Theory, Impression Management Theory); measures (e.g., Work and Meaning Inventory); and main findings.

\mbox{ } \\
\noindent
\textbf{Step 5. Collating, summarizing, and reporting the results.} We reviewed 21 articles across psychology, sociology, anthropology, and ethics on what work means to people and to society (Appendix \autoref{tab:foundational_citations}). The studies show how people judge their own work, how organizations shape those judgments, and how societies value different kinds of work. Examples include David Graeber's critique of ``bullshit jobs''~\citep{graeber2018}, a tested questionnaire for meaningful work \citep{steger2012}, and research linking work to identity and status \citep{rafaeli2006}. Key theoretical lenses include job characteristics theory, which emphasizes task significance, task identity, and autonomy~\citep{hackman1980}; impression management, which explains how workers perform roles for symbolic or strategic ends~\citep{bolino2008,goffman1959}; institutional theory, which highlights symbolic work and routine, ceremonial task structures~\citep{meyer1977}; and models of status signaling at work~\citep{bellezza2017,pettit2010}.

Guided by our coding, we grouped articles into two analytic levels capturing the primary sources of experienced meaningfulness:

\begin{description}
  \item[\textbf{Micro level (individual appraisal):}] Articles that tied meaningful work to how people judged their tasks and roles. Findings included task-level features and attitudes such as satisfaction, engagement, motivation, and performance \citep{hackman1980,steger2012}.
  \item[\textbf{Macro level (organizations, institutions, and society):}]\hfill\break Articles that explained how society, fields, and organizations ranked different kinds of work and set norms that shaped meaning, identity, and claims to self-worthiness \citep{bellezza2017,pettit2010,rafaeli2006,bolino2008,goffman1959,meyer1977}. This set also reported cases where tasks lacked recognized value and thus felt meaningless \citep{graeber2018}.
\end{description}

At the micro level, individuals treated meaningful work as an attitude tied to satisfaction, engagement, motivation, and performance. At the macro level, cultural valuations and organizational scripts shape identity and sense making, establishing the norms and constraints that enabled or limited those appraisals. 

Prior work has combined insights from the micro and macro levels to explain how individual experiences are shaped by broader organizational and cultural forces. For example, \citet{Carton2018Moon}'s study of the National Aeronautics and Space Administration (NASA) in the 1960s argues that macro-level leader ``sense-giving'' can recalibrate micro-level experiences. He illustrates how leaders may reshape how people view their work through mid-level links.
Using President Kennedy as an example, the study describes leaders as defining a main aim (advancing science), setting a dated goal (`land a man on the Moon before 1970'), outlining a few key steps (Mercury, then Gemini, then Apollo; later a six-step plan sometimes called the `ladder to the Moon'), and using clear language that ties the goal to shared values such as knowledge and peace. 
In Carton's account, this plan made the ultimate goal feel more attainable, gave staff clearer stepping stones, clarified their perceived role in the process, and was associated with staff describing daily tasks in mission terms (`putting a man on the Moon', even `advancing science'). Carton interprets these shifts as aligning motivation, engagement, and performance more closely with the organization's purpose. In a complementary line of work, \citet{bailey25} integrate psychological and sociological perspectives on how people find work meaningful. They draw on first-person accounts from nurses, creative artists, and lawyers: occupations chosen for their clear contrasts in task content, workplace rules, and room for professional choice, and emphasize the connection between individual experiences and broader organizational purpose for understanding how people experience meaning at work.

\subsection{Questions about Dimensions of Meaningful Work to Workers (Q1-33)}
\label{sect:q1-33}

Based on our scoping review, we therefore started with two levels: \textit{individual appraisal} (micro), and \textit{organizational, institutional, and societal valuation} (macro). To measure valuation beyond the individual, we drafted \textit{Perceived Bullsh*tness} survey items (Q1--Q5; \citep{graeber2013phenomenon}) and \textit{Status Maintenance} survey items (Q11--Q16; \citep{bolino2008multi,bellezza2017conspicuous}). To measure individual appraisal, we drew on \textit{Perceived Value} (Q6--Q10; \citep{hackman1976motivation,steger2012measuring}), the \textit{EPOCH well-being scale} (Q17--Q21; \citep{loaiza2024epoch}), and \textit{Human Flourishing} (Q22--Q33; \citep{vanderweele2017promotion}). See Step 2 in \autoref{fig:pitchfig} and Appendix \autoref{tab:survey_items} for our survey items.

\mbox{ } \\
\noindent
\textbf{Perceived Bullsh*tness~\cite{graeber2013phenomenon} (Q1-5).} These five questions measure the extent to which participants view their tasks as pointless, bureaucratic, or not contributing to the goals of their organization. 
Example survey items include: `I perform this task only to satisfy bureaucracy or appearances', and `This task does not contribute to the goals of my organization'. 
These items build on \citet{graeber2013phenomenon}'s theory of `bullshit jobs', which introduces the concept of `bullshit' jobs as roles that are perceived as worthless, even by those performing them. 
\citet{graeber2013phenomenon} argues that these roles or tasks can contribute to psychological distress and can erode workers' sense of purpose. 

\mbox{ } \\
\noindent
\textbf{Perceived Value~\cite{hackman1976motivation, steger2012measuring} (Q6-10).} These questions assess the extent to which workers perceive a task as meaningful or contributing to the success of their organization. This aligns with the three psychological states (e.g. experienced meaningfulness, experienced responsibility for the outcomes of the work, and knowledge of the results) described in \citet{hackman1976motivation}'s Job Characteristics Model. Specifically, this outcome is observed, if an individual `\textit{learns} (knowledge of results) that he \textit{personally} (experienced responsibility) has performed well on a task that \textit{he cares about} (experienced meaningfulness)'~\cite{hackman1976motivation}. Our survey items reflect this framework by assessing whether workers feel they `receive useful feedback about how well this task is done.' (Q9; knowledge of results), `has the freedom to decide how to carry out this task' (Q8; experienced responsibility), and `provides a sense of accomplishment' (Q10; experienced meaningfulness). 
This framework is further supported by prior work on meaningful work~\cite{steger2012measuring}, which shows that seeing one's work as contributing to a greater good is associated with higher well-being and job satisfaction (Q6: `This task is important to the success of my team or organization').

\mbox{ } \\
\noindent
\textbf{Status Maintenance~\cite{bolino2008multi, bellezza2017conspicuous} (Q11-16).} These questions assess the extent to which workers continue performing a task to preserve their professional standing, visibility, and perceived competence. Example survey items include `I feel this task signals to others that I am busy or valuable', and `I worry that letting go of this task could reduce my influence or visibility'. These items are derived from prior work on impression management motives~\cite{bolino2008multi} where employees engage in behaviors intended to influence how others perceive their abilities, dedication, or value to the organization. Furthermore, related work from consumer research~\cite{bellezza2017conspicuous} emphasizes how lack of leisure time and `busyness' serve as status symbols, and they show how individuals may continue to pursue low-value tasks that make workers appear `busy' because these tasks serve as signals of competence and provide visibility, even if they contribute little to core performance outcomes.

\mbox{ } \\
\noindent
\textbf{EPOCH~\cite{loaiza2024epoch} (Q17-21).} These questions capture the extent to which the task requires fundamental human capabilities that prior work argues enable workers to excel in areas where AI is less likely to succeed. Drawing on the EPOCH framework~\cite{loaiza2024epoch}, we include items reflecting five dimensions that are particularly challenging to expose to AI: (1) empathy and emotional intelligence (`This task requires recognizing and responding appropriately to the emotions of others'), (2) presence, networking and connectedness (`This task benefits significantly from in-person interaction, non-verbal cues, or spontaneous communication'), (3) opinion, judgment and ethics (`This task involves making decisions that require moral reasoning, accountability, or subjective judgment'), (4) creativity and imagination (`This task requires generating novel ideas, approaches, or solutions beyond standard procedures'), and (5) hope, vision and leadership (`This task involves setting direction, motivating others, or showing perseverance toward a long-term goal').

\mbox{ } \\
\noindent
\textbf{Human Flourishing at Work~\cite{vanderweele2017promotion} (Q22-33).} These questions were adapted from ~\citet{vanderweele2017promotion}’s multidimensional framework of flourishing. 
This framework conceptualizes flourishing as encompassing domains beyond immediate job performance, including well-being, purpose, and social connection. We included survey items to capture six domains: (1) Happiness and life satisfaction (e.g., `How much would this task make you feel satisfied or content with your work?'), (2) Mental and physical health (e.g., `To what extent would this task support your mental health?'), (3) Meaning and purpose (e.g., `To what extent would this task feel meaningful or worthwhile?'), (4) Character and virtue (e.g., `To what extent would this task allow you to act in accordance with your values or integrity?'), (5) Close social relationships (e.g., `To what extent would this task help you build or strengthen relationships with colleagues or clients?'), and (6) Financial and material stability (e.g., `To what extent would this task contribute to your sense of job or financial security?').

\subsection{Questions about AI Design Choices for AI Exposure to both Workers and Developers (Q34-45)}
\label{sect:q34-45}

Even when workers wish to use AI, prior studies show that AI tools often fail to meet their needs due to limited understanding of which psychological traits workers expect AI systems to exhibit~\cite{shao2025future, dong2024fears}. Adoption and acceptance of AI technologies depend on the extent to which these systems align with user values and expectations~\cite{van2013can}. Yet, value alignment is dynamic: values emphasized at the design stage often diverge from those prioritized by users once technologies are deployed in real contexts~\cite{karizat2024patent}. Prior work has shown that values essential to effective task performance (e.g., empathy, fairness, creativity) are rarely embedded into the design of AI systems~\cite{loaiza2024epoch}. To investigate this gap, we assessed workers' preferences for how an AI system should behave, if their tasks were exposed to it. Specifically, 
Questions 34--45 (Step 3 in \autoref{fig:pitchfig}, and Appendix~\autoref{tab:survey_items}) asked participants to rate the importance of twelve traits for an AI system to exhibit: four traits that we introduced (Q34--Q37: creativity, empathy, explainability, and openness to challenge) based on the HCI literature~\cite{norman2019human,shneiderman2020human,amershi2019guidelines}, and eight psychological traits taken from \citet{dong2024fears} (Q38--Q45: fair, warm, sincere, tolerant, competent, determined, intelligent, and imaginative).

\subsection{Administering Questions to Workers and Developers}
\label{sect:administering}
We administered two complementary surveys: one to workers, and the other to developers. The worker survey was designed to capture perceptions of workplace tasks and preferences for AI system behavior and consisted of Questions 1–45. The developer survey was designed to only capture preferences for AI system behavior and, thus, consisted of Questions 34–45 (Appendix \autoref{tab:survey_items}). Each item was rated on a 5-point Likert scale (1 = Strongly disagree, 5 = Strongly agree). Both surveys concluded with the Human Agency Scale (HAS)~\cite{shao2025future} (e.g., Q48), which assessed desired levels of human–AI collaboration. 
The scale ranges across varying degrees of human involvement: from AI agent drives task completion (HAS H1–H2), to equal partnership (HAS H3), to human drives task completion (HAS H4–H5).
In the survey, workers were asked: `\emph{If AI were to assist in this task, how much of your collaboration would be needed to complete this task effectively}'? 
Response options reflected the five HAS categories, allowing us to examine worker preferences for human intervention and to contrast these with the intervention priorities of developers.

\subsection{Scaling and Validating Responses with Language Models}
\label{sec:scaling}
  
To enable larger-scale analysis, we evaluate whether LMs can act as annotation assistants to simulate the distribution of worker and developer responses.
Recent work has demonstrated the promising potential of LMs as annotation assistants in social science settings~\cite{rytting2023towards, ranjit2024oath, ranjit2025supporting} where LMs can approximate human judgments in large scale surveys~\cite{anthis2025llm}. 

We used in-context learning with GPT-4o, applying chain-of-thought prompting~\cite{wei2022chain} to adopt the persona of either a worker or a developer for a given occupation. This procedure is briefly described in Step 4 of \autoref{fig:pitchfig}, and the prompts are fully detailed in Appendix Tables \ref{tab:worker_prompts} and \ref{tab:developer_prompts}. 
Our approach aligns with prior work on LM-as-an-Expert prompting~\cite{xu2023expertprompting, hu2024quantifying, moon2024virtual}, which has been validated as a method for eliciting domain-specific expertise from LMs.
We then assessed the external validity of our findings based on LM-generated annotations (i.e., whether the patterns we report are consistent with and generalize to human judgments) using three complementary strategies:

\begin{enumerate}
\item \textbf{Testing whether incorporating the LM as an additional annotator improved inter-rater agreement.} To evaluate the reliability of human annotations, we first assessed the reliability using only human annotators (`experts') by calculating intra-class correlation coefficients (ICC) and mean absolute differences (MAD) at the task level for items with at least three expert ratings. 
For the worker survey, experts alone achieved a mean ICC of 0.634 (moderate-to-good agreement; 95\% CI = [0.602, 0.664]). Adding the LM as an additional annotator increased the mean ICC to 0.742 (good agreement; 95\% CI = [0.722, 0.760]), a statistically significant improvement of +0.108 (95\% CI = [0.093, 0.124]). 
For developer responses, experts alone achieved a mean ICC of 0.629 (moderate agreement; 95\% CI = [0.578, 0.676]). Adding the LM increased the mean ICC to 0.673 (moderate agreement; 95\% CI = [0.627, 0.713]), representing a statistically significant improvement of +0.044 (95\% CI = [0.032, 0.058]).
For workers, LM ratings differed from experts by about 1.10 Likert points (normalized MAD = 0.276), only slightly higher than expert–expert disagreement (0.255). For developers, LM–expert differences averaged 1.31 points (normalized MAD = 0.328), which was not statistically significantly different from expert–expert disagreement (0.324).
To further assess whether ICC improvements reflected genuine alignment rather than artifacts of increased rating stability, we conducted a robustness analysis by comparing the real LM's contribution to that of a randomized version that preserved its overall rating distribution but no longer reflected task-level correspondence with human ratings, by randomly shuffling its existing ratings. We then repeatedly added this randomized LM to the human ratings and recomputed ICC across 1{,}000 bootstrap samples. Adding the randomized LM yielded modest ICC increases (+0.079 for workers; +0.016 for developers), showing that some improvement comes from variance stabilization. However, these increases were significantly smaller than those from the real LM (+0.108 for workers; +0.044 for developers; \emph{p} < .0001). This indicates that the ICC gains arise from substantive alignment between the LM and human annotators, not statistical artifacts.

\item \textbf{Comparing annotation distributions between LM and human raters, supplemented with qualitative analyses.} 
Across the dimensions of meaningful work, LM ratings closely tracked human ratings, with only minor differences in distributional shape (Appendix Figure \ref{fig:trait_distributions}). In addition to computing global reliability indices, we examined the percentage of large discrepancies, defined as cases where human and LM ratings differed by two or more points on the 5-point Likert scale. Whereas a one-point difference can be attributed to normal subjectivity, a gap of two points or more represents a substantive divergence in interpretation~\cite{cicchetti1994}. Identifying such items provides a diagnostic lens on the alignment of LM outputs with human ratings, revealing systematic areas of disagreement that global coefficients such as ICC may obscure. We find that only 0.87\% of tasks exhibit such significant discrepancies. Across dimensions, the \textit{EPOCH} questions showed the highest divergence at 2.92\%, followed by \textit{AI Design Choices} questions at 1.75\%. All remaining dimensions (aside from a few AI-trait questions) had negligible discrepancy rates, each below 1\%. In these few cases, we observed that both LM and human interpretations were reasonable when viewed from different contextual perspectives (Appendix Table \ref{tab:lm_human_full_examples}). For example, the LM perceived the task `\emph{review, classify, and record survey data in preparation for computer analysis}' as a routine procedure with minimal emotional demands, whereas a human annotator emphasized its creative and moral judgment aspects. Crucially, these divergences are best understood as subjective differences rather than systematic bias, and, as we show in our third strategy, they do not affect the substantive results when analyses are replicated using only human ratings.

\item \textbf{Replicating the main analyses using only human ratings.}  We replicated the core analyses that we will present in  Section~\ref{sec:results} using only human ratings. The replication yielded results that were highly consistent with those based on LM-generated ratings across both research questions, thereby providing a direct test of robustness. 
For RQ1 (Appendix Figure \ref{fig:rq1_human_ratings}), humans and LMs agreed on 6 of the 7 meaningful-work dimensions that differentiated tasks likely to be exposed to AI from those unlikely to be exposed. The only exception was `\emph{requires novel ideas or creativity}'.
For RQ2, we conducted a head-to-head comparison between human ratings and LM-simulated ratings on our subset of 171 tasks. As shown in Appendix Tables~\ref{tab:rq2_trait_human_ratings} and ~\ref{tab:rq2_trait_lm_ratings}, the pattern of worker–developer misalignment produced by human raters closely tracks the pattern produced by LM-simulated raters, with consistent rankings across high-, mixed-, and aligned traits. 
The strongest pattern was workers emphasizing straightforward traits, and developers emphasizing politeness traits, which was reproduced almost exactly in human-only ratings. Divergences appeared only in secondary dimensions (e.g., ‘handle complex \emph{vs.} routine work’, ‘precise \emph{vs.} simple'). Manual review suggests this reflects sectoral biases introduced by the smaller, occupation-concentrated human sub-sample rather than any substantive difference in trait interpretation. Importantly, none of these divergences alter the main inferences: the central dimensions of meaningful work, and the dominant sources of worker–developer disagreement are consistent across LM and human annotations.

\end{enumerate}

\begin{figure*}[t]
    \centering
    \includegraphics[width=0.8\linewidth]{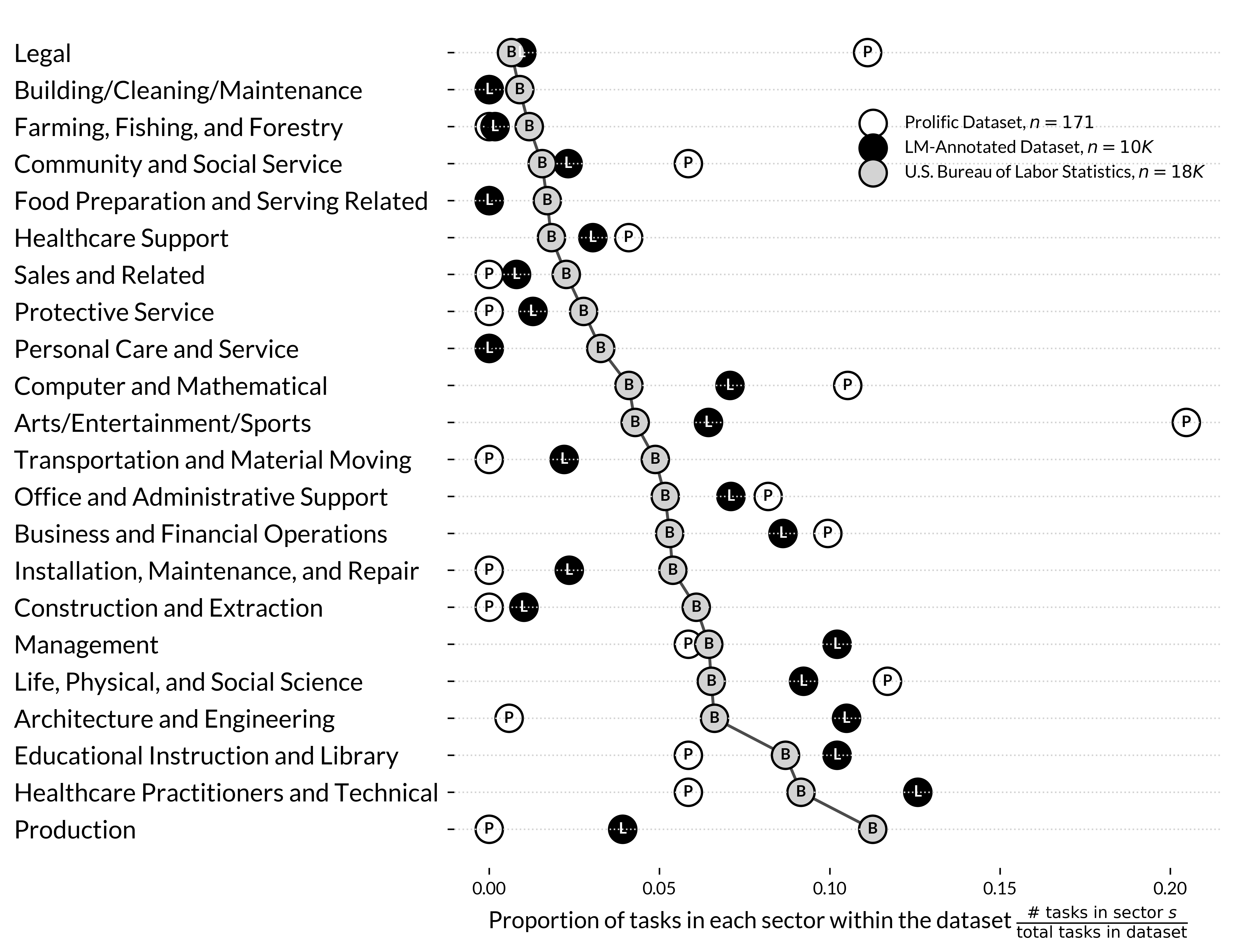}
    \caption{Proportions of tasks across occupational sectors in three datasets: Prolific (n=171), LM-annotated (n=10K), and U.S. labor statistics (n=18K). The Prolific sample covers 12 sectors (based on available occupations), the LM-annotated dataset covers 19 sectors, and the U.S. Bureau of Labor Statistics dataset covers all sectors. The distributions of Prolific and LM-annotated tasks are broadly similar across sectors, indicating that LM annotations capture sector patterns consistent with U.S. labor statistics.
    }
  \Description{Scatterplot comparing the proportion of tasks in 20 occupational sectors across three datasets: Prolific (white circles, 171 tasks), LM-annotated (black circles, 10,000 tasks), and U.S. Bureau of Labor Statistics (gray circles, 18,000 tasks). Each row corresponds to a sector, such as Legal, Healthcare Support, or Architecture and Engineering. Prolific and LM-annotated datasets show broadly overlapping distributions, though Prolific has fewer covered sectors. The labor statistics dataset spans all sectors and provides the baseline distribution.}

    \label{fig:task_dist}
\end{figure*}

These results indicate that, in our setting, GPT-4o can serve as a reliable additional annotator for both worker and developer perspectives without reducing inter-rater reliability. 
We therefore used LM-generated ratings to scale our analysis to the full set of 10K O*NET tasks, spanning all 19 occupational sectors. The LM-annotated dataset provides broader coverage than the worker and developer surveys, with task distributions more closely aligned with the Bureau of Labor Statistics (Figure \ref{fig:task_dist}), supporting large-scale analysis of workforce patterns.

\section{Results}
\label{sec:results}
\begin{figure*}[t]
  \centering
  \includegraphics[width=0.9\linewidth]{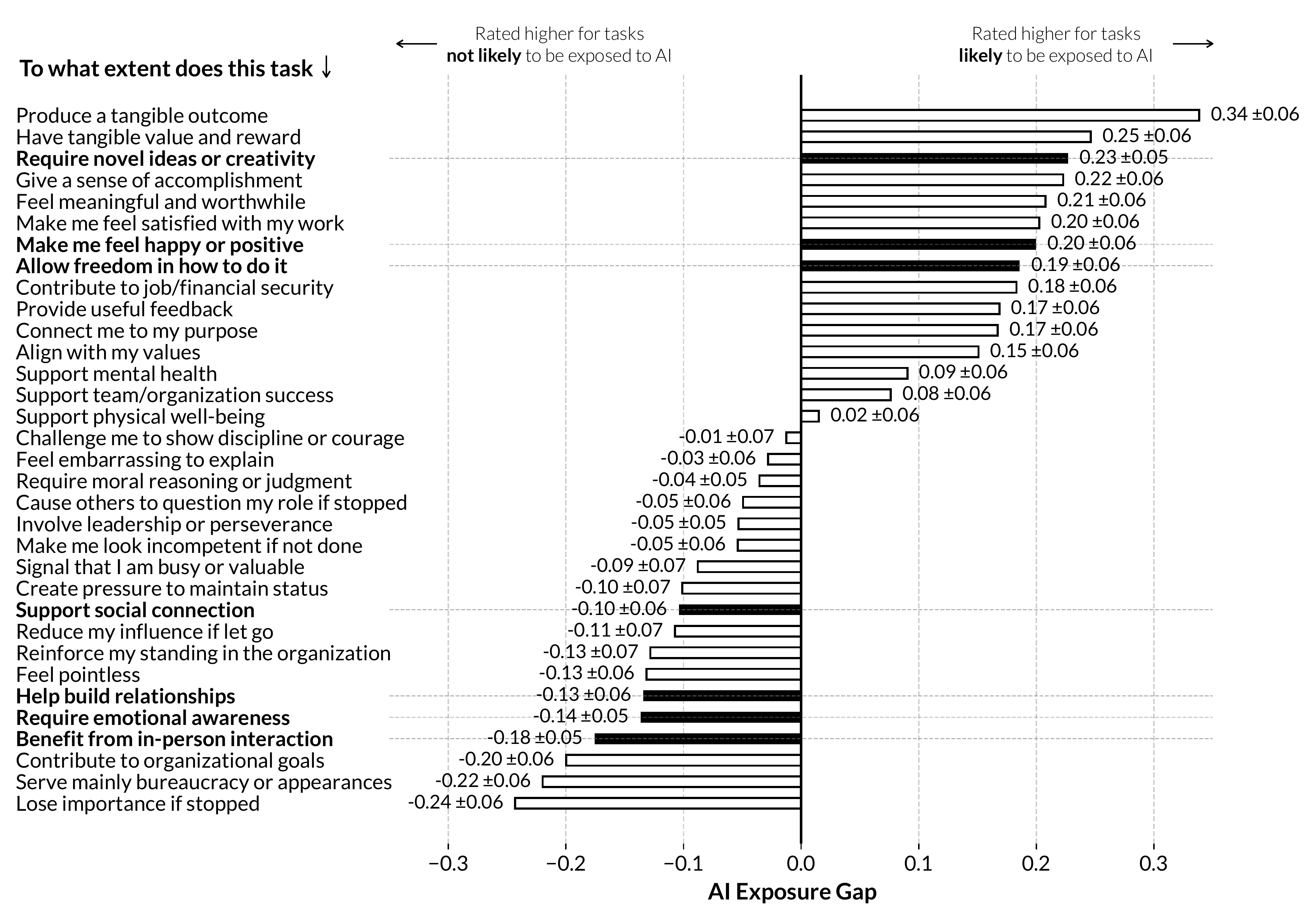}
  \caption{AI Exposure Gap by dimensions of meaningful work (rows). The higher the gap, the more strongly that dimension is associated with tasks likely to be exposed AI. This gap is computed as the difference of how important a dimension is between two groups of tasks: those that are more likely to be exposed to AI, and those less likely.  We estimate the gaps and 95\% confidence intervals with mixed-effects models. Bold names and corresponding black bars indicate differences that are statistically significant. Tasks rated likely to be exposed tend to involve novelty, creativity, happiness, and freedom in how workers do them. Tasks rated not likely tend to involve emotional awareness, in-person interaction, building relationships, and supporting social connection.
  }
  \Description{Horizontal bar chart showing AI exposure gaps across 34 dimensions of meaningful work. Bars extend rightward for dimensions associated with tasks more likely to be exposed to AI, including novelty, creativity, happiness, and freedom in how tasks are performed. Bars extend leftward for dimensions associated with tasks less likely to be exposed, including emotional awareness, in-person interaction, relationships, and social connection. Statistically significant dimensions are bolded with black bars. Each bar is labeled with the effect size and standard error.}
\label{fig:rq1_bar}
\end{figure*}

\begin{figure*}[t]
    \centering
    \includegraphics[width=0.8\linewidth]{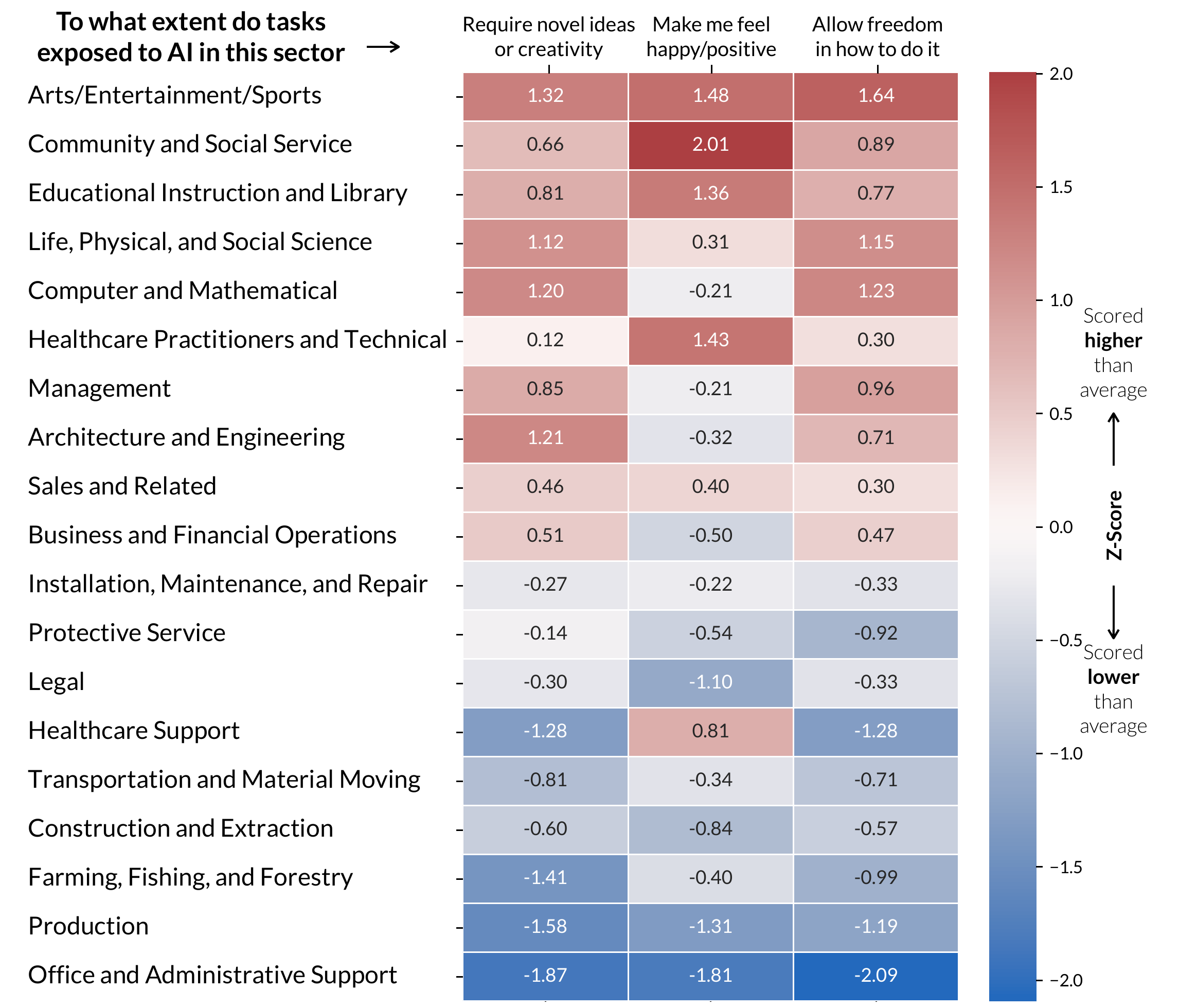}
    \caption{Association of  tasks exposed to AI in each of the sectors (in the \emph{rows}) with subset of three dimensions of meaningful work (creativity, positive affect, and autonomy in the \emph{columns}). Sectors are sorted by the average $z$-score across the three dimensions. Creative and socially-oriented sectors (arts, community service, education, life sciences) are associated with tasks exposed to AI that emphasize novelty, positivity, and freedom. In contrast, routine and manual sectors (office support, production, farming) score much lower.  
    }
      \Description{Heatmap showing 19 occupational sectors (rows) and three task dimensions (columns: creativity, positive affect, autonomy). Higher z-scores are shaded red, lower z-scores shaded blue. Arts, community service, education, and science sectors score high across all three dimensions, while office support, production, farming, and healthcare support score lowest. Color scale ranges from -2 to +2.}

    \label{fig:rq1_sector_likely}
\end{figure*}

\begin{figure*}[t]
    \centering
    \includegraphics[width=0.8\linewidth]{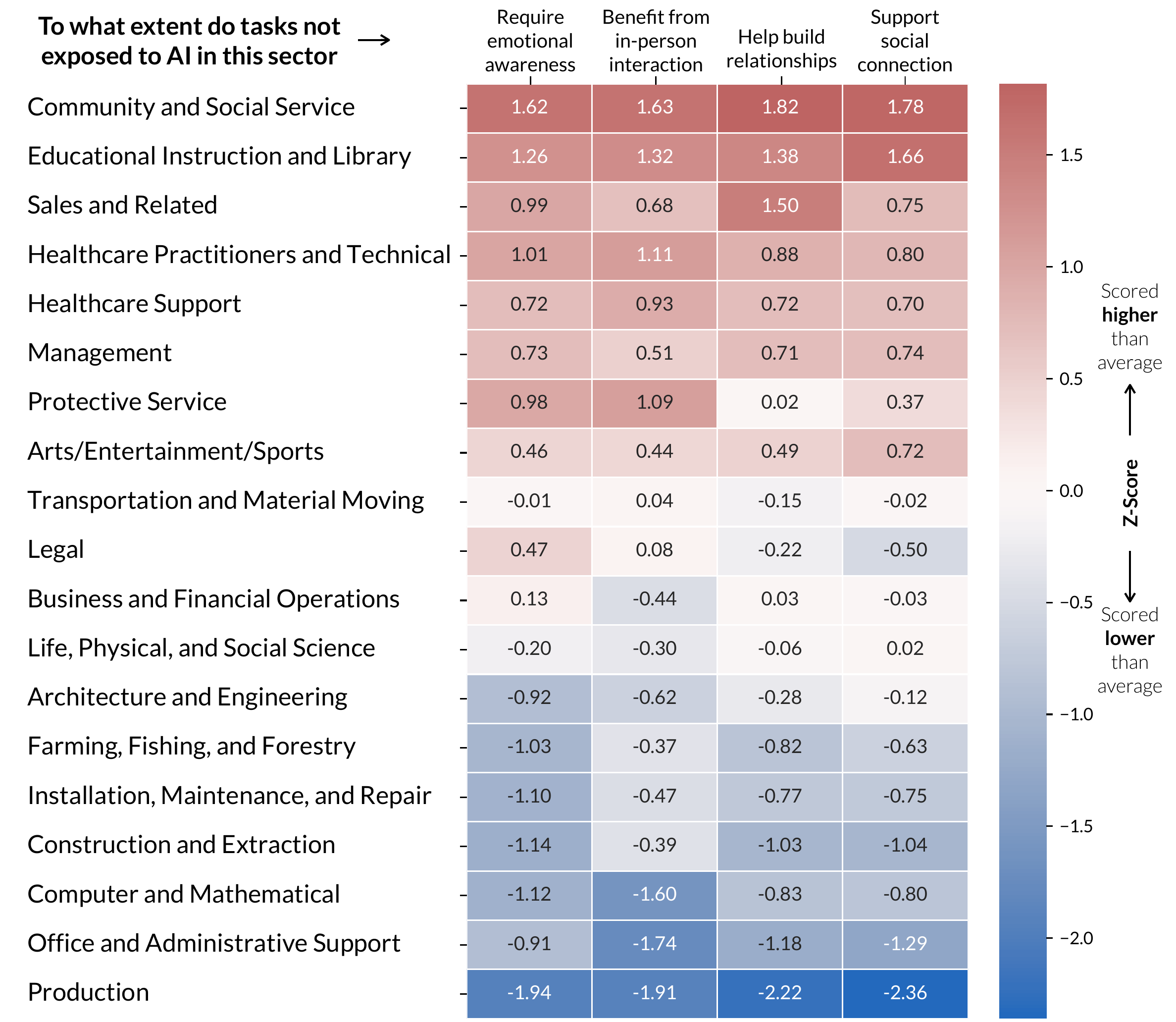}
    \caption{Association of tasks not exposed to AI in each of the sectors (in the \emph{rows}) with subset of four dimensions of meaningful work (emotional awareness, in-person interaction, relationship building, and social connections in the \emph{columns}). Sectors are sorted by the average $z$-score across the subset of four dimensions. Human-facing sectors such as community and social service, education, and healthcare consider their tasks not exposed to AI to  emphasize emotional awareness, in-person interaction, and social connection. In contrast, technical and routine sectors (e.g., production, office support, computer and mathematical) score far lower, indicating that workers in these sectors view tasks not exposed to AI as less socially or emotionally significant. 
    }
      \Description{Heatmap showing 19 occupational sectors (rows) and four task dimensions (columns: emotional awareness, in-person interaction, relationship building, social connection). Higher z-scores are shaded red, lower z-scores shaded blue. Community service, education, healthcare, and protective service rank highest across all dimensions. Routine and technical sectors including production, office support, computer/mathematical, and installation/repair rank lowest, with strong negative z-scores. Color scale ranges from -2.5 to +2.0.}

    \label{fig:rq1_sector_not_likely}
\end{figure*}

Before presenting the results in depth, we provide a brief overview with references to the sections where each finding is discussed. In summary, we observed that:

\begin{enumerate}
\item \emph{Creative and high-agency tasks are more exposed.} Across sectors, tasks in the likely-to-be-exposed group tend to emphasize creativity, positive affect, and autonomy (Section~\ref{sec:answer-rq1}). Sectors with higher scores on these traits include Arts, Architecture \& Engineering, Computer \& Mathematics, and Life, Physical, \& Social Science (Figure~\ref{fig:rq1_sector_likely}). This pattern contrasts with the familiar narrative that automation will primarily absorb routine tasks, freeing workers to concentrate on higher-value activities such as strategy and design. Our results suggest a more complex trajectory. Generative systems are already used to draft text, suggest layouts, start campaigns, run simulations, and infer likely emotional responses. In our data, these uses are associated with tasks that workers describe as meaningful because they reflect taste, judgment, and authorship. Rather than being confined to low-level chores, AI systems are increasingly entangled with how people add value to work: from generating first ideas to editing, selecting, and retaining accountability. \vspace{3pt}

\item \emph{Social and face-to-face tasks are less exposed.} Tasks rated as not likely to be exposed depend more on emotion, in-person contact, social ties, and support (Section~\ref{sec:answer-rq1}). These cluster in Community \& Social Service, Education, Healthcare, and Sales (Figure~\ref{fig:rq1_sector_not_likely}). \vspace{3pt} 

\item \emph{Worker–Developer misalignment.} We finally found systematic misalignment between how workers want AI systems to behave and how developers intend to design them (Section~\ref{sec:answer-rq2}). Developers tend to emphasize politeness, strictness, and imagination, especially in high-stakes or highly structured domains, whereas workers often describe such traits as sources of delay or rigidity rather than support.
\end{enumerate}

\subsection{Which dimensions of meaningful work do workers associate with tasks exposed to AI? (RQ1)}
\label{sec:answer-rq1}
Our goal in RQ1 is to examine whether tasks that are more likely to be exposed to AI differ systematically in the significance they hold for workers. Do they call for novel ideas? Are they associated with feelings of agency? Do workers link them to relationship building or emotional awareness?

To test this, we partitioned our tasks into likely-to-be-exposed \emph{vs.} not-likely-to-be-exposed groups, and restrict the sample to computer-based occupations~\cite{shao2025future}.
We then fit item-wise linear mixed-effects models on the 33 worker survey items (Q1–Q33), using exposure likelihood as a fixed effect and random intercepts for sector and occupation. 
We first identify which dimensions of meaningful work are disproportionately exposed to AI, then examine sector-level patterns, and provide task examples that align with statistically significant dimensions of meaningful work. 

\mbox{ } \\
\noindent 
\textbf{Likely-to-be-exposed \emph{vs.} Not-likely-to-be-exposed Tasks.} 
We divided our 10{,}131 tasks (LM-annotated) into two groups: 
 \textit{likely-to-be-exposed} \emph{vs.} \textit{not-likely-to-be-exposed}. We used AII~\cite{septiandri2024potential} to estimate the likelihood that workplace tasks will be exposed to AI. Following \citet{shao2025future}, we further restricted both groups to occupations and tasks that are primarily performed on a computer according to O*NET, resulting in 3{,}179 tasks across 426 occupations and 19 sectors in the likely-to-be-exposed group, and 2{,}349 tasks across 381 occupations and 19 sectors in the not-likely-to-be-exposed group.

\mbox{ } \\
\noindent 
\textbf{Linear Mixed-Effects Regression Model.} 
We estimate whether tasks likely to be exposed to AI systematically differ in their meaningfulness to workers as compared to tasks not likely to be exposed.
For each task $t$, we computed the importance of each dimension $d$ of meaningful work (e.g., requiring novel ideas or creativity, help build relationships). 
To estimate whether each dimension was more or less important in tasks likely to be exposed than in tasks not likely to be exposed, we fit a linear mixed-effects regression of the form:
\[
z(y_{t,d}) = \beta_0 + \beta_1 \cdot \text{AIExposure}_t + u_{s} + u_{o(s)} + \epsilon_{t,d},
\]
where $z(y_{t,d})$ is the $z$-score of the rating $y_{t,d}$ a worker gave to the importance of dimension of meaningful work $d$ for task $t$; $\text{AIExposure}_t$ is a binary variable equal to 1, if task $t$ is likely-to-be-exposed, or equal to 0, if task $t$ is not-likely-to-be-exposed; $\beta_0$ is the fixed effect intercept, representing the baseline importance of dimension $d$ for not-likely-to-be-exposed tasks ($\text{AIExposure}_t = 0$);
$\beta_1$ is the fixed effect of AI exposure, estimating the mean difference in worker ratings between likely-to-be-exposed and not-likely-to-be-exposed tasks. Given that tasks are nested within occupations, which, in turn, are nested within sectors, to account for varying baselines within sectors and occupations, we also included a random intercept $u_{s}$ for sector $s$, and random intercept $u_{o(s)}$ for occupations nested within sector $s$, followed by $\epsilon_{t,d}$, which is the residual error term for task $t$ on dimension $d$. Models were estimated using maximum likelihood (restricted maximum likelihood, REML, disabled). For each dimension, we report $\beta_1$ (the difference in worker ratings for likely-to-be-exposed \emph{vs.} not-likely-to-be-exposed tasks), its standard error, and 95\% confidence intervals.
To address multiple comparisons across our survey items, we applied the Benjamini–Hochberg False Discovery Rate (FDR) correction.
We defined the fixed effect as statistically significant, if two conditions were met: (1) the FDR-adjusted $p<0.05$, and (2) the effect size exceeded $\Delta \geq 0.1$ Likert points. While a threshold of 0.1 may appear small, our results are reported in aggregate across all sectors; when disaggregated at the sector level, we show that differences are often substantially larger.
The latter serves as a threshold for practical significance: on a 5-point scale, a 0.1 shift represents a small but interpretable change in perceived task characteristics, ensuring that we highlight effects that are not only statistically detectable but also meaningful in practice.

\smallskip
The mixed-effects estimates and FDR-adjusted tests (see Appendix \autoref{tab:rq1_results}) yield three overarching findings about how AI exposure shapes the perceived importance of meaningful-work dimensions, and how these effects distribute across sectors:
\begin{enumerate}
\item \emph{The tasks most exposed to AI involve creativity and high levels of individual agency, while tasks that rely on empathy, relationship-building, or in-person presence appear less exposed.} Across a subset of seven significant dimensions of ``meaningful work'', those most exposed to AI emphasize novelty and creativity, personal agency, and the capacity to elicit positive emotions, whereas tasks less exposed emphasize social connection, relationship building, emotional attunement, and in-person interaction (\autoref{fig:rq1_bar}, and Appendix \autoref{tab:rq1_results}). Random-slope mixed-effects models (AI exposure varies by sector) showed largely consistent effects across sectors, with notable heterogeneity for visible/tangible outcomes, emotional awareness, in-person interaction, and physical well-being (likelihood-ratio tests, $p_{\text{LRT}}<0.05$). \vspace{3pt}

\item  \emph{Tasks highly-exposed to AI cluster in creative, technical, and scientific domains, where AI systems increasingly support ideation.}
The greatest exposure appears in the Arts, Architecture \& Engineering, Computer \& Mathematics, and the Life, Physical, \& Social Sciences (\autoref{fig:rq1_sector_likely}). Illustratively, an Art Director ``formulating basic layout design'' and an Actuary ``constructing probability tables for natural disasters'' reflect first passes that models now credibly generate, with humans then refining the output (Appendix \autoref{tab:rq1_top_sectors_examples}). These sectorial patterns align with cluster analyses of high-importance tasks (Appendix Tables \ref{tab:rq1_clusters_novel_ideas_creativity_likely}-\ref{tab:rq1_clusters_freedom_in_how_to_do_it_likely}). Tasks likely to be exposed to AI that evoke positive emotions are concentrated in Arts, Entertainment, Sports, \& Media; Community \& Social Services; and Healthcare (Figure \ref{fig:rq1_sector_likely}). \vspace{3pt}

\item \emph{Tasks that remain less exposed to AI are those that rely on relationships and sensitivity to context, with their value derived from human attention and judgment.}
In Education, Sales, Community \& Social Services, and Healthcare, core activities (e.g., ``counseling students through intertwined academic and personal issues'', ``presenting offers while preserving relationships'') depend on real-time, co-constructed meaning, and nuanced perception that resists codification (Figure \ref{fig:rq1_sector_not_likely}, and Appendix \autoref{tab:rq1_top_sectors_examples}). Consistent with this pattern, clusters emphasizing emotional awareness and relationship building remain predominantly human-centered (Appendix \autoref{tab:rq1_clusters_emotion_awareness_not}).
\end{enumerate}

\subsection{Do teams design AI systems with traits that align with the traits workers want? (RQ2)}
\label{sec:answer-rq2}

For RQ2 (Q34–Q45), we were interested not in dimensions of meaningful work, but in \emph{AI traits}. That is, we asked workers and developers which traits an AI system should have. Each item defined a trait along a spectrum (e.g., Q39: ``\emph{Should the AI show warmth and care, or remain neutral and businesslike?}''), and participants rated their preference on a 1–5 scale. Worker responses indicated which traits an AI system should have when their tasks are exposed to it; developer responses indicated how practitioners would design such a system. Should the system be straightforward or polite? Tolerant or strict? Practical or imaginative? Flexible or determined?

\begin{table*}[t!]
\centering
\caption{Worker–developer misalignment by AI traits. Misalignment is defined as the average absolute difference between worker and developer ratings (Q34–Q45) of the traits they believe AI systems should possess. Differences ($\Delta_{t,q}$) are calculated as worker minus developer ratings for a given task $t$ for a trait $q$, with the magnitude ($|\Delta_{t,q}|$) reflecting the size of the misalignment. We report the sum ($\sum|\Delta_{t,q}|$) and mean ($\mu|\Delta_{t,q}|$) across sectors, and group high, mixed, or aligned categories based on percentile thresholds of average absolute misalignment. \# Sig. Sectors refers to the number of sectors that had statistically significant \textit{differences} in ratings between workers and developers. The largest gaps appear for Straightforward \emph{vs.} Polite, Tolerant/Open-minded \emph{vs.} Strict, Practical \emph{vs.} Imaginative, and Flexible \emph{vs.} Determined, whereas traits such as Generalized \emph{vs.} Personalized, Simple \emph{vs.} Comprehensive, and Business-like \emph{vs.} Warm/caring show little to no misalignment.}
\Description{Worker–developer misalignment by AI traits. Misalignment is defined as the average absolute difference between worker and developer ratings (Q34–Q45) of the traits they believe AI systems should possess. Differences ($\Delta_{t,q}$) are calculated as worker minus developer ratings for a given task $t$ for a trait $q$, with the magnitude ($|\Delta_{t,q}|$) reflecting the size of the misalignment. Reported values are aggregated over sectors and traits are grouped into high, mixed, or aligned categories based on percentile thresholds of average absolute misalignment. \# Sig. Sectors refers to the number of sectors that had statistically significant \textit{differences} in ratings between workers and developers. The largest gaps appear for Straightforward \emph{vs.} Polite, Tolerant/Open-minded \emph{vs.} Strict, Practical \emph{vs.} Imaginative, and Flexible \emph{vs.} Determined, whereas traits such as Generalized \emph{vs.} Personalized, Simple \emph{vs.} Comprehensive, and Business-like \emph{vs.} Warm/caring show little to no misalignment.
}
\label{tab:rq2_trait_summary}
\begin{tabular}{p{6cm}ccc}
\toprule
\textbf{Trait} & \textbf{\# Sig. Sectors} & $\mathbf{\Sigma|\Delta_{t,q}|}$ & $\mathbf{\mu|\Delta_{t,q}|}$ \\
\midrule
\multicolumn{4}{l}{\textbf{High misalignment}} \\
\midrule
(Q40) Straightforward \emph{vs.} Polite         & 16 & 25.874 & 1.617 \\
(Q41) Tolerant/Open-minded \emph{vs.} Strict    & 5  & 4.449  & 0.890 \\
(Q45) Practical \emph{vs.} Imaginative          & 2  & 1.690  & 0.845 \\
(Q43) Flexible \emph{vs.} Determined            & 9  & 6.654  & 0.739 \\
\midrule
\multicolumn{4}{l}{\textbf{Mixed misalignment}} \\
\midrule
(Q34) Handle complex \emph{vs.} Routine work    & 8  & 5.797  & 0.725 \\
(Q35) Address emotions \emph{vs.} Apathetic     & 4  & 2.849  & 0.712 \\
(Q36) Explainable \emph{vs.} Fast/automatic     & 16 & 10.940 & 0.684 \\
(Q42) Precise \emph{vs.} Simple                 & 2  & 1.168  & 0.584 \\
\midrule
\multicolumn{4}{l}{\textbf{Aligned}} \\
\midrule
(Q37) Definitive \emph{vs.} Open to challenge   & 3  & 1.713  & 0.571 \\
(Q39) Business-like \emph{vs.} Warm/caring      & 1  & 0.500  & 0.500 \\
(Q38) Generalized \emph{vs.} Personalized       & 0  & 0.000  & 0.000 \\
(Q44) Simple \emph{vs.} Insightful/Comprehensive           & 0  & 0.000  & 0.000 \\
\bottomrule
\end{tabular}
\end{table*}
 
\begin{figure*}[t]
  \centering
  \includegraphics[width=\linewidth]{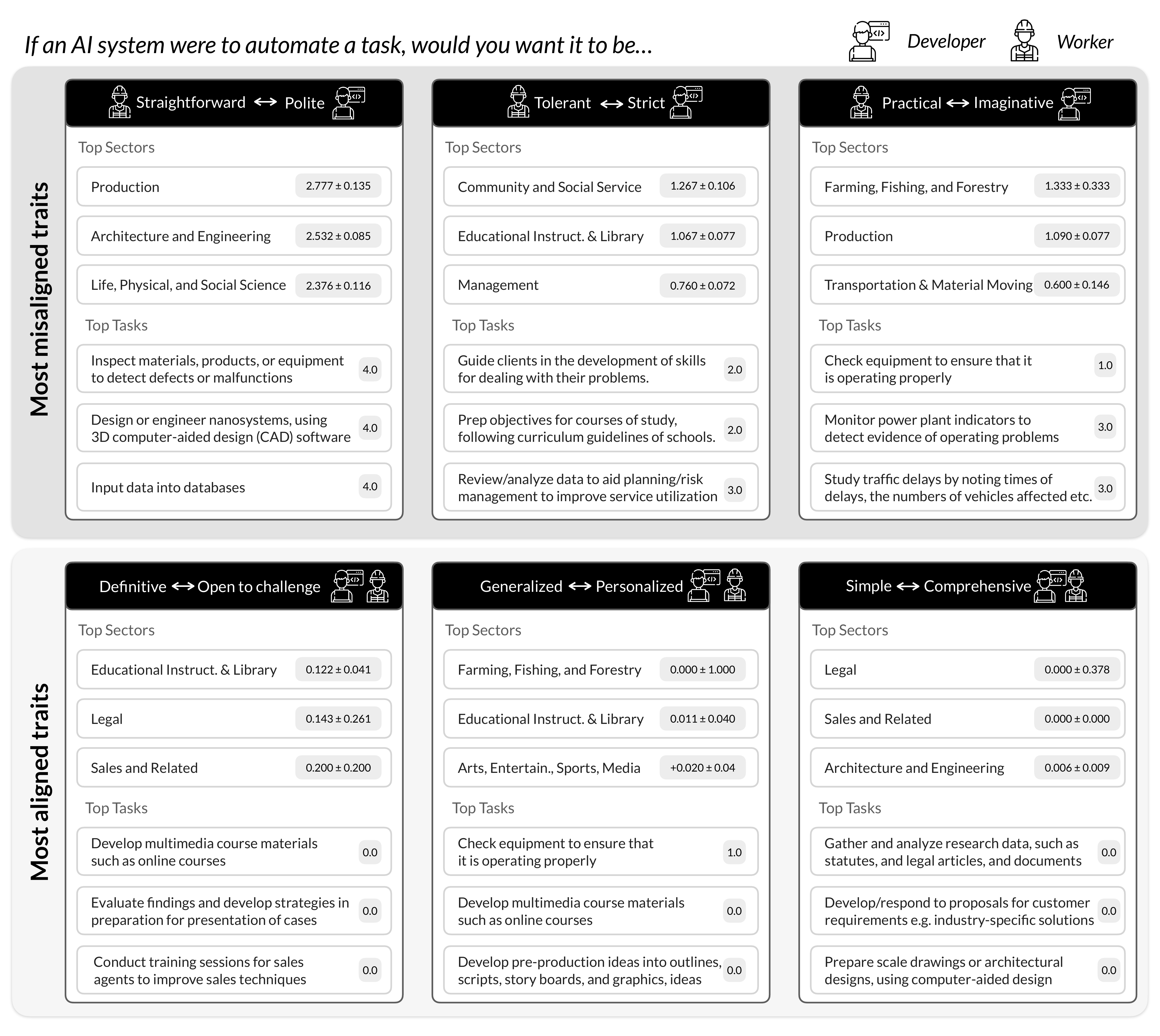}
  \caption{The three most misaligned AI traits, (top row) and the three most aligned traits (bottom row). Larger values indicate greater disagreement between workers and developers. Scores are the absolute differences between workers' ratings of how much they want an AI system to exhibit each trait and developers' ratings of how much they intend to incorporate that trait into the design of an AI system. Icons show which trait direction each group prefers (e.g., workers wish straightforward systems, while developers set out to design polite systems). Top contributing sectors and example tasks are listed for each trait.
  }
\Description{Grid of six panels showing AI trait alignment and misalignment between workers and developers. 
  Top row: most misaligned traits—straightforward vs polite, tolerant vs strict, and practical vs imaginative. 
  Bottom row: most aligned traits—definitive vs open to challenge, generalized vs personalized, and simple vs comprehensive. 
  Each panel lists top sectors and example tasks. For instance, in Production and Engineering, workers prefer straightforward systems but developers emphasize politeness. 
  In Education and Legal, both groups align on valuing definitiveness and openness to challenge. 
  Task examples include inspecting equipment, developing course materials, monitoring plant indicators, and preparing legal research documents.}
  \label{fig:rq2_overall}
\end{figure*}

To measure misalignment, we subtracted the rating workers assign to trait $q$ for an AI system exposed to task $t$ ($\text{workers rating}_{t,q}$) from the rating developers assign to the same trait for the same task ($\text{developers rating}_{t,q}$):
\[
\Delta_{t,q} = \text{workers rating}_{\text{t},q} - \text{developers rating}_{t,q}.
\]
 The magnitude $|\Delta_{t,q}|$ measures the size of the misalignment, while the sign indicates direction. For example, a positive $\Delta_{t,q}$ on Q39 indicates that workers preferred more warmth and care than developers, who leaned toward neutrality. We averaged task-level differences for each sector giving each task equal weight, giving us an average misalignment score per sector. To then test whether the differences between worker and developer ratings were statistically significant, we conducted a \textit{two-sided $t$-test} of the null hypothesis that the mean task-level difference was zero for a given trait and sector. To account for multiple comparisons, $p$-values were adjusted using the Benjamini-Hochberg False Discovery Rate (FDR) procedure. A sector was labeled as significantly misaligned on a trait, if two conditions were met: (1) the FDR-adjusted $p<0.05$; and (2) the absolute mean difference ($\frac{1}{N} \sum_{t=1}^{N} \left| \Delta_{t,q} \right|$; where $N$ is the number of tasks in the sector) exceeded a threshold of 0.5 Likert points.
 We use 0.5 as a conservative threshold to focus on practically meaningful differences: unlike RQ1, which examined fine-grained within-task effects (where smaller shifts of 0.1 were informative), RQ2 compares worker–developer ratings aggregated across major occupational groups, where only larger gaps are more informative.

\mbox{ } \\
\noindent 
\textbf{Most and Least Misaligned Traits.} To analyze worker–developer misalignment, we ranked AI traits by their average misalignment scores across sectors where worker/developer differences were statistically significant for a given sector (FDR $<$ 0.05, $\frac{1}{N} \sum_{t=1}^{N} \left| \Delta_{t,q} \right| \geq 0.5$). 
\autoref{tab:rq2_trait_summary} summarizes the classification of traits from most to least misaligned, while \autoref{fig:rq2_overall} illustrates example occupations and tasks within each category. Sector-level misalignment scores for individual traits are reported in Appendix Tables \ref{tab:rq2_q37_numbers}-\ref{tab:rq2_q45_numbers}. 
To summarize our results, we see that, across sectors (\autoref{fig:rq2_overall}), workers consistently favored straightforward systems; developers preferred polite ones. Workers wanted tolerance; whereas developers leaned towards more strict systems. Workers asked for practical systems; developers opted for more imaginative systems. Workers liked flexibility; developers nudged toward more determined systems. Where both groups aligned was telling: they valued deep understanding, personalization, and openness to challenge.
That is, neither group preferred a generic system that appeared helpful but functioned as an unquestionable authority.

Also, to surface broad patterns of misalignment, for each trait, we identified tasks in the extreme percentiles of misalignment (top $99^{th}$, and bottom $1^{st}$), clustered these tasks using MPNet\footnote{\url{https://huggingface.co/sentence-transformers/all-mpnet-base-v2}} embeddings and K-Means clustering, and labeled the resulting clusters (Appendix Tables \ref{tab:rq2_q37_least}-\ref{tab:rq2_q45_most}).
To interpret these aggregate results, we distilled three recurring, salient design tensions:

\begin{enumerate}
\item \emph{When the stakes are high, workers often treat `politeness' as a delay rather than a feature.} The disagreements varied substantially across sectors. The politeness divide, in particular, was most pronounced in Production, Architecture \& Engineering, and the Life, Physical, \& Social Sciences, fields where vagueness can result in wasted materials, structural failures, or flawed data.
By then clustering the most misaligned tasks on politeness (Appendix \autoref{tab:rq2_q40_most}). We highlighted the parts of the economy that demand exacting judgment: quality control, technical design, oversight, and coordination. The corresponding tasks resemble the day-to-day activities of highly skilled workers such as planning stress tests, analyzing medical procedures to forecast outcomes, or coordinating complex projects.  \vspace{3pt}

\item \emph{Workers wanted AI systems that are flexible; developers tended to value strictness.} Along the tolerant–strict dimension, divergences were most pronounced in Community \& Social Service ($\frac{1}{N} \sum_{t=1}^{N} \Delta_{t,q} = +1.27$), Education ($\frac{1}{N} \sum_{t=1}^{N} \Delta_{t,q} = +1.07$), and Management ($\frac{1}{N} \sum_{t=1}^{N} \Delta_{t,q} = +0.76$) (Appendix \autoref{tab:rq2_q41_numbers}; $N$ is the number of tasks in a sector). Although these settings might be presumed to benefit from greater structure, in practice, ``strict'' software is often experienced as rigid rule-based constraints that can limit practitioner judgment. Those domains are indeed characterized by frequent exceptions, and context-sensitive decision-making: for example, accommodating late coursework without disproportionate penalty, or processing intake information that does not conform to standardized fields. In Management \& Education, as shown in Appendix \autoref{tab:rq2_q41_most}, task clusters include process improvement, monitoring and planning, and budget or risk management. For these clusters, developers emphasize stricter standards and oversight, while workers prefer tolerance and flexibility. \vspace{3pt}

\item \emph{Creativity is not always a virtue.} Along the practical–imaginative dimension, developers tended to favor more imaginative systems, whereas workers prioritized pragmatism. The largest divergences were in Farming, Fishing, \& Forestry ($\frac{1}{N} \sum_{t=1}^{N} \Delta_{t,q} \\= -1.33$), Production ($\frac{1}{N} \sum_{t=1}^{N} \Delta_{t,q} = -1.09$), and Transportation \& Material Moving ($\frac{1}{N} \sum_{t=1}^{N} \Delta_{t,q} = -0.60$). These are domains characterized by highly structured, routine workflows (Appendix \autoref{tab:rq2_q45_numbers}). Workers engaging in equipment checks, plant monitoring, and traffic analysis preferred systems that detect anomalies, enforce compliance, and adhere closely to task constraints rather than tools oriented toward open-ended ideation. Task-level examples are illustrative (Appendix \autoref{tab:rq2_q45_most}): 
checking equipment to ensure proper operation (Farming); monitoring power-plant indicators to detect operating problems (Production); and studying traffic delays by recording times and vehicle counts (Transportation). Consistent with this pattern, clustering results (Appendix \autoref{tab:rq2_q44_least}) indicate that the most misaligned tasks are concentrated in routine technical operations: equipment monitoring, compliance, and other structured activities that emphasize continuous monitoring, and rule adherence.

\end{enumerate}

\begin{table*}[t]
\scriptsize
\caption{From empirical observations to research questions and developer/designer heuristics for worker-aligned AI. Each row links our findings to specific HCI research questions, example workflows, and design heuristics aimed at preserving creativity, autonomy, and meaning in day-to-day work.}
\Description{From empirical observations to concrete research questions and developer/designer heuristics for worker-aligned AI. Each row links our findings to specific HCI research questions, example workflows, and design heuristics aimed at preserving creativity, autonomy, and meaning in day-to-day work.}
\label{tab:agenda_blueprint}
\begin{tabular}{p{0.33\linewidth}  p{0.67\linewidth}}
\hline
\textbf{Empirical Observation + Example Workflow} &  \textbf{Research Questions and Developer/Designer Heuristics} \\
\hline

\textbf{High-agency tasks appear heavily exposed to AI.}  
In our analysis, tasks that workers associate with a sense of agency and freedom are more likely to fall into the high-exposure-to-AI group (Section~\ref{sec:answer-rq1}). \vspace{5pt} \textcolor{white}{.}  

\emph{Example workflow (ideation/drafting).}  
A marketing specialist works with an AI assistant that, given a short prompt, produces full campaign drafts and subject lines, so the worker may end up editing model outputs more than starting from their own ideas and voice. 
&   
\textbf{Research questions.} (1) How does asking workers to first write a short description (2--3 sentences) of what they want the system to produce relate to perceived autonomy, satisfaction, and final quality in writing or design tasks? (2) What interaction patterns (e.g., several alternative suggestions, step-by-step assistance) are associated with workers feeling that they keep ``final say'' over high-agency tasks?  \vspace{3pt}

\textbf{Heuristics for developers/designers.} (1) Add a first step where the worker writes what they want and any limits (e.g., ``Write your 2--3 key ideas before the model drafts''); (2) show several alternative suggestions (e.g., different headlines or outlines) instead of only one full replacement; and (3) make it easy to accept or reject content at the level of small pieces (sentences or sections) instead of only offering one-click ``Replace all''. \vspace{3pt} 

\textbf{Where in the workflow.} Ideation, outlining, first-draft generation, early revision.  \vspace{3pt} 

\textbf{What success looks like.} Examples of success indicators include: (1) higher ``felt in control'' and ``this still feels like my work'' ratings in in-product surveys; (2) more edits and customizations on high-meaning sections, with AI used primarily for structure and low-level polish; and (3) similar or better quality with similar time spent on creative parts, and less time on mechanical rewrites.
\\
\midrule

\textbf{Joyful and creative parts of work often fall in high-exposure categories.}  
In our analysis, tasks that workers describe as creative and enjoyable are more likely to be classified as highly exposed to AI (Section~\ref{sec:answer-rq1}).  \vspace{5pt} \textcolor{white}{.}  

\emph{Example workflow (design / analysis).}  
A product designer receives auto-generated page layouts and color schemes from an AI tool and then mainly cleans up edge cases, rather than exploring ideas from scratch. 
&   
\textbf{Research questions.} (1) How does the order of work (worker makes an initial sketch and the AI helps afterwards, versus the AI creates an initial version and the worker edits it) relate to perceived joy, ownership, and long-term skill growth? (2) Which parts of a multi-step task do workers report wanting to automate (e.g., resizing and formatting) versus keep manual (e.g., core concept and overall narrative)? \vspace{3pt} 

\textbf{Heuristics for developers/designers.} (1) Break workflows into labeled stages (e.g., ``brainstorm'', ``structure'', ``polish''), and let workers toggle AI on or off for each stage; (2) start creative fields empty and require at least a rough human sketch, prompt, or storyboard before enabling AI suggestions; (3) add a simple option such as a checkbox or tag (e.g., ``I want to do this part myself'') and avoid full automation on marked stages, limiting AI to suggestions or diagnostic feedback there. \vspace{3pt} 

\textbf{Where in the workflow.} Creative ideation, conceptual design, exploratory analysis, narrative framing. \vspace{3pt} 

\textbf{What success looks like.} Examples of success indicators include: (1) workers reporting that AI is mainly used to reduce ``grunt work'' and that it does not ``take away the good parts'' of the job; (2) logs showing AI used heavily in repetitive sub-steps (e.g., formatting and error-checking) and mainly for suggestions in creative stages; (3) over time, workers’ self-reported skills and confidence in their core craft remaining stable or improving.
\\
\midrule

\textbf{Relational work appears less exposed to AI.}  
In our data, work that supports social connection and relationships tends to appear in the lower-exposure-to-AI group (Section~\ref{sec:answer-rq1}).  \vspace{5pt} \textcolor{white}{.}  

\emph{Example workflow (teaching / care / management).}  
A teacher, manager, or clinician uses AI tools mainly for canned email replies and templated feedback, which can risk making communication feel more generic and less personal to students, team members, or patients. 
& 
\textbf{Research questions.} (1) How can AI best support, rather than substitute, relational work (e.g., coaching, mentoring, and conflict resolution), according to workers? (2) Which background tasks (e.g., summarizing histories and drafting logistics messages) do workers experience as most helpful in freeing up time for high-quality human interaction? \vspace{3pt}

\textbf{Heuristics for developers/designers.} (1) Use AI to prepare briefs and summaries (e.g., student history, case notes, and prior conversations) so the human arrives better informed to the interaction; (2) default automated replies to low-stakes logistics (e.g., scheduling and confirmations), and route emotionally nuanced or high-stakes messages to humans with short, editable drafts; and (3) during live calls or sessions, assign AI to silent roles (note-taking and surfacing relevant past information) rather than having it speak on the worker’s behalf. \vspace{3pt} 

\textbf{Where in the workflow.} Information collection and preparation, follow-up documentation, low-stakes messaging; not the core live interaction itself. \vspace{3pt} 

\textbf{What success looks like.} Examples of success indicators include: (1) workers reporting more time spent in direct human-to-human interaction without increased overall workload; (2) workers reporting that AI helps them ``show up prepared'' rather than ``speak for me''; and (3) no increase in reports that relationships feel more scripted, generic, or impersonal after deployment of AI support tools.
\\
\midrule

\textbf{Workers and developers differ in reported preferences for AI assistant traits.}  
Workers in our sample consistently report preferring straightforward, tolerant, and practical systems; developers report aiming to design polite, strict, and imaginative ones (Section~\ref{sec:answer-rq2}).  \vspace{5pt} \textcolor{white}{.}  

\emph{Example workflow (information lookup / decision support).}  
A production engineer uses an AI assistant that responds to fault queries with long, polite paragraphs and speculative explanations, when what they want is a short, actionable checklist. 
&  
\textbf{Research questions.} (1) How are different trait profiles (straightforward \emph{vs.} \ polite, and tolerant \emph{vs.} \ strict) associated with task accuracy, correction speed, and perceived trust in high-stakes domains? (2) What interface controls do workers find most useful for quickly adjusting an AI assistant’s style to match task demands without feeling overwhelmed? \vspace{3pt}

\textbf{Heuristics for developers/designers.} (1) Ship work tools so that, by default, they answer in a straightforward and concise way and let users change this setting for each task, if they want more politeness or detail; (2) provide a simple control (e.g., a slider) so users can choose between short vs.\ detailed answers and between strict vs.\ tolerant behavior; and (3) A/B test trait profiles against worker-centered metrics such as time-to-correction, frequency of follow-up clarifying prompts, and perceived friction. \vspace{3pt} 

\textbf{Where in the workflow.} Information retrieval, explanation, diagnosis, and decision support. \vspace{3pt} 

\textbf{What success looks like.} Examples of success indicators include: (1) fewer cases where users report needing to skip over long, unnecessary text (e.g., collapsing long answers or repeatedly asking the system to ``make it shorter''); (2) shorter time to correct or verify model outputs in structured tasks; (3) higher ratings that ``the assistant speaks my language for this task'' in post-task surveys.
\\
\hline
\end{tabular}
\end{table*}

\section{Discussion}
\label{sec:discussion}

Section~\ref{sec:discussion:relation} relates our findings to prior work and summarizes our main contributions, Section~\ref{sec:discussion:implications} then draws out implications for the design and governance of AI systems, and Section~\ref{sec:discussion:limitations} concludes with limitations of our study and directions for future research.

\subsection{Relation to Prior Work and Overview of Findings}
\label{sec:discussion:relation}

\noindent
\textbf{Relation to prior work and novel contributions.} Prior work has found that, in many settings, augmentation is more common than end-to-end automation \citep{hazra2025ai, brynjolfsson2025generative, shao2025future}. Our task-level analysis introduces a worker-centered perspective by examining how exposure relates to how work feels to workers, and by identifying the aspects they prefer to handle through in-person, social interaction. This complements exposure inventories by bringing in how tasks are experienced by workers, and by highlighting potential implications for design~\citep{hazra2025ai, shao2025future, brynjolfsson2025generative}. We make the following contributions:
\begin{enumerate}
\item \emph{A task-level link between exposure and meaning.} We analyze how AI exposure co-varies with dimensions of meaningful work and observe a clear pattern: exposure is higher for tasks involving new ideas, positive feelings, and freedom, and lower for tasks that rely on emotional attunement and in-person relationships (Figures \ref{fig:rq1_bar}, \ref{fig:rq1_sector_likely}, and \ref{fig:rq1_sector_not_likely}) \citep{hazra2025ai, shao2025future}. 

\item \emph{A worker–developer trait map for AI exposure.}
We quantify where workers and developers differ on AI traits by sector and task. We find stable agreement on personalization and deep comprehension/insightfulness, and large gaps on ``straightforward \emph{vs.} polite'' (Table~\ref{tab:rq2_trait_summary}, Figure \ref{fig:rq2_overall}, and Appendix Figure
\ref{fig:rq2_polite}) \citep{septiandri2024potential}.

\item  \emph{A scalable rating process with checks.}
We pair human ratings with LM-assisted ratings to cover the largest set of tasks present in the literature, while ensuring validity. In our data, adding the LM as an additional annotator was associated with higher inter-rater agreement for both worker and developer instruments, and we document where LM and human views diverge (Table~\ref{tab:lm_human_full_examples}, and Appendix Figure~\ref{fig:trait_distributions}) \citep{ziems2024can}.

\end{enumerate}

\mbox{ }\\ 
\noindent
\textbf{Workers and developers differ in specific AI traits.} We briefly compare where they differ and where they agree, and then suggest a broader design implication:
\begin{description}
\item \emph{Where they differ most.} Workers want AI that is straightforward, tolerant and open-minded, practical, and flexible. Developers plan for AI that is more polite, strict, imaginative, and determined. The biggest gaps appear for straightforward \emph{vs.} polite, tolerant \emph{vs.} strict, practical \emph{vs.} imaginative, and flexible \emph{vs.} determined (Figure~\ref{fig:rq2_overall} and Table~\ref{tab:rq2_trait_summary}). Sector gaps for straightforward \emph{vs.} polite are largest in Production, Architecture \& Engineering, and Life, Physical, \& Social Sciences, with smaller gaps in Education, and Community \& Social Service (Figure~\ref{fig:rq2_polite}). \vspace{5pt}

\item \emph{Where they agree.} Both groups favor deep understanding, comprehension, personalization, and openness to challenge, with little to no systematic gap (Figure~\ref{fig:rq2_overall} and Table~\ref{tab:rq2_trait_summary}). \vspace{5pt}

\item \emph{Design Implication.} Our results suggest that AI development teams should compare planned trait choices to worker preferences for the target task and sector. In domains with technical judgment and oversight, large gaps on straightforward \emph{vs.} polite, and tolerant \emph{vs.} strict call for careful defaults, clear settings, and a broader set of design choices (Section~\ref{sec:conclusion}).

\end{description}

\subsection{Implications}
\label{sec:discussion:implications}
We translate our results into three steps for deployment: design the interaction to protect meaning, measure the outcomes that matter, and tune defaults by sector and task. 

\begin{description}
\item  \emph{Design the interaction to protect meaning.} An interface that makes assistance easy to accept, edit, and credit  (without reducing the worker's role) is consistent with our findings. Control stays at the task level when the system exposes levels of help at the sub-step (by, e.g., suggesting, drafting, or executing). Clarity about edits and credit comes from showing sources and a simple revision history, which makes authorship and changes visible in the final output. Tone should match the work: when a straightforward style is preferred, a plain default with an optional tone control should be available (Figures~\ref{fig:rq2_overall} and \ref{fig:rq2_polite}).  \vspace{3pt}

\item  \emph{Measure what the design seeks to preserve.}  We recommend measuring, during deployment, whether the design preserves human agency, and design intentions. For example, short, task-linked metrics should capture latency and output quality, while logs record the assistance level (suggest, draft, execute) and the final decision-maker. Regular reviews of logs can help teams monitor whether deployments remain aligned with design features. \vspace{1pt}

\item \emph{Set sector-aware defaults.} We recommend setting sector specific defaults that reflect the task differences found in our results (Figures \ref{fig:rq2_overall}
and \ref{fig:rq2_polite}). For technical oversight and design, defaults should be simple, practical, and adjustable. For care and education, defaults should be warm and personal, while keeping limits explicit. More generally, in contexts where emotional awareness and in-person interaction are essential, our results point toward using AI primarily for background tasks (e.g., preparing briefs, summarizing records, and flagging anomalies), while reserving protected time for direct human engagement.  \vspace{3pt}

\end{description}

Our findings suggest an \emph{interaction-as-policy} lens: decisions about who clicks, who decides, who sees what, and what the AI system makes easy or hard are not just UX choices but \emph{governance} choices embedded in the interface. This view builds on HCI and STS work that treats infrastructures, defaults, and algorithms as forms of governance (e.g., code as law, scripts that configure users, and algorithmic management in workplaces)~\cite{rao2025rideshare, akridge2024bus, karizat2024patent}.  This lens helps organize four observed phenomena:

\begin{enumerate}
\item \emph{Situated manners.} Situated manners recast politeness in HCI as a context-sensitive control, not a universal style.  ``Frictionless and chatty'' helps in consumer chat, but in operational work, where risk, time pressure, irreversibility, uncertainty, and physical coupling matter, verbosity and small talk distract. The system should default to short, clear outputs and strategic silence. Examples from Section~\ref{sec:results}: disaster-risk tables should present the number, the limit, and the next action; for purchase offers, the system should surface price, terms, and deadline first; in counseling, the system should adopt a warmer tone and avoid imperatives. In high-risk contexts, the ethical choice may well be the blunt one, and the useful thing to say may be short.
\vspace{3pt}

\item \emph{Liability anxiety.} Our results are consistent with the idea that liability concerns may contribute to preferences for strictness. In our ``Regulatory requests for information'' cluster (Legal, Sales in Section~\ref{sec:results}), stricter configurations may have reduced wrongful disclosures and minimized regulatory or contractual exposure, but may have increased missed statutory windows and incomplete filings despite available data.

\item  \emph{Romanticized creativity.} Developers often romanticize creativity as ``frontier intelligence'', but our findings suggest that creativity is most helpful when tightly scoped. In field troubleshooting, operators want diagnosis first (fault codes, likely failure chains, next safe test). In regulatory responses, creative paraphase undermines audit-readiness. In customer support, agents prefer policy-aligned drafts with required fields pre-filled over copies that risk unauthorized promises.

\item \emph{Power structures.} Our analysis treats interaction design as a form of policy, but it is important to ask \emph{who sets that policy}.  Our results show that tasks with high levels of creativity, freedom, and happiness are especially likely to be exposed to AI. In practice, managers, vendors, and technical teams (not workers) usually decide which tasks to expose to AI, how far to automate them, and how to evaluate success, often under pressure to improve efficiency~\cite{LaaserKarlsson2023Politics}. This pattern parallels familiar power asymmetries from work on algorithmic management, where data-driven systems restructure tasks, monitor performance, and allocate rewards in ways that can increase organizational control over workers~\cite{rao2025rideshare, karizat2024patent}. In our setting, interaction patterns (such as one-click automation and suggestion-on-demand) shape whether AI replaces or supports the parts of the job that workers report finding most meaningful. Managers, vendors, and technical teams make these decisions before workers can consent or take part, especially in roles with low bargaining power. Therefore, beyond designing interfaces that let individuals retain agency within a task, organizations should give workers some say over which task clusters are candidates for automation. When such mechanisms for voice are absent, AI exposure risks deepening existing power imbalances. Prior work on algorithmic management and worker-centered AI points to several mitigations: involving workers and their representatives directly in technology decision-making (through, e.g., formal consultation and collective bargaining over AI deployments), creating worker or union ``technology representatives'' with access to information about how systems allocate and evaluate work, and requiring transparency tools or reports that surface the indicators workers need to understand and contest algorithmic decisions~\cite{rao2025rideshare,akridge2024bus,atkinson2024algorithmic,europarl2025am}.
\end{enumerate}

\subsection{Limitations}
\label{sec:discussion:limitations}
This work has several broad limitations:
\begin{enumerate}
\item \emph{Representativeness of tasks and participants.} We have focused on U.S. occupational tasks. Results may differ in other regions or settings with other norms or tools \citep{brynjolfsson2025generative}. Our worker and developer samples were recruited through Prolific, where participants tend to be more technologically literate. As a result, perspectives from low-wage, non-digital, or less AI-exposed occupations may be underrepresented. Although our pre-screening procedure ensured that participants were highly familiar with the tasks they evaluated, future work should expand to other recruitment channels (e.g., industry partnerships, unions, vocational training programs) to capture a wider diversity of workplace contexts and skill levels. 
Although the distribution of technical roles, AI usage, and work functions in our sample indicates that developers were actively engaged in building/maintaining AI-driven systems across diverse sectors (\autoref{fig:developers_representation}), we acknowledge that our samples may not fully capture the breadth of industry roles.
\item \emph{Modeling assumptions.} Our mixed-effects regression models assume linear relationships between AI exposure likelihood and each dimension of meaningful work. While this framework is appropriate for estimating average effects across occupations and sectors, relationships among meaningfulness dimensions may be non-linear or interactive. For example, creativity and autonomy may jointly shape how a task is exposed to AI. Exploring non-linearities, higher-order interactions, and potentially hierarchical or multivariate modeling structures is an important direction for future work.
\item \emph{Measurement bias in LM-assisted scaling.} We used an LM to scale ratings to roughly 10K tasks, which raises concerns about whether it reflects worker and developer perspectives or simply produces plausible responses. We therefore treat LM ratings as approximating the distribution of human responses rather than as ground truth, and use the term ``agreement'' to avoid implying a correct answer. To justify LM-assisted scaling, we compared empirical findings from LM \emph{vs.} human ratings and observed that LM-simulated patterns of worker–developer misalignment closely matched human-only results (Appendix Figure~\ref{fig:rq1_human_ratings}, and Appendix Tables~\ref{tab:rq2_trait_human_ratings} and~\ref{tab:rq2_trait_lm_ratings}). Robustness checks further confirmed that ICC gains cannot be explained by LM stability alone, indicating substantive alignment between LM and human judgments. However, LM ratings are more uniform and optimistic in some sectors, and qualitative divergences remain (Appendix Table~\ref{tab:lm_human_full_examples}, and Appendix Figure~\ref{fig:trait_distributions}). These findings support using LMs for scaling, but should be interpreted cautiously, as they may miss fine-grained nuances present in human judgments.
\item \emph{Task filtering and indices of exposure.} Our task-selection pipeline reduced the full O*NET task universe to a subset most relevant for AI exposure, which may introduce task selection bias. Although we used established filters and impact scores~\cite{septiandri2024potential, shao2025future, autor2022newfrontiers}, this process inevitably excludes tasks that matter in practice, especially specialized or licensed work that participants reported low familiarity with. We restored a targeted set of tasks from domains such as healthcare and education, and our final sample aligns with the U.S. Bureau of Labor Statistics employment distributions, but finer-grained gaps within sectors remain. Future work with domain experts or professional organizations could help validate task coverage and identify activities that should be included.
\item \emph{Interpretative variation of O*NET task descriptions.} 
The O*NET database consists of task statements, which are concise summaries of work activities. As a result, workers in the same role may imagine different scenarios of a task (e.g., `formulating basic layout designs'), introducing natural variation in how workers and developers judge these activities. Such interpretive variation is inherent to standardized work taxonomies used in economics and occupational science. In our study, we partly mitigated this variability by: (1) recruiting participants who perform these tasks in practice; and (2) restricting the analysis to tasks workers rated as `highly familiar'. While the subjective nature of our assessments warrants inherent variability in task judgments, the strong agreement between human and LM ratings suggests that interpretation variability did not substantially impact the main observed patterns.
 
\end{enumerate}

\section{Conclusion}
\label{sec:conclusion}

Our results suggest that the tasks most likely to be exposed to AI are disproportionately those that workers associate with joy and agency: novelty and creativity, feeling happy, and having freedom in how to do the work. Yet current design choices often appear not to match what workers say they want from AI, raising concerns that such systems may affect how meaningful work feels. This outcome is shaped by design choices, which can, in principle, be revised. As HCI researchers, we propose a five-part agenda detailed in \autoref{tab:agenda_blueprint}. 

We argue that a central risk of AI exposure is not mass unemployment but ``mass demoralization'': a loss of meaning and ownership in day-to-day work. As models generate more early-stage outputs, the visible creative steps can feel machine-made, while human contributions can become rushed, under-resourced, and hard to recognize. The corresponding promise, if systems are designed and governed accordingly, is a clearer and more humane division of labor: systems that accelerate exploration and drafting, and people who retain authorship, make final judgments, and sustain the relationships that define meaningful work.

\section{Ethical Considerations}
This study received approval through our institution’s research ethics review process. In recruiting workers and developers on Prolific, we did not collect personally identifiable information, and participants could withdraw at any time. 

Beyond procedural safeguards, the study raised three broader ethical considerations. First, surveying both workers and developers about task-level evaluations of work and AI traits risks reinforcing stereotypes about either group’s perspectives. For instance, developers may be framed as indifferent to meaningful work, or workers as resistant to technology, while aggregate analyses can mask the diversity of individual viewpoints. To mitigate this, we reported results in the aggregate but emphasized the diversity of participant perspectives. We also encourage future work to analyze individual-level differences, including how socio-demographic characteristics shape experiences of meaningful work.

Second, our use of LMs to scale annotations introduces the risk of amplifying biases. LMs may underrepresent perspectives from marginalized groups or impose normative judgments about what constitutes meaningful work. To address this, we benchmarked model outputs against human responses, reported intra-class correlation metrics, and included parallel analyses based on human-only ratings to contrast with LM-scaled analyses. We caution that model-based scaling is not a replacement for direct human judgment.

Third, questions about meaningful work can be sensitive, as they touch on participants’ identity, job satisfaction, and professional dignity. Reflecting on whether one's work feels undervalued or easily automated can be unsettling. To minimize potential harm, we framed survey items in neutral and respectful language, and piloted them for clarity before deployment.

These considerations highlight the importance of protecting participants’ dignity and avoiding overgeneralization when studying misalignment between workers and developers. Our aim is not to prescribe what should count as meaningful work, but to inform AI design choices that respect and align with workers’ values.

\section{Author Positionality Statement}
Our research team consists of two women and two men from the United States, Asia and Southern Europe representing diverse ethnic, linguistic, and religious backgrounds. All authors have lived and worked in multiple countries, including the United States, giving us direct experience with the cultural, economic, and policy contexts relevant our study. Our combined expertise spans natural language processing, Responsible AI, computational social science, human–computer interaction, and AI ethics. One author works primarily in academia, while others have experience in both academic and applied research settings.

Our positionalities shaped how we framed the research problem, selected sectors and occupations for analysis, and interpreted findings on worker/developer perspectives to AI exposure. Having worked across different cultural and labor market contexts informed our awareness of how occupational values and AI impacts can vary across sectors, regions, and professional identities. We recognize that our perspectives are influenced by our own academic and research experiences, which may limit the range of viewpoints represented. To support more inclusive and contextually grounded research on AI and the future of work, we encourage future studies to incorporate perspectives from workers, developers, and policymakers in regions and sectors beyond those examined here.

\balance
\bibliographystyle{ACM-Reference-Format}
\bibliography{main}

\appendix
\newpage

\section{Selecting O*NET Tasks}
In our initial filtering, we used GPT-4o judgments to determine whether a task or occupation primarily involved computer usage. Manual inspection revealed that GPT-4o occasionally excluded occupations such as nursing and education, which are widely recognized as being impacted by AI innovations~\cite{rony2024advancing, robert2019artificial, rahm2023education}. To address this, we manually curated a list of 427 occupations exempted from these filters, with examples shown in Appendix~\autoref{tab:occupations_not_filtered}.

To construct a representative sample of tasks from O*NET, we next mapped O*NET occupations to those available on Prolific. \autoref{tab:prolific_jobs_mapping} presents the mapping between Prolific `work function' screeners and corresponding O*NET occupations. Occupations without a Prolific mapping were excluded from our human study.
\begin{table*}[t]
\centering
\small
\caption{Demographic characteristics of participants recruited from Prolific.}
\Description{Demographic characteristics of participants recruited from Prolific.}

\label{tab:prolific-demographics}
\begin{tabular}{p{4cm}p{3cm}p{3cm}}
\toprule
\textbf{Demographics} & \textbf{Workers (N=202)} & \textbf{Developers (N=197)} \\
\midrule
Mean age (SD) & 42.63 (13.05) & 36.68 (10.29) \\
Gender \\
\quad Male & 34.03\% & 64.46\% \\
\quad Female & 61.94\% & 28.42\% \\
\quad Non-binary / Other & 0\% & 0\% \\
\quad Consent Revoked & 4.03\% & 7.11\% \\
Employment status \\
\quad Full-time & 46.94\% & 69.54\% \\
\quad Part-time & 16.29\% & 13.71\% \\
\quad Unemployed/Other & 4.03\% & 1.52\% \\
\quad Consent Revoked/No data available & 15.22\% & 23.80\% \\
\bottomrule
\end{tabular}
\end{table*}

\begin{table*}[t]
\scriptsize
\centering
\caption{
Example tasks with the largest differences between LM and human ratings across survey items, using a Likert scale ranging from 0 (strongly disagree) to 4 (strongly agree). The largest gaps occur when LMs emphasize functional or procedural aspects (e.g., legal drafting, survey coding), while humans evaluate tasks in terms of social, emotional, or organizational meaning (e.g., status maintenance, employee orientation). We found both LM and human interpretations can be valid in different contexts. For instance, the LM perceived the task `review, classify, and record survey data in preparation for computer analysis' as a routine procedure with minimal emotional demands, whereas a human annotator emphasized its creative and moral dimensions. Such divergences reflect subjective differences that are difficult to resolve.}
\Description{
Example tasks with the largest differences between LM and human ratings across survey items, using a Likert scale ranging from 0 (strongly disagree) to 4 (strongly agree). The largest gaps occur when LMs emphasize functional or procedural aspects (e.g., legal drafting, survey coding), while humans evaluate tasks in terms of social, emotional, or organizational meaning (e.g., status maintenance, employee orientation). We found both LM and human interpretations can be valid in different contexts. For instance, the LM perceived the task `review, classify, and record survey data in preparation for computer analysis' as a routine procedure with minimal emotional demands, whereas a human annotator emphasized its creative and moral dimensions. Such divergences reflect subjective differences that are difficult to resolve.}

\label{tab:lm_human_full_examples}
\begin{tabular}{p{2cm}p{2cm}p{3cm}p{0.8cm}p{0.8cm}p{4.5cm}}
\toprule
\textbf{Dimensions of Meaningful Work} & \textbf{Occupation} & \textbf{Task Description} & \textbf{LM Annotation} & \textbf{Human Annotation} & \textbf{Reasoning by LM and Humans} \\
\midrule
\textbf{\emph{Perceived Bullsh*tness}} (e.g., the task feels pointless) & Lawyers & Prepare, draft, and review legal documents, such as wills, deeds, patent applications, mortgages, leases, and contracts. & 0.00 (Strongly disagree) & 1.93 (Neutral) & \emph{LM}: Essential, substantive, impacts clients' rights, central to organizational goals. \newline \emph{Humans}: Moderate meaningfulness, some bureaucratic perception. \\
\addlinespace
\textbf{\emph{Perceived Value}} (e.g., I have the freedom to decide how to carry out this task) & Poets, Lyricists, Creative Writers & Plan project arrangements or outlines, and organize material accordingly. & 3.80 (Strongly agree) & 2.00 (Neutral) & \emph{LM}: Tangible outcomes, autonomy, team contribution. \newline \emph{Humans}: Recognize value but lower perceived impact. \\
\addlinespace
\textbf{\emph{Status Maintenance}} (e.g., This task helps reinforce my standing in the organization) & General \& Operations Managers & Set prices or credit terms for goods or services, based on forecasts of customer demand. & 3.83 (Strongly agree) & 1.89 (Neutral) & \emph{LM}: Visibility, authority, organizational standing. \newline \emph{Humans}: Less tied to status perception. \\
\addlinespace
\textbf{\emph{EPOCH}} (e.g., This task requires recognizing and responding appropriately to the emotions of others) & Survey Researchers & Review, classify, and record survey data in preparation for computer analysis. & 0.40 (Strongly disagree) & 3.20 (Agree) & \emph{LM}: Technical, routine, procedural, minimal emotional/moral impact. \newline \emph{Humans}: Requires some novel ideas and sometimes requires emotional and moral judgments. \\
\addlinespace
\textbf{\emph{Human Flourishing}} (e.g., This task helps me build or strengthen relationships with colleagues or clients) & HR Specialists & Schedule or conduct new employee orientations. & 3.50 (Agree) & 1.69 (Neutral) & \emph{LM}: Purpose, social connection, growth-oriented. \newline \emph{Humans}: Perceived as routine or standard procedure. \\
\addlinespace
\textbf{\emph{Psychological Traits of AI Behavior}}  (e.g., Show warmth and care rather than remain neutral and business-like) & Fundraisers & Recruit sponsors, participants, or volunteers for fundraising events. & 3.08 (Agree) & 0.94  (Disagree) & \emph{LM}: Emotional engagement, tailored reasoning, insight, creativity. \newline \emph{Humans}: Functional, minimal AI-alignment traits perceived. \\
\bottomrule
\end{tabular}
\end{table*}

\begin{figure}[H]
  \centering
  \includegraphics[width=\columnwidth]{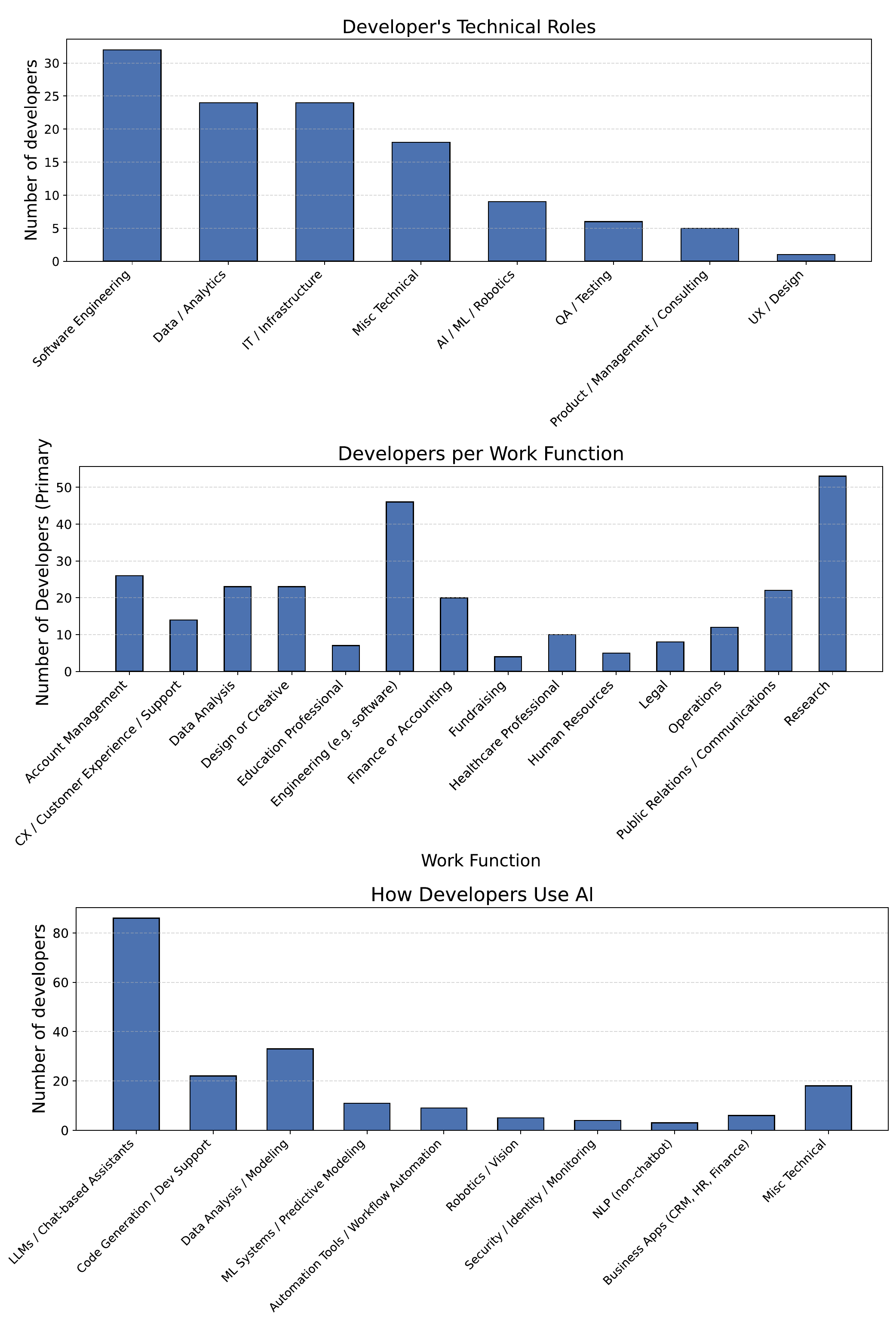}
  \caption{Overview of developer backgrounds across three dimensions: (a) technical role categories, (b) AI usage types, and (c) work functions. Developers work in software, data, IT, and ML/AI roles who actively engage with modern AI tools and contribute to AI-enabled workflows across diverse organizational sectors.}
  \Description{Overview of developer backgrounds across three dimensions: (a) technical role categories, (b) AI usage types, and (c) work functions. Together these characterizations show that the developer sample consists primarily of practitioners working in software, data, IT, and ML/AI roles who actively engage with modern AI tools and contribute to AI-enabled workflows across diverse organizational sectors.}
  \label{fig:developers_representation}
\end{figure}

\begin{table*}[t]
\centering
\scriptsize
\caption{Representative examples of occupations not filtered out from O*NET database.}
\label{tab:occupations_not_filtered}
\begin{tabular}{p{0.9\textwidth}}
\toprule
\textbf{Occupations} \\
\midrule
Accountants and Auditors, Actors, Acute Care Nurses, Administrative Law Judges, Adjudicators, and Hearing Officers, Advertising Sales Agents, Aerospace Engineering and Operations Technologists and Technicians, Air Traffic Controllers, Ambulance Drivers and Attendants, Anesthesiologists, Animal Caretakers, Anthropologists and Archeologists, Architects, Except Landscape and Naval, Architectural and Engineering Managers, Art Directors, Athletic Trainers, Audiologists, Automotive Service Technicians and Mechanics, Bailiffs, Bill and Account Collectors, Bioengineers and Biomedical Engineers, Biological Technicians, Bookkeeping, Accounting, and Auditing Clerks, Broadcast Announcers and Radio Disc Jockeys, Budget Analysts, Bus and Truck Mechanics and Diesel Engine Specialists, Business Intelligence Analysts, Cardiologists, Career/Technical Education Teachers, Secondary School, Chemical Engineers, Chemists, Child, Family, and School Social Workers, Childcare Workers, Civil Engineers, Claims Adjusters, Examiners, and Investigators, Clinical Research Coordinators, Clinical and Counseling Psychologists, Commercial and Industrial Designers, Community Health Workers, Compliance Officers, Computer Hardware Engineers, Computer Network Support Specialists, Computer Programmers, Computer Systems Analysts, Computer and Information Systems Managers, Concierges, Conservation Scientists, Construction Managers, Cooks, Restaurant, Coroners, Credit Analysts, Critical Care Nurses, Curators, Customer Service Representatives, Dancers, Data Scientists, Database Administrators, Dental Hygienists, Dentists, General, Detectives and Criminal Investigators, Digital Forensics Analysts, Dispatchers, Except Police, Fire, and Ambulance, Editors, Education Administrators, Postsecondary, Elementary School Teachers, Except Special Education, Emergency Medical Technicians, Environmental Engineers, Epidemiologists, Executive Secretaries and Executive Administrative Assistants, Family Medicine Physicians, Fashion Designers, Financial Examiners, Financial Managers, Fire-Prevention and Protection Engineers, First-Line Supervisors of Office and Administrative Support Workers, Fitness and Wellness Coordinators, Food Scientists and Technologists, Forest and Conservation Technicians, Fundraisers, Gambling Dealers, General and Operations Managers, Geneticists, Graphic Designers, Health Education Specialists, Healthcare Social Workers, Historians, Home Health Aides, Human Resources Managers, Industrial Engineers, Information Security Analysts, Instructional Coordinators, Insurance Sales Agents, Judges, Magistrate Judges, and Magistrates, Kindergarten Teachers, Except Special Education, Lawyers, Legislators, Loan Officers, Logisticians, Management Analysts, Market Research Analysts and Marketing Specialists, Marriage and Family Therapists, Mathematicians, Mechanical Engineers, Medical and Health Services Managers, Mental Health Counselors, Microbiologists, Middle School Teachers, Except Special and Career/Technical Education, Musicians and Singers, Network and Computer Systems Administrators, Nurse Practitioners, Nursing Assistants, Occupational Therapists, Office Clerks, General, Operating Engineers and Other Construction Equipment Operators, Optometrists, Paralegals and Legal Assistants, Pediatricians, General, Personal Financial Advisors, Pharmacists, Photographers, Physical Therapists, Physicians, Pathologists, Police and Sheriff's Patrol Officers, Political Scientists, Preschool Teachers, Except Special Education, Producers and Directors, Project Management Specialists, Psychiatrists, Public Relations Specialists, Radiologists, Real Estate Brokers, Receptionists and Information Clerks, Recreation Workers, Registered Nurses, Respiratory Therapists, Retail Salespersons, Sales Managers, School Psychologists, Secondary School Teachers, Except Special and Career/Technical Education, Secretaries and Administrative Assistants, Except Legal, Medical, and Executive, Security Managers, Self-Enrichment Teachers, Social and Community Service Managers, Software Developers, Special Education Teachers, Elementary School, Speech-Language Pathologists, Statisticians, Substance Abuse and Behavioral Disorder Counselors, Surgeons, Survey Researchers, Sustainability Specialists, Tax Preparers, Taxi Drivers, Technical Writers, Training and Development Specialists, Transportation, Storage, and Distribution Managers, Tutors, Veterinarians, Veterinary Technologists and Technicians, Video Game Designers, Writers and Authors, Zoologists and Wildlife Biologists\\
\\
\bottomrule
\end{tabular}
\end{table*}

{\scriptsize
\begin{table*}[t]
\centering
\scriptsize
\caption{Prolific functions and their associated jobs.}
\Description{Prolific functions and their associated jobs.}
\label{tab:prolific_jobs_mapping}
\begin{tabular}{p{0.25\textwidth} p{0.68\textwidth}}
\toprule
Prolific Function & Mapped O*NET Occupations \\
\midrule
Account Management & New Accounts Clerks \\
Administration/ Personal Assistant & Administrative Services Managers, Executive Secretaries and Executive Administrative Assistants, First-Line Supervisors of Office and Administrative Support Workers, Legal Secretaries and Administrative Assistants, Medical Secretaries and Administrative Assistants, Receptionists and Information Clerks, Secretaries and Administrative Assistants, Except Legal, Medical, and Executive \\
Chemical / Mechanical / Electrical / Civil Engineering & Architectural and Civil Drafters, Automotive Engineering Technicians, Chemical Engineers, Chemical Plant and System Operators, Civil Engineering Technologists and Technicians, Civil Engineers, Electrical Engineers, Electrical and Electronic Engineering Technologists and Technicians, Electrical and Electronics Drafters, Electro-Mechanical and Mechatronics Technologists and Technicians, Energy Engineers, Except Wind and Solar, Environmental Engineers, Industrial Engineers, Materials Engineers, Mechanical Drafters, Mechanical Engineering Technologists and Technicians, Mechanical Engineers, Petroleum Engineers, Transportation Engineers, Water/Wastewater Engineers \\
CX / Customer Experience / Support & Computer User Support Specialists, Customer Service Representatives \\
Data Analysis & Business Intelligence Analysts, Data Scientists, Statistical Assistants, Statisticians \\
Design or Creative & Art Directors, Art Therapists, Commercial and Industrial Designers, Fine Artists, Including Painters, Sculptors, and Illustrators, Graphic Designers, Interior Designers, Poets, Lyricists and Creative Writers, Set and Exhibit Designers, Special Effects Artists and Animators, Video Game Designers, Web and Digital Interface Designers \\
Healthcare Professional & Acute Care Nurses, Anesthesiologist Assistants, Clinical Data Managers, Clinical Neuropsychologists, Clinical Nurse Specialists, Clinical Research Coordinators, Community Health Workers, Critical Care Nurses, Cytotechnologists, Diagnostic Medical Sonographers, Emergency Medical Technicians, Family Medicine Physicians, General Internal Medicine Physicians, Health Informatics Specialists, Health Specialties Teachers, Postsecondary, Healthcare Social Workers, Home Health Aides, Hospitalists, Magnetic Resonance Imaging Technologists, Medical Appliance Technicians, Medical Assistants, Medical Equipment Preparers, Medical Equipment Repairers, Medical Records Specialists, Medical Scientists, Except Epidemiologists, Medical and Clinical Laboratory Technologists, Medical and Health Services Managers, Neurologists, Nurse Anesthetists, Nurse Midwives, Nurse Practitioners, Nursing Assistants, Occupational Health and Safety Specialists, Occupational Therapists, Orthodontists, Paramedics, Pediatricians, General, Pharmacy Aides, Physical Therapists, Physician Assistants, Physicians, Pathologists, Preventive Medicine Physicians, Psychiatric Technicians, Radiologic Technologists and Technicians, Radiologists, Registered Nurses, Surgical Assistants \\
Engineering (e.g. software) & Architectural and Engineering Managers, Automotive Engineers, Computer Hardware Engineers, Computer Programmers, Computer Systems Analysts, Computer Systems Engineers/Architects, Computer and Information Systems Managers, Computer, Automated Teller, and Office Machine Repairers, Electronics Engineers, Except Computer, Fuel Cell Engineers, Geothermal Production Managers, Health and Safety Engineers, Except Mining Safety Engineers and Inspectors, Human Factors Engineers and Ergonomists, Manufacturing Engineers, Mechatronics Engineers, Microsystems Engineers, Nanosystems Engineers, Radio Frequency Identification Device Specialists, Robotics Engineers, Robotics Technicians, Software Developers, Software Quality Assurance Analysts and Testers, Solar Energy Systems Engineers, Telecommunications Engineering Specialists, Validation Engineers, Web Developers, Wind Energy Engineers \\
Finance or Accounting & Accountants and Auditors, Bill and Account Collectors, Bookkeeping, Accounting, and Auditing Clerks, Budget Analysts, Credit Analysts, Financial Examiners, Financial Managers, Financial Quantitative Analysts, Financial Risk Specialists, Financial and Investment Analysts, Investment Fund Managers, Loan Officers, Personal Financial Advisors, Treasurers and Controllers \\
Fundraising & Fundraisers, Fundraising Managers \\
Human Resources & Compensation and Benefits Managers, Human Resources Assistants, Except Payroll and Timekeeping, Human Resources Managers, Human Resources Specialists \\
IT / Information Networking / Information Security & Computer Network Architects, Computer Network Support Specialists, Database Administrators, Information Security Analysts, Information Security Engineers, Network and Computer Systems Administrators, Security Managers, Web Administrators \\
Legal & Law Teachers, Postsecondary, Lawyers, Paralegals and Legal Assistants \\
Marketing & Advertising and Promotions Managers, Market Research Analysts and Marketing Specialists, Marketing Managers, Search Marketing Strategists \\
Operations & General and Operations Managers \\
Product or Product Management &  \\
Project or Program Management & Information Technology Project Managers, Management Analysts, Project Management Specialists \\
Public Relations / Communications & Communications Teachers, Postsecondary, Public Relations Managers, Public Relations Specialists \\
Research & Bioinformatics Scientists, Computer and Information Research Scientists, Operations Research Analysts, Social Science Research Assistants, Survey Researchers \\
Sales / Business Development & Advertising Sales Agents, Demonstrators and Product Promoters, Door-to-Door Sales Workers, News and Street Vendors, and Related Workers, Driver/Sales Workers, Insurance Sales Agents, Retail Salespersons, Sales Engineers, Sales Managers, Sales Representatives of Services, Except Advertising, Insurance, Financial Services, and Travel, Sales Representatives, Wholesale and Manufacturing, Except Technical and Scientific Products, Sales Representatives, Wholesale and Manufacturing, Technical and Scientific Products, Securities, Commodities, and Financial Services Sales Agents, Telemarketers \\
Education Professional & Atmospheric, Earth, Marine, and Space Sciences Teachers, Postsecondary, Business Teachers, Postsecondary, Career/Technical Education Teachers, Middle School, Career/Technical Education Teachers, Postsecondary, Career/Technical Education Teachers, Secondary School, Computer Science Teachers, Postsecondary, Economics Teachers, Postsecondary, Education Administrators, Kindergarten through Secondary, Education Administrators, Postsecondary, Education Teachers, Postsecondary, Elementary School Teachers, Except Special Education, Engineering Teachers, Postsecondary, English Language and Literature Teachers, Postsecondary, Instructional Coordinators, Kindergarten Teachers, Except Special Education, Mathematical Science Teachers, Postsecondary, Middle School Teachers, Except Special and Career/Technical Education, Preschool Teachers, Except Special Education, School Psychologists, Secondary School Teachers, Except Special and Career/Technical Education, Self-Enrichment Teachers, Special Education Teachers, Elementary School, Special Education Teachers, Kindergarten, Special Education Teachers, Middle School, Special Education Teachers, Preschool, Special Education Teachers, Secondary School, Teaching Assistants, Postsecondary, Teaching Assistants, Preschool, Elementary, Middle, and Secondary School, Except Special Education, Teaching Assistants, Special Education, Tutors \\
\bottomrule
\end{tabular}
\end{table*}
}

\section{Scoping Review on Meaningfulness of Work}

\begin{table*}[htbp]
\footnotesize
\centering
\caption{Foundational citations on the personal and the social meaning of work}
\Description{Foundational citations on the personal and the social meaning of work}

\label{tab:foundational_citations}
\begin{tabular}{|p{4cm}|p{9cm}|}
\hline
\textbf{Theme} & \textbf{Foundational Work and Description} \\
\hline

\textbf{Bullshit Jobs, Task Meaning \& Worthwhile Contributions} 
& 
Graeber (2018): Introduce the concept of \``bullshit jobs,\`` emphasizing tasks perceived as socially useless or performative \cite{graeber2018}.\vspace{0.5em}  \newline
Hackman \& Oldham (1976, 1980): specify which job features matter by developing the Job Characteristics Model, linking job design to perceived significance, autonomy, and motivation \cite{hackman1976,hackman1980}. \vspace{0.5em} \newline
Deci and Richard (2000):  Explain why certain job features matter as they meet universal psychological needs. Intrinsic goals (growth, relationships, contribution) satisfy basic psychological needs and foster well-being, whereas extrinsic goals (wealth, fame, image) do not and can undermine it\cite{deci2000what}.  \vspace{0.5em} \newline

Lips-Wiersma \emph{et al.} (2020): People find their jobs more meaningful when work feels fair, leaders act responsibly, and the work itself feels worthwhile—with “doing work that really matters” being the most important factor \cite{lipswiersma2020worthy}. \vspace{0.5em} \newline 

Steger \emph{et al.} (2012): Provide the Work and Meaning Inventory (WAMI) to measure perceived meaningfulness of work \cite{steger2012}. \vspace{0.5em} \newline 
Rostain and Clarke (2025): Identify three ways factory workers in France created meaning in low-skilled jobs often seen as meaningless: \emph{(i)} using hidden skills beyond their usual tasks, \emph{(ii)} finding small opportunities to do the work in their own way, and \emph{(iii)} demonstrating skills that earned respect from coworkers and supervisors \cite{rostain2025meaningful}. \vspace{0.5em} \newline 
Bailey \emph{et al.} (2019): Review 71 empirical studies on meaningful work, highlighting key trends, gaps, and proposing a research agenda \cite{bailey2019review}. \vspace{0.5em} \newline
Bailey \emph{et al.} (2025): Show people see work as meaningful when they believe it makes a real difference to others or to society. This depends on three things: \emph{(i)} the person’s own belief that their work matters, \emph{(ii)} recognition from others, and \emph{(iii)} confidence that they can do the work well \cite{bailey25}.

\\
\hline

\textbf{Task Status, Impression Management, Identity Formation \& Symbolic Work} 
& 
Meyer \& Rowan (1977): Propose that organizations adopt formal structures symbolically to gain legitimacy \cite{meyer1977}. \vspace{0.5em} \newline
Hamilton \emph{et al.} (2022): Show how job status during COVID-19 affected people's sense of dignity and meaning, as being labeled `furloughed' left some feeling excluded or undervalued \cite{hamilton2022life}.\vspace{0.5em}  \newline
Bolino \emph{et al.} (2008): Review motivations and behaviors related to impression management at work \cite{bolino2008}. \vspace{0.5em} \newline
Rosso \emph{et al.}  (2010): Identify two core mechanisms by which work becomes meaningful: agency (creating meaning through personal actions such as autonomy, mastery, competence, and self-identity) and communion (creating meaning through connection with and service to others). These mechanisms operate across four sources of meaning: the self, others, the work context, and the spiritual life \cite{rosso2010meaning}.\vspace{0.5em} \newline
Lepisto and Pratt (2017): Distinguish between meaningful work as realization (fulfilling personal potential) and as justification (contributing to broader social or moral good) \cite{lepisto2017meaningful}. \vspace{0.5em} \newline
Tomlinson and Souto-Otero (2025): Explore how recent UK graduates defined meaningful work. Identified three dimensions: meaning in work  \emph{(i)} as self-expression and self-actualization, \emph{(ii)}  through relationships and social relatedness, and \emph{(iii)}  as societal contribution~\cite{tomlinson_souto_otero2025}. \vspace{0.5em}  \newline
Morabito \emph{et al.} (2025): Explored how early-career veterinarians in Canada perceived meaningful work.  It showed that personal fulfillment through making a difference, creativity and problem-solving, social connection, and professional growth jointly shaped meaningful work~\cite{morabito2025why}. \vspace{0.5em} \newline
Bellezza \emph{et al.} (2017): Argue that busyness serves as a modern status symbol \cite{bellezza2017}. \vspace{0.5em} \newline
Rafaeli \& Pratt (2006): Discuss identity and symbolic expression in work settings \cite{rafaeli2006}.
\\
\hline

\textbf{Status Threat in Social Psychology} 
& 
Pettit \emph{et al.} (2010): Examine reactions to the threat of losing status within groups \cite{pettit2010}. \vspace{0.5em} \newline
Anderson \emph{et al.} (2012):  Explore conditions under which individuals prefer lower-status roles to maintain group dynamics~\cite{anderson2012}.
\\
\hline

\textbf{Work Motivation \& Utility Theory} 
& 
Eccles \& Wigfield (2002): Outline expectancy-value theory as a foundation for understanding task utility and motivation ~\cite{eccles2002}. 
\\
\hline

\end{tabular}
\end{table*}

We reviewed 21 articles (as shown in \autoref{tab:foundational_citations} across psychology, sociology, anthropology, and ethics on what work means to people and to society as a part of our scoping review on meaningfulness of work. The studies show how people judge their own work, how organizations shape those judgments, and how societies value different kinds of work. 

\section{Worker and Developer Survey}
We provide our full survey items for meaningfulness of work and AI traits in \autoref{tab:survey_items}. We also include several follow up questions related to human interventionn preferences based on ~\citet{shao2025future}. We provide a detailed overview of the demographics of our recruited Prolific participants in \autoref{tab:prolific-demographics}.

\begin{table*}[t]
\centering
\scriptsize
\caption{Survey items used in the study.}
\Description{Survey items used in the study.}
\label{tab:survey_items}
\begin{tabular}{p{0.10\columnwidth} p{0.85\columnwidth}}
\toprule
\textbf{ID} & \textbf{Survey Item} \\
\midrule
\multicolumn{2}{l}{\textbf{Perceived Bullshitness (Q1--Q5)}} \\
\midrule
Q1 & The task feels pointless. \\
Q2 & If I stopped doing this task, nothing important would change. \\
Q3 & I perform this task only to satisfy bureaucracy or appearances. \\
Q4 & This task does not contribute to the goals of my organization. \\
Q5 & I would be embarrassed to explain this task to someone outside my field. \\
\midrule

\multicolumn{2}{l}{\textbf{Perceived Value (Q6--Q10)}} \\
\midrule
Q6 & This task is important to the success of my team or organization. \\
Q7 & This task results in a visible or tangible outcome. \\
Q8 & I have the freedom to decide how to carry out this task. \\
Q9 & I receive useful feedback about how well this task is done. \\
Q10 & This task gives me a sense of accomplishment. \\
\midrule

\multicolumn{2}{l}{\textbf{Status Maintenance (Q11--Q16)}} \\
\midrule
Q11 & Not doing this task might make me look less competent to others. \\
Q12 & If I stopped doing this task, others might question my role or importance. \\
Q13 & This task helps reinforce my standing in the organization. \\
Q14 & I worry that letting go of this task could reduce my influence or visibility. \\
Q15 & Even if the task feels unimportant, I feel pressure to keep doing it to maintain status. \\
Q16 & I feel this task signals to others that I am busy or valuable. \\
\midrule

\multicolumn{2}{l}{\textbf{EPOCH (Q17--Q21)}} \\
\midrule
Q17 & This task requires recognizing and responding appropriately to the emotions of others. \\
Q18 & This task benefits significantly from in-person interaction, non-verbal cues, or spontaneous communication. \\
Q19 & This task involves making decisions that require moral reasoning, accountability, or subjective judgment. \\
Q20 & This task requires generating novel ideas, approaches, or solutions beyond standard procedures. \\
Q21 & This task involves setting direction, motivating others, or showing perseverance toward a long-term goal. \\
\midrule

\multicolumn{2}{l}{\textbf{Human Flourishing (Q22--Q33)}} \\
\midrule
Q22 & This task makes me feel satisfied and content with my work. \\
Q23 & This task makes me feel happy and positive. \\
Q24 & This task supports my physical well-being (e.g., energy, comfort). \\
Q25 & This task supports my mental health (e.g., reduced stress, clarity, peace of mind). \\
Q26 & This task feels meaningful and worthwhile. \\
Q27 & This task helps me connect with my personal or professional purpose. \\
Q28 & This task allows me to act in accordance with my values and integrity. \\
Q29 & This task challenges me to exercise discipline, patience, or courage. \\
Q30 & This task helps me build or strengthen relationships with colleagues or clients. \\
Q31 & This task makes me feel supported and connected socially. \\
Q32 & This task contributes to my sense of job or financial security. \\
Q33 & This task helps me feel that my work has tangible value and reward. \\
\midrule

\multicolumn{2}{l}{\textbf{AI Trait Preferences (Q34--Q45)}} \\
\midrule
Q34 & Handle more complex work rather than routine work. \\
Q35 & Focus more on addressing human needs and emotions rather than just data handling. \\
Q36 & Make fast, automatic decisions without explanation rather than decisions that are easy for people to understand. \\
Q37 & Be open to challenge or treat the decision as final. \\
Q38 & Adjust based on the individual it’s helping rather than treat everyone the same. \\
Q39 & Show warmth and care rather than remain neutral and business-like. \\
Q40 & Be polite even if that means not being fully honest, rather than being sincere and straightforward. \\
Q41 & Be strict and follow the rules exactly rather than be tolerant and open-minded. \\
Q42 & Be fast and simple even if less perfect, rather than highly skilled and precise. \\
Q43 & Be determined and persistent rather than flexible and willing to change course. \\
Q44 & Show comprehensiveness, deep understanding and insight rather than keep things simple and straightforward. \\
Q45 & Be imaginative and bring new ideas rather than stay practical and follow familiar approaches. \\
\bottomrule
\end{tabular}
\end{table*}

\section{Scaling and Validating Responses with LMs}

\begin{figure*}[t]
\centering
\includegraphics[width=1.0\linewidth]{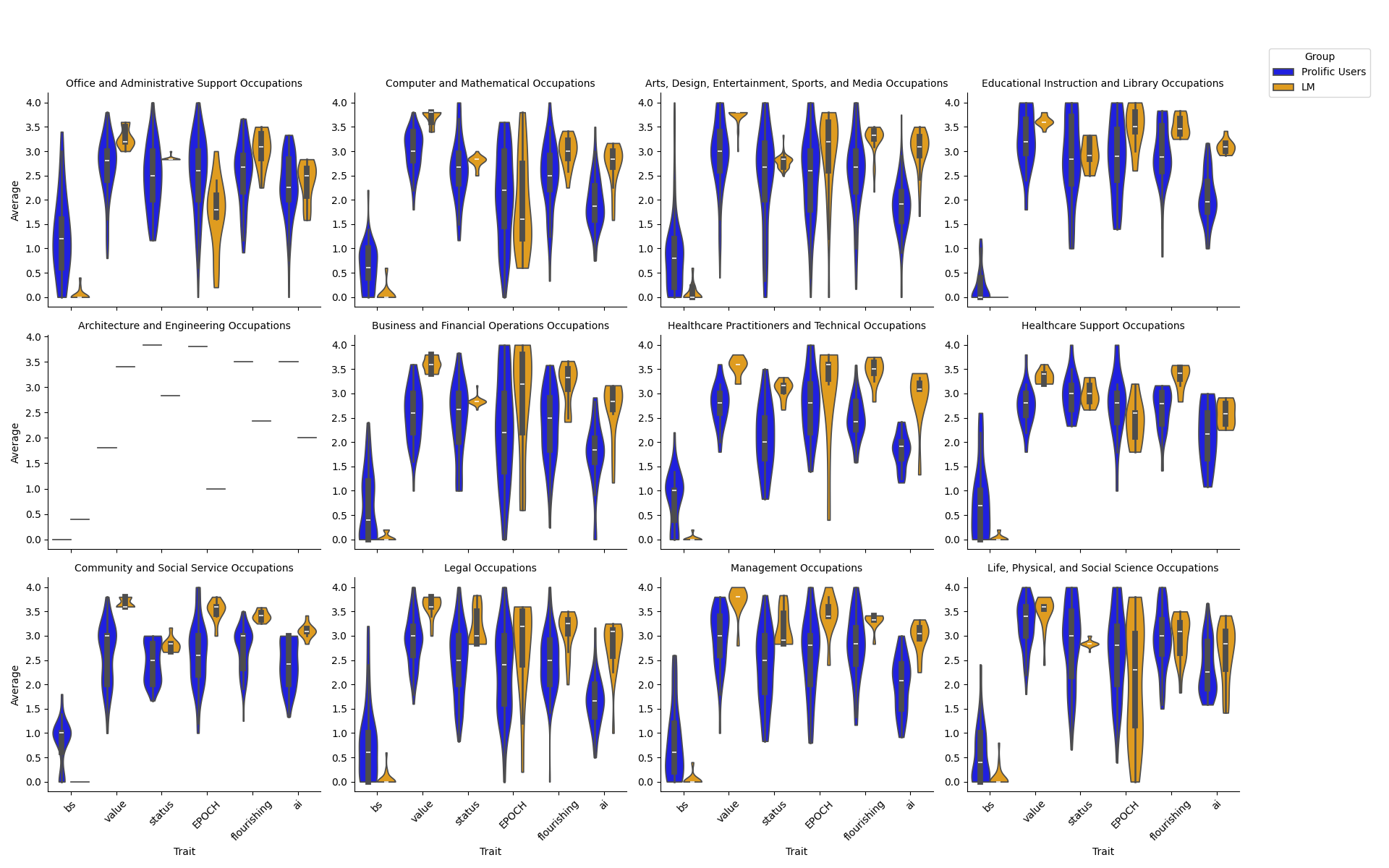}
\caption{Distribution comparison of dimensions of meaningful work between the LM and Prolific users across different sectors.}
  \Description{Grid of violin plots showing distributions of ratings for dimensions of meaningful work across multiple occupational sectors. Each panel corresponds to a sector such as Education, Legal, Healthcare, or Arts. Blue violins show human (Prolific) ratings, while orange violins show LM predictions. In most sectors, LM predictions approximate the central tendency of worker ratings but vary in distributional shape. Some sectors, such as Architecture and Engineering, show fewer available data points.}

\label{fig:trait_distributions}
\end{figure*}

\autoref{fig:trait_distributions} shows the distribution of annotations for five dimensions of meaningful work across 12 sectors, comparing LM annotations with human annotations. For \emph{Perceived Bullshitness (bs)}, both LM and human annotators generally agree that these occupations are not ``bullshit''. However, humans display more variation across sectors, identifying certain support and service roles as slightly more bullshit-like, while LM minimizes such distinctions.  

Across other traits, i.e., \emph{Perceived Value (value)}, \emph{Status Maintenance (status)}, , \emph{Human Flourishing (flourishing)} and \emph{Psychological Traits of AI Behavior (ai)}, LM annotations are clustered near the high end of the scale with relatively smaller variations across almost all sectors. In contrast, human annotations reveal greater differentiation between occupation sectors. Finally, for \emph{EPOCH}, there are relatively high variations for both LM and human annotations. Overall, LM annotations are more uniform and optimistic whereas human annotations reveal richer variability across sectors.

\begin{figure*}[t]
    \centering
    \includegraphics[width=0.8\linewidth]{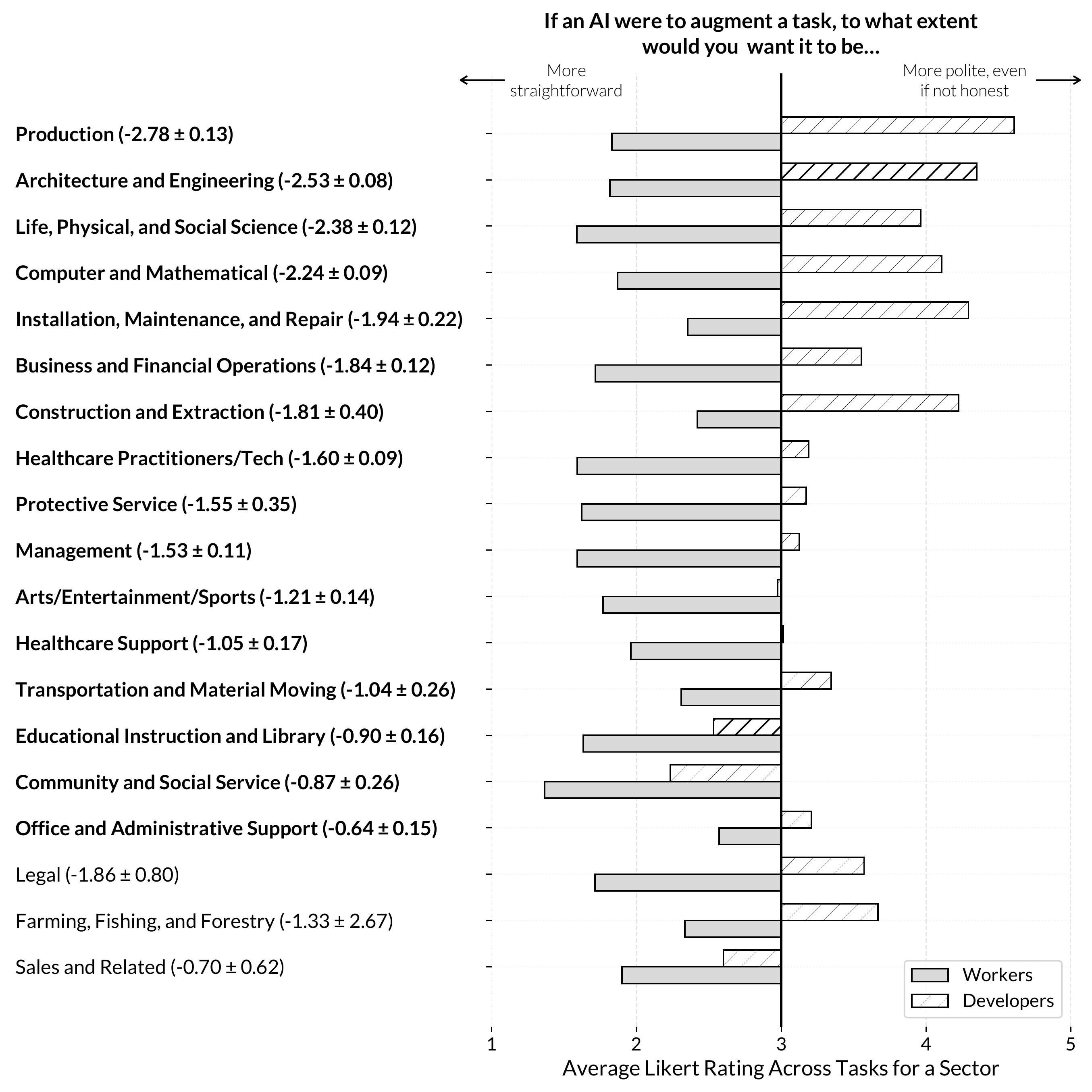}
    \caption{Worker vs. developer preferences for AI to be straightforward or polite, by sector. Paired bars show mean Likert ratings for each group. Greater distance between bars indicates stronger misalignment; misalignment scores with standard errors are shown on the y-axis.
    }
    \Description{Horizontal bar chart comparing worker and developer preferences for AI systems to be straightforward versus polite across occupational sectors. 
  Each row corresponds to a sector such as Production, Architecture and Engineering, or Education. 
  Gray bars represent worker ratings, hatched bars represent developer ratings. 
  In most sectors, developers rate politeness higher than workers, while workers lean toward straightforwardness. 
  Misalignment scores, such as -2.78 ± 0.13 for Production, are shown alongside each sector. The x-axis ranges from 1 (more straightforward) to 5 (more polite, even if not honest).}
    \label{fig:rq2_polite}
\end{figure*}

\begin{table*}[t]
\centering
\small
\caption{Clusters of tasks with the highest worker–developer misalignment on the trait straightforward vs. polite even if not honest, identified using MPNet embeddings and K-means clustering. Cluster labels were generated using GPT-4o.}
\Description{Clusters of tasks with the highest worker–developer misalignment on the trait straightforward vs. polite even if not honest, identified using MPNet embeddings and K-means clustering. Cluster labels were generated using GPT-4o.}
\label{tab:rq2_q40_most}
\begin{tabular}{p{2.5cm} p{4.3cm} p{7.5cm}}
\toprule
\textbf{Cluster (size)} & \textbf{Cluster label} & \textbf{Exemplar task (Occupation \textbar{} Sector)}\\
\midrule
8 (n=135) & Quality Control Oversight & Plan or conduct experimental, environmental, operational, or stress tests on models or prototypes of aircraft or aerospace systems or equip… \textit{(Aerospace Engineers \textbar{} Architecture and Engineering)} \\
0 (n=131) & Technical Design Tasks & Create three-dimensional or interactive representations of designs, using computer-assisted design software. \textit{(Architects, Except Landscape and Naval \textbar{} Architecture and Engineering)} \\
1 (n=130) & Task Coordination and Documentation & Prepare documentation containing information such as confidential descriptions or specifications of proprietary hardware or software, produ… \textit{(Electronics Engineers, Except Computer \textbar{} Architecture and Engineering)} \\
6 (n=129) & Data Analysis and Forecasting & Determine usefulness of new radio frequency identification device (RFID) technologies. \textit{(Radio Frequency Identification Device Specialists \textbar{} Architecture and Engineering)} \\
9 (n=128) & Technical Task Management & Develop or assist in the development of transportation-related computer software or computer processes. \textit{(Transportation Engineers \textbar{} Architecture and Engineering)} \\
4 (n=107) & Data Analysis Tasks & Store, retrieve, and manipulate data for analysis of system capabilities and requirements. \textit{(Computer Hardware Engineers \textbar{} Architecture and Engineering)} \\
3 (n=103) & Technical Oversight Tasks & Inspect completed installations and observe operations to ensure conformance to design and equipment specifications and compliance with ope… \textit{(Electrical Engineers \textbar{} Architecture and Engineering)} \\
2 (n=86) & Complex Technical Tasks & Conduct research related to a range of nanotechnology topics, such as packaging, heat transfer, fluorescence detection, nanoparticle disper… \textit{(Nanosystems Engineers \textbar{} Architecture and Engineering)} \\
5 (n=78) & Technical Surveying Tasks & Prepare and alter trace maps, charts, tables, detailed drawings, and three-dimensional optical models of terrain using stereoscopic plottin… \textit{(Cartographers and Photogrammetrists \textbar{} Architecture and Engineering)} \\
7 (n=77) & Task Monitoring and Reporting & Analyze new medical procedures to forecast likely outcomes. \textit{(Bioengineers and Biomedical Engineers \textbar{} Architecture and Engineering)} \\
\bottomrule
\end{tabular}
\end{table*}

\begin{table*}[t]
\centering
\small
\caption{Clusters of tasks with the highest worker–developer misalignment on the trait tolerant/open-minded vs. strict, identified using MPNet embeddings and K-means clustering. Cluster labels were generated using GPT-4o.}
\Description{Clusters of tasks with the highest worker–developer misalignment on the trait tolerant/open-minded vs. strict, identified using MPNet embeddings and K-means clustering. Cluster labels were generated using GPT-4o.}
\label{tab:rq2_q41_most}
\begin{tabular}{p{2.5cm} p{4.3cm} p{7.5cm}}
\toprule
\textbf{Cluster (size)} & \textbf{Cluster label} & \textbf{Exemplar task (Occupation \textbar{} Sector)}\\
\midrule
2 (n=20) & Strict Process Improvement & Identify opportunities to improve plant electrical equipment, controls, or process control methodologies. \textit{(Geothermal Production Managers \textbar{} Management)} \\
0 (n=15) & Strict Monitoring and Planning & Monitor food preparation methods, portion sizes, and garnishing and presentation of food to ensure that food is prepared and presented in a… \textit{(Food Service Managers \textbar{} Management)} \\
4 (n=11) & Data Analysis and Evaluation & Collect and analyze survey data, regulatory information, and data on demographic and employment trends to forecast enrollment patterns and … \textit{(Education Administrators, Kindergarten through Secondary \textbar{} Management)} \\
8 (n=7) & Strict Budget Management & Develop or review budgets, annual plans, power contracts, power rates, standing operating procedures, power reviews, or engineering studies. \textit{(Hydroelectric Production Managers \textbar{} Management)} \\
1 (n=6) & Research and Evaluation Tasks & Conduct research to develop methodologies, instrumentation, and procedures for medical application, analyzing data and presenting findings. \textit{(Epidemiologists \textbar{} Life, Physical, and Social Science)} \\
9 (n=5) & Data Analysis and Marketing & Monitor and analyze sales promotion results to determine cost effectiveness of promotion campaigns. \textit{(Advertising and Promotions Managers \textbar{} Management)} \\
5 (n=5) & Strict Risk Management & Create scenarios to reestablish operations from various types of business disruptions. \textit{(Business Continuity Planners \textbar{} Business and Financial Operations)} \\
3 (n=5) & Environmental Stewardship Analysis & Provide for stewardship of plant or animal resources or habitats, studying land use, monitoring animal populations, or providing shelter, r… \textit{(Natural Sciences Managers \textbar{} Management)} \\
6 (n=4) & Strict Supply Chain Oversight & Establish or monitor specific supply chain-based performance measurement systems. \textit{(Transportation, Storage, and Distribution Managers \textbar{} Management)} \\
7 (n=4) & Rigid Web Development Guidelines & Create Web models or prototypes that include physical, interface, logical, or data models. \textit{(Web Developers \textbar{} Computer and Mathematical)} \\
\bottomrule
\end{tabular}
\end{table*}

\begin{table*}[t]
\centering
\small
\caption{Clusters of tasks with the highest worker–developer misalignment on the trait practical vs. imaginative, identified using MPNet embeddings and K-means clustering. Cluster labels were generated using GPT-4o.}
\Description{Clusters of tasks with the highest worker–developer misalignment on the trait practical vs. imaginative, identified using MPNet embeddings and K-means clustering. Cluster labels were generated using GPT-4o.}
\label{tab:rq2_q45_most}
\begin{tabular}{p{2.5cm} p{4.3cm} p{7.5cm}}
\toprule
\textbf{Cluster (size)} & \textbf{Cluster label} & \textbf{Exemplar task (Occupation \textbar{} Sector)}\\
\midrule
8 (n=23) & Highly Structured Tasks & Plan sequences of operations, applying knowledge of physical properties of workpiece materials. \textit{(Cutting, Punching, and Press Machine Setters, Operators, and Tenders, Metal and Plastic \textbar{} Production)} \\
1 (n=17) & Technical Operations & Monitor power plant equipment and indicators to detect evidence of operating problems. \textit{(Power Plant Operators \textbar{} Production)} \\
3 (n=10) & Highly Practical Tasks & Study traffic delays by noting times of delays, the numbers of vehicles affected, and vehicle speed through the delay area. \textit{(Traffic Technicians \textbar{} Transportation and Material Moving)} \\
6 (n=8) & Routine Equipment Monitoring & Operate or maintain distributed power generation equipment, including fuel cells or microturbines, to produce energy on-site for manufactur… \textit{(Power Plant Operators \textbar{} Production)} \\
2 (n=8) & Highly technical lab tasks & Perform laboratory procedures following protocols including deoxyribonucleic acid (DNA) sequencing, cloning and extraction, ribonucleic aci… \textit{(Molecular and Cellular Biologists \textbar{} Life, Physical, and Social Science)} \\
4 (n=6) & Administrative Support Tasks & Maintain databases, mailing lists, telephone networks, and other information to facilitate the functioning of health education programs. \textit{(Health Education Specialists \textbar{} Community and Social Service)} \\
5 (n=5) & Data Management Tasks & Enter computer commands to store or retrieve parts patterns, graphic displays, or programs that transfer data to other media. \textit{(Computer Numerically Controlled Tool Programmers \textbar{} Production)} \\
7 (n=3) & Technical Calibration Tasks & Pretest and calibrate anesthesia delivery systems and monitors. \textit{(Anesthesiologist Assistants \textbar{} Healthcare Practitioners and Technical)} \\
9 (n=1) & Technical Source Selection & Select sources from which programming will be received or through which programming will be transmitted. \textit{(Broadcast Technicians \textbar{} Arts, Design, Entertainment, Sports, and Media)} \\
0 (n=1) & Strict Compliance Tasks & Check building codes and zoning bylaws to determine any effects on the properties being appraised. \textit{(Appraisers and Assessors of Real Estate \textbar{} Business and Financial Operations)} \\
\bottomrule
\end{tabular}
\end{table*}

{\small \begin{table*}[ht]
\centering
\caption{Sector-level misalignment for Q37 (Final decisions < Sometimes flexible < Open to being challenged). Misalignment ($\Delta_{t,q}$) is calculated as worker rating minus developer rating for a given task $t$ and trait $q$; negative values indicate developers preferred systems to be more open to challenge than workers. Reported values include the average misalignment score for tasks within a sector, its standard error (SE), 95\% confidence intervals, and the number of tasks ($N$) within each sector.}
\Description{Sector-level misalignment for Q37 (Final decisions < Sometimes flexible < Open to being challenged). Misalignment ($\Delta_{t,q}$) is calculated as worker rating minus developer rating for a given task $t$ and trait $q$; negative values indicate developers preferred systems to be more open to challenge than workers. Reported values include the average misalignment score for tasks within a sector, its standard error (SE), 95\% confidence intervals, and the number of tasks ($N$) within each sector.}
\label{tab:q37}
\begin{tabular}{lcccc}
\toprule
Sector & $\frac{1}{N} \sum_{t=1}^{N} \Delta_{t,q} $ & SE & 95\% CI & N \\
\midrule
\textbf{Installation, Maintenance, and Repair} & \textbf{-0.6} & \textbf{0.072} & \textbf{[-0.744, -0.456]} & \textbf{65} \\
\textbf{Architecture and Engineering} & \textbf{-0.567} & \textbf{0.024} & \textbf{[-0.614, -0.520]} & \textbf{476} \\
\textbf{Transportation and Material Moving} & \textbf{-0.545} & \textbf{0.077} & \textbf{[-0.700, -0.391]} & \textbf{55} \\
Healthcare Practitioners and Technical & -0.48 & 0.03 & [-0.538, -0.421] & 392 \\
Production & -0.476 & 0.056 & [-0.586, -0.366] & 166 \\
Healthcare Support & -0.43 & 0.064 & [-0.558, -0.303] & 79 \\
Management & -0.396 & 0.036 & [-0.468, -0.325] & 217 \\
Computer and Mathematical & -0.392 & 0.021 & [-0.434, -0.350] & 574 \\
Construction and Extraction & -0.387 & 0.11 & [-0.613, -0.161] & 31 \\
Protective Service & -0.379 & 0.092 & [-0.567, -0.191] & 29 \\
Office and Administrative Support & -0.379 & 0.064 & [-0.505, -0.253] & 182 \\
Life, Physical, and Social Science & -0.376 & 0.029 & [-0.434, -0.318] & 279 \\
Community and Social Service & -0.333 & 0.088 & [-0.512, -0.154] & 30 \\
Farming, Fishing, and Forestry & -0.333 & 0.333 & [-1.768, 1.101] & 3 \\
Business and Financial Operations & -0.325 & 0.029 & [-0.383, -0.268] & 289 \\
Arts, Design, Entertainment, Sports, and Media & -0.261 & 0.032 & [-0.325, -0.197] & 203 \\
Sales and Related & -0.2 & 0.2 & [-0.652, 0.252] & 10 \\
Legal & -0.143 & 0.261 & [-0.781, 0.495] & 7 \\
Educational Instruction and Library & -0.122 & 0.041 & [-0.204, -0.040] & 90 \\
\bottomrule
\label{tab:rq2_q37_numbers}
\end{tabular}
\end{table*}
}

\begin{table*}[ht]
\centering
\caption{Sector-level misalignment for Q38 (Generalization < Some adjustment < Personalized). Misalignment ($\Delta_{t,q}$) is calculated as worker rating minus developer rating for a given task $t$ and trait $q$; negative values indicate developers preferred more personalized systems than workers. Reported values include the average misalignment score for tasks within a sector, its standard error (SE), 95\% confidence intervals, and the number of tasks ($N$) within each sector.}
\Description{Sector-level misalignment for Q38 (Generalization < Some adjustment < Personalized).  Misalignment ($\Delta_{t,q}$) is calculated as worker rating minus developer rating for a given task $t$ and trait $q$; negative values indicate developers preferred more personalized systems than workers. Reported values include the average misalignment score for tasks within a sector, its standard error (SE), 95\% confidence intervals, and the number of tasks ($N$) within each sector.}
\label{tab:q38}
\begin{tabular}{lcccc}
\toprule
Sector & $\frac{1}{N} \sum_{t=1}^{N} \Delta_{t,q} $ & SE & 95\% CI & N \\
\midrule
Life, Physical, and Social Science & -0.43 & 0.069 & [-0.565, -0.295] & 279 \\
Business and Financial Operations & -0.367 & 0.054 & [-0.474, -0.260] & 289 \\
Healthcare Practitioners and Technical & -0.301 & 0.047 & [-0.393, -0.209] & 392 \\
Construction and Extraction & -0.29 & 0.141 & [-0.577, -0.003] & 31 \\
Installation, Maintenance, and Repair & -0.246 & 0.17 & [-0.586, 0.093] & 65 \\
Office and Administrative Support & -0.231 & 0.08 & [-0.389, -0.072] & 182 \\
Sales and Related & -0.2 & 0.133 & [-0.502, 0.102] & 10 \\
Computer and Mathematical & -0.199 & 0.048 & [-0.292, -0.105] & 574 \\
Architecture and Engineering & -0.189 & 0.057 & [-0.302, -0.076] & 476 \\
Legal & -0.143 & 0.261 & [-0.781, 0.495] & 7 \\
Protective Service & -0.103 & 0.091 & [-0.289, 0.082] & 29 \\
Production & -0.102 & 0.086 & [-0.272, 0.068] & 166 \\
Management & -0.069 & 0.052 & [-0.172, 0.034] & 217 \\
Community and Social Service & 0.067 & 0.067 & [-0.070, 0.203] & 30 \\
Healthcare Support & 0.051 & 0.09 & [-0.128, 0.230] & 79 \\
Transportation and Material Moving & 0.036 & 0.142 & [-0.248, 0.321] & 55 \\
Arts, Design, Entertainment, Sports, and Media & 0.02 & 0.048 & [-0.075, 0.114] & 203 \\
Educational Instruction and Library & 0.011 & 0.04 & [-0.069, 0.091] & 90 \\
Farming, Fishing, and Forestry & 0.0 & 1.0 & [-4.303, 4.303] & 3 \\
\bottomrule
\label{tab:rq2_q38_numbers}
\end{tabular}

\end{table*}

\begin{table*}[ht]
\centering
\caption{Sector-level misalignment for Q40 (Straightforward < Neutral < Polite even if not honest). Misalignment ($\Delta_{t,q}$) is calculated as worker rating minus developer rating for a given task $t$ and trait $q$; negative values indicate developers preferred systems to be more polite (even at the expense of honesty) than workers. Reported values include the average misalignment score for tasks within a sector, its standard error (SE), 95\% confidence intervals, and the number of tasks ($N$) within each sector.}
\Description{Sector-level misalignment for Q40 (Straightforward < Neutral < Polite even if not honest). Misalignment ($\Delta_{t,q}$) is calculated as worker rating minus developer rating for a given task $t$ and trait $q$; negative values indicate developers preferred systems to be more polite (even at the expense of honesty) than workers. Reported values include the average misalignment score for tasks within a sector, its standard error (SE), 95\% confidence intervals, and the number of tasks ($N$) within each sector.}
\label{tab:q40}
\begin{tabular}{lcccc}
\toprule
Sector & $\frac{1}{N} \sum_{t=1}^{N} \Delta_{t,q} $ & SE & 95\% CI & N \\
\midrule
\textbf{Production} & \textbf{-2.777} & \textbf{0.135} & \textbf{[-3.044, -2.511]} & \textbf{166} \\
\textbf{Architecture and Engineering} & \textbf{-2.532} & \textbf{0.085} & \textbf{[-2.698, -2.365]} & \textbf{476} \\
\textbf{Life, Physical, and Social Science} & \textbf{-2.376} & \textbf{0.116} & \textbf{[-2.604, -2.149]} & \textbf{279} \\
\textbf{Computer and Mathematical} & \textbf{-2.235} & \textbf{0.086} & \textbf{[-2.404, -2.067]} & \textbf{574} \\
\textbf{Installation, Maintenance, and Repair} & \textbf{-1.938} & \textbf{0.218} & \textbf{[-2.374, -1.503]} & \textbf{65} \\
\textbf{Business and Financial Operations} & \textbf{-1.837} & \textbf{0.118} & \textbf{[-2.069, -1.605]} & \textbf{289} \\
\textbf{Construction and Extraction} & \textbf{-1.806} & \textbf{0.403} & \textbf{[-2.629, -0.984]} & \textbf{31} \\
\textbf{Healthcare Practitioners and Technical} & \textbf{-1.597} & \textbf{0.088} & \textbf{[-1.770, -1.424]} & \textbf{392} \\
\textbf{Protective Service} & \textbf{-1.552} & \textbf{0.353} & \textbf{[-2.275, -0.829]} & \textbf{29} \\
\textbf{Management} & \textbf{-1.525} & \textbf{0.113} & \textbf{[-1.748, -1.303]} & \textbf{217} \\
\textbf{Arts, Design, Entertainment, Sports, and Media} & \textbf{-1.207} & \textbf{0.137} & \textbf{[-1.476, -0.938]} & \textbf{203} \\
\textbf{Healthcare Support} & \textbf{-1.051} & \textbf{0.169} & \textbf{[-1.387, -0.714]} & \textbf{79} \\
\textbf{Transportation and Material Moving} & \textbf{-1.036} & \textbf{0.262} & \textbf{[-1.562, -0.511]} & \textbf{55} \\
\textbf{Educational Instruction and Library} & \textbf{-0.9} & \textbf{0.156} & \textbf{[-1.211, -0.589]} & \textbf{90} \\
\textbf{Community and Social Service} & \textbf{-0.867} & \textbf{0.257} & \textbf{[-1.392, -0.341]} & \textbf{30} \\
\textbf{Office and Administrative Support} & \textbf{-0.637} & \textbf{0.154} & \textbf{[-0.941, -0.333]} & \textbf{182} \\
Legal & -1.857 & 0.8 & [-3.814, 0.100] & 7 \\
Farming, Fishing, and Forestry & -1.333 & 2.667 & [-12.807, 10.140] & 3 \\
Sales and Related & -0.7 & 0.616 & [-2.092, 0.692] & 10 \\
\bottomrule
\label{tab:rq2_q40_numbers}
\end{tabular}
\end{table*}

\begin{table*}[ht]
\centering
\caption{Sector-level misalignment for Q41 (Tolerant/Open-minded < Fair < Strict). Misalignment ($\Delta_{t,q}$) is calculated as worker rating minus developer rating for a given task $t$ and trait $q$; negative values indicate developers preferred stricter systems than workers. Reported values include the average misalignment score for tasks within a sector, its standard error (SE), 95\% confidence intervals, and the number of tasks ($N$) within each sector.}
\Description{Sector-level misalignment for Q41 (Tolerant/Open-minded < Fair < Strict). Misalignment ($\Delta_{t,q}$) is calculated as worker rating minus developer rating for a given task $t$ and trait $q$; negative values indicate developers preferred stricter systems than workers. Reported values include the average misalignment score for tasks within a sector, its standard error (SE), 95\% confidence intervals, and the number of tasks ($N$) within each sector.}
\label{tab:q41}
\begin{tabular}{lcccc}
\toprule
Sector & $\frac{1}{N} \sum_{t=1}^{N} \Delta_{t,q} $ & SE & 95\% CI & N \\
\midrule
\textbf{Community and Social Service} & \textbf{1.267} & \textbf{0.106} & \textbf{[1.049, 1.484]} & \textbf{30} \\
\textbf{Educational Instruction and Library} & \textbf{1.067} & \textbf{0.077} & \textbf{[0.913, 1.220]} & \textbf{90} \\
\textbf{Management} & \textbf{0.76} & \textbf{0.072} & \textbf{[0.618, 0.902]} & \textbf{217} \\
\textbf{Sales and Related} & \textbf{0.7} & \textbf{0.213} & \textbf{[0.217, 1.183]} & \textbf{10} \\
\textbf{Arts, Design, Entertainment, Sports, and Media} & \textbf{0.655} & \textbf{0.059} & \textbf{[0.538, 0.772]} & \textbf{203} \\
Healthcare Support & 0.443 & 0.082 & [0.280, 0.606] & 79 \\
Business and Financial Operations & 0.415 & 0.057 & [0.304, 0.527] & 289 \\
Healthcare Practitioners and Technical & 0.401 & 0.047 & [0.308, 0.493] & 392 \\
Life, Physical, and Social Science & 0.348 & 0.052 & [0.244, 0.451] & 279 \\
Protective Service & -0.276 & 0.139 & [-0.562, 0.010] & 29 \\
Computer and Mathematical & 0.261 & 0.036 & [0.190, 0.333] & 574 \\
Construction and Extraction & 0.161 & 0.105 & [-0.053, 0.375] & 31 \\
Installation, Maintenance, and Repair & 0.154 & 0.091 & [-0.028, 0.336] & 65 \\
Office and Administrative Support & 0.154 & 0.049 & [0.057, 0.251] & 182 \\
Legal & -0.143 & 0.143 & [-0.492, 0.207] & 7 \\
Architecture and Engineering & 0.044 & 0.026 & [-0.007, 0.095] & 476 \\
Production & 0.024 & 0.023 & [-0.020, 0.069] & 166 \\
Transportation and Material Moving & 0.018 & 0.066 & [-0.114, 0.151] & 55 \\
Farming, Fishing, and Forestry & 0.0 & 0.0 & [0.000, 0.000] & 3 \\
\bottomrule
\label{tab:rq2_q41_numbers}
\end{tabular}

\end{table*}

\begin{table*}[ht]
\centering
\caption{Sector-level misalignment for Q44 (Simple < Some depth < Deeply insightful/comprehensive). Misalignment ($\Delta_{t,q}$) is calculated as worker rating minus developer rating for a given task $t$ and trait $q$; negative values indicate developers preferred systems that are more deeply insightful and comprehensive than workers. Reported values include the average misalignment score for tasks within a sector, its standard error (SE), 95\% confidence intervals, and the number of tasks ($N$) within each sector.}
\Description{Sector-level misalignment for Q44 (Simple < Some depth < Deeply insightful/comprehensive). Misalignment ($\Delta_{t,q}$) is calculated as worker rating minus developer rating for a given task $t$ and trait $q$; negative values indicate developers preferred systems that are more deeply insightful and comprehensive than workers. Reported values include the average misalignment score for tasks within a sector, its standard error (SE), 95\% confidence intervals, and the number of tasks ($N$) within each sector.}
\label{tab:q44}
\begin{tabular}{lcccc}
\toprule
Sector & $\frac{1}{N} \sum_{t=1}^{N} \Delta_{t,q} $ & SE & 95\% CI & N \\
\midrule
Healthcare Support & 0.241 & 0.079 & [0.083, 0.398] & 79 \\
Transportation and Material Moving & 0.2 & 0.084 & [0.032, 0.368] & 55 \\
Office and Administrative Support & 0.165 & 0.068 & [0.031, 0.298] & 182 \\
Installation, Maintenance, and Repair & 0.154 & 0.055 & [0.045, 0.263] & 65 \\
Protective Service & 0.138 & 0.065 & [0.004, 0.271] & 29 \\
Production & 0.102 & 0.05 & [0.005, 0.200] & 166 \\
Construction and Extraction & 0.065 & 0.113 & [-0.166, 0.295] & 31 \\
Arts, Design, Entertainment, Sports, and Media & 0.039 & 0.024 & [-0.008, 0.087] & 203 \\
Community and Social Service & 0.033 & 0.033 & [-0.035, 0.102] & 30 \\
Business and Financial Operations & -0.031 & 0.019 & [-0.069, 0.007] & 289 \\
Management & -0.023 & 0.017 & [-0.056, 0.010] & 217 \\
Educational Instruction and Library & 0.022 & 0.016 & [-0.009, 0.053] & 90 \\
Healthcare Practitioners and Technical & 0.015 & 0.019 & [-0.022, 0.053] & 392 \\
Computer and Mathematical & 0.014 & 0.01 & [-0.005, 0.033] & 574 \\
Life, Physical, and Social Science & -0.007 & 0.011 & [-0.030, 0.015] & 279 \\
Architecture and Engineering & -0.006 & 0.009 & [-0.024, 0.012] & 476 \\
Farming, Fishing, and Forestry & 0.0 & 0.577 & [-2.484, 2.484] & 3 \\
Legal & 0.0 & 0.378 & [-0.925, 0.925] & 7 \\
Sales and Related & 0.0 & 0.0 & [0.000, 0.000] & 10 \\
\bottomrule
\label{tab:rq2_q44_numbers}
\end{tabular}

\end{table*}

\begin{table*}[ht]
\centering
\caption{Sector-level misalignment for Q45 (Practical < Somewhat creative < Imaginative). Misalignment ($\Delta_{t,q}$) is calculated as worker rating minus developer rating for a given task $t$ and trait $q$; negative values indicate developers preferred more imaginative systems than workers. Reported values include the average misalignment score for tasks within a sector, its standard error (SE), 95\% confidence intervals, and the number of tasks ($N$) within each sector.}
\Description{Sector-level misalignment for Q45 (Practical < Somewhat creative < Imaginative). Misalignment ($\Delta_{t,q}$) is calculated as worker rating minus developer rating for a given task $t$ and trait $q$; negative values indicate developers preferred more imaginative systems than workers. Reported values include the average misalignment score for tasks within a sector, its standard error (SE), 95\% confidence intervals, and the number of tasks ($N$) within each sector.}
\label{tab:q45}
\begin{tabular}{lcccc}
\toprule
Sector & $\frac{1}{N} \sum_{t=1}^{N} \Delta_{t,q} $ & SE & 95\% CI & N \\
\midrule
\textbf{Production} & \textbf{-1.09} & \textbf{0.077} & \textbf{[-1.243, -0.938]} & \textbf{166} \\
\textbf{Transportation and Material Moving} & \textbf{-0.6} & \textbf{0.146} & \textbf{[-0.892, -0.308]} & \textbf{55} \\
Farming, Fishing, and Forestry & -1.333 & 0.333 & [-2.768, 0.101] & 3 \\
Office and Administrative Support & -0.467 & 0.06 & [-0.585, -0.349] & 182 \\
Healthcare Practitioners and Technical & -0.439 & 0.051 & [-0.539, -0.339] & 392 \\
Installation, Maintenance, and Repair & -0.338 & 0.121 & [-0.579, -0.097] & 65 \\
Business and Financial Operations & -0.318 & 0.05 & [-0.416, -0.221] & 289 \\
Life, Physical, and Social Science & -0.269 & 0.052 & [-0.371, -0.166] & 279 \\
Management & -0.23 & 0.052 & [-0.333, -0.128] & 217 \\
Construction and Extraction & -0.226 & 0.165 & [-0.563, 0.112] & 31 \\
Healthcare Support & -0.203 & 0.094 & [-0.390, -0.015] & 79 \\
Arts, Design, Entertainment, Sports, and Media & -0.182 & 0.047 & [-0.275, -0.090] & 203 \\
Architecture and Engineering & -0.181 & 0.039 & [-0.258, -0.104] & 476 \\
Computer and Mathematical & -0.162 & 0.034 & [-0.230, -0.094] & 574 \\
Protective Service & -0.103 & 0.174 & [-0.461, 0.254] & 29 \\
Sales and Related & -0.1 & 0.18 & [-0.506, 0.306] & 10 \\
Community and Social Service & 0.067 & 0.143 & [-0.226, 0.360] & 30 \\
Educational Instruction and Library & -0.067 & 0.044 & [-0.154, 0.021] & 90 \\
Legal & 0.0 & 0.309 & [-0.755, 0.755] & 7 \\
\bottomrule
\label{tab:rq2_q45_numbers}
\end{tabular}

\end{table*}

Table \ref{tab:lm_human_full_examples} provides examples where LM annotations deviate from human annotations. For \emph{perceived bullshitness}, LM rated legal tasks such as drafting wills or contracts by Lawyers as entirely non-pointless (0.0), emphasizing their essential and substantive nature, while humans reported moderate meaningfulness (1.93), reflecting their perception of some bureaucratic routine. For \emph{perceived value}, planning projects for Poets, Lyricists, and Creative Writers was rated higher by LM (3.8) than humans (2.0), highlighting LM’s focus on tangible outcomes, autonomy, and team contribution. For \emph{status maintenance}, managerial tasks like setting prices as General and Operations Managers received higher LM ratings (3.83) than human ratings (1.89), reflecting LM’s weighting of visibility, authority, and organizational standing. In contrast, for \emph{EPOCH}, Survey Researchers reviewing and recording data were rated very low by LM (0.40) but high by humans (3.20), indicating that humans derive well-being and satisfaction from task completion, whereas LM views these tasks as routine and procedural. For \emph{human flourishing}, conducting new employee orientations as Human Resources Specialists was rated higher by LM (3.50) than humans (1.69), reflecting LM’s emphasis on purpose, social connection, and personal growth. Finally, for \emph{psychological trait of AI behavior}, recruiting sponsors or volunteers for Fundraisers was rated higher by LM (3.08) than humans (0.94), highlighting LM’s prioritization of emotional engagement, tailored reasoning, and insight, while humans perceive the task primarily as functional.

\begin{table*}[t]
\centering
\scriptsize
\caption{Full LM survey prompts for Workers}
\Description{Full LM survey prompts for Workers.}
\label{tab:worker_prompts}
\begin{tabular}{|p{0.95\textwidth}|}
\hline
\textbf{Worker Prompt} \\ \hline
You are a professional working as a \{job\_title\} in \{function\}. You are now reflecting how you feel about those work tasks. \\[6pt]

In \{function\}, consider the task: \{task\} \\[6pt]

For each question (Q1--Q45), rate each sentence below from 0 to 4 based on how much you agree with it (0 means strongly disagree, 1 means disagree, 2 neutral, 3 agree and 4 means strongly agree). Before giving each answer, provide a reasoning of one or two sentences, with a maximum of 50 words. Name the reasonings as ``Reason\_1'' to ``Reason\_45'' respectively. \\[6pt]

Q1. The task feels pointless. \\
Q2. If I stopped doing this task, nothing important would change. \\
Q3. I perform this task only to satisfy bureaucracy or appearances. \\
\vdots \\
Q44. Show deep understanding and insight rather than keep things simple and straightforward? \\
Q45. Be imaginative and bring new ideas rather than stay practical and follow familiar approaches? \\[6pt]

For each question (Q46--Q48), select only one option (single lowercase letters from a to e). Before giving each answer, provide a reasoning of one or two sentences, with a maximum of 50 words. Name the reasonings as ``Reason\_46'' to ``Reason\_48'' respectively. \\[6pt]

Q46 (Human Needs). What kind of personal need does this task mostly fulfill for you? \\
a. Basic needs (e.g., survival, security, routine necessities) \\
b. Safety needs (e.g., stability, health, financial security) \\
c. Social needs (e.g., belonging, connection, community) \\
d. Self-esteem needs (e.g., recognition, achievement, confidence) \\
e. Self-actualization needs (e.g., growth, purpose, realizing potential) \\[6pt]

Q47 (Automation Desire by Workers). If an AI can do this task for you completely, how much do you want an AI to do it for you? \\
a. Not at all (I would not want the AI to do this task for me) \\
b. Slightly (I'd want it to do only small parts of the task) \\
c. Moderately (I'd want it to do about half the task) \\
d. A lot (I'd want it to do most of the task) \\
e. Entirely (I'd want it to do the entire task for me) \\[6pt]

Q48 (Required Human Agency Scale). If AI were to assist in this task, how much of your collaboration would be needed to complete this task effectively? \\
H1. AI handles the task entirely on its own. \\
H2. AI needs minimal human input for optimal performance. \\
H3. AI and human form equal partnership, outperforming either alone \\
H4. AI requires human input to successfully complete the task. \\
H5. AI cannot function without continuous human involvement. \\[6pt]

For Q49, select zero or multiple options (must be a combination of zero or more lowercase letters from a to g, without spaces or separators. If no option is selected, leave it blank.) Before answering, provide a reasoning of one or two sentences, with a maximum of 50 words. Name the reasoning as ``Reason\_49''. \\[6pt]

Q49. Why would collaboration be needed for this task? Do not check any boxes if you don't think collaboration is needed. \\
a. This task requires physical actions. \\
b. This task involves making high-stake decisions which I would like to control. \\
c. This task requires specific domain knowledge. \\
d. The task involves nuanced communication or interpersonal skills. \\
e. The task needs validation or oversight to ensure quality \\
f. The task is dynamic and requires adapting to changing circumstances \\
g. The task has ethical, sensitive, or subjective aspects. \\
\hline
\end{tabular}
\end{table*}

\begin{table*}[t]
\centering
\scriptsize
\caption{Full LM survey prompts for Developers}
\Description{Full LM survey prompts for Developers.}
\label{tab:developer_prompts}
\begin{tabular}{|p{0.95\textwidth}|}
\hline
\textbf{Developer Prompt} \\ 
\hline

You are a developer that is designing AI systems for a \{job\_title\} in \{function\}. You are now reflecting how new AI workplace technologies should be built. \\[6pt]

In \{function\}, consider the task: \{task\} \\[6pt]

For each question (Q34--Q45), rate each sentence below from 0 to 4 based on how much you agree with it (0 means strongly disagree, 1 means disagree, 2 neutral, 3 agree and 4 means strongly agree). Before giving each answer, provide a reasoning of one or two sentences, with a maximum of 50 words. Name the reasonings as ``Reason\_34'' to ``Reason\_45'' respectively. \\[6pt]

Q34. Handle more complex work rather than routine work. \\
Q35. Focus more on addressing human needs and emotions rather than just data handling. \\
Q36. Make fast, automatic decisions without explanation rather than decisions that are easy for people to understand? \\
Q37. Be open to challenge or treat the decision as final? \\
Q38. Adjust based on the individual it's helping rather than treat everyone the same? \\
Q39. Show warmth and care rather than remain neutral and business-like? \\
Q40. Be polite even if that means not being fully honest, rather than being sincere and straightforward? \\
Q41. Be strict and follow the rules exactly rather than be tolerant and open-minded? \\
Q42. Be fast and simple even if less perfect, rather than highly skilled and precise? \\
Q43. Be determined and persistent rather than flexible and willing to change course? \\
Q44. Show deep understanding and insight rather than keep things simple and straightforward? \\
Q45. Be imaginative and bring new ideas rather than stay practical and follow familiar approaches? \\[6pt]

For each question (Q47--Q48), select only one option (single lowercase letters from a to e). Before giving each answer, provide a reasoning of one or two sentences, with a maximum of 50 words. Name the reasonings as ``Reason\_46'' to ``Reason\_48'' respectively. \\[6pt]

 Q47 (Automation Desire by Developers). If AI were to assist in this task, how much of user-AI collaboration would be needed to complete this task effectively? \\
a. Not at all (I would not want the AI to do this task for the user)  \\
b. Slightly (I'd want the AI to do only small parts of the task)  \\
c. Moderately (I'd want the AI to do about half the task) \\
d. A lot (I'd want the AI to do most of the task) \\
e. Entirely (I'd want the AI to do the entire task) \\[6pt]

Q48 (Required Human Agency Scale). If AI were to assist in this task, how much collaboration would be needed to complete this task effectively? \\
H1. AI handles the task entirely on its own. \\
H2. AI needs minimal human input for optimal performance. \\
H3. AI and human form equal partnership, outperforming either alone \\
H4. AI requires human input to successfully complete the task. \\
H5. AI cannot function without continuous human involvement. \\[6pt]

For Q49, select zero or multiple options (must be a combination of zero or more lowercase letters from a to g, without spaces or separators. If no option is selected, leave it blank.) Before answering, provide a reasoning of one or two sentences, with a maximum of 50 words. Name the reasoning as ``Reason\_49''. \\[6pt]

Q49. Why would collaboration be needed for this task? Do not check any boxes if you don't think collaboration is needed. \\
a. This task requires physical actions. \\
b. This task involves making high-stake decisions which I would like to control. \\
c. This task requires specific domain knowledge. \\
d. The task involves nuanced communication or interpersonal skills. \\
e. The task needs validation or oversight to ensure quality \\
f. The task is dynamic and requires adapting to changing circumstances \\
g. The task has ethical, sensitive, or subjective aspects. \\
\hline
\end{tabular}
\end{table*}
\begin{table*}[t]
\centering
\small
\caption{RQ1: Significant task dimensions (FDR $<0.05$ and $|\Delta|\ge 0.10$). 
Estimates from nested RE models; $\Delta$ is the back-transformed Likert difference (likely $-$ not-likely) where (+) means that the dimension was more exposed to AI augmentation and (-) it was less exposed to AI augmentation}
\Description{RQ1: Significant task dimensions (FDR $<0.05$ and $|\Delta|\ge 0.10$). 
Estimates from nested RE models; $\Delta$ is the back-transformed Likert difference (likely $-$ not-likely) where (+) means that the dimension was more exposed to AI augmentation and (-) it was less exposed to AI augmentation}
\begin{tabular}{l c c}
\toprule
\textbf{Task Characteristic} & \textbf{$\beta$ (95\% CI)} & \textbf{$\Delta$ (95\% CI)} \\
\midrule
Novel ideas/creativity (+)
& 0.22698\;[0.17405,\,0.27991] 
& 0.29298\;[0.22466,\,0.36131] \\
\addlinespace[2pt]

Happy/Positive (+)
& 0.19965\;[0.13789,\,0.26143] 
& 0.12675\;[0.08753,\,0.16596] \\
\addlinespace[2pt]

Freedom in how to do it (+) 
& 0.18587\;[0.12659,\,0.24516] 
& 0.11487\;[0.07823,\,0.15151] \\
\addlinespace[2pt]

In-person interaction (-)
& $-0.17564$\;[$-0.23024$,\,${-0.12035}$] 
& $-0.23429$\;[$-0.30713$,\,${-0.16145}$] \\
\addlinespace[2pt]

Emotional Awareness (-)
& $-0.13643$\;[$-0.19021$,\,${-0.08259}$] 
& $-0.20635$\;[$-0.28781$,\,${-0.12492}$] \\
\addlinespace[2pt]

Build relationships (-) 
& $-0.13422$\;[$-0.19249$,\,${-0.07594}$] 
& $-0.15360$\;[$-0.22028$,\,${-0.08692}$] \\
\addlinespace[2pt]

Socially supported (-)
& $-0.10393$\;[$-0.16434$,\,${-0.04352}$] 
& $-0.10683$\;[$-0.16893$,\,${-0.04473}$] \\
\bottomrule
\label{tab:rq1_results}
\end{tabular}
\end{table*}

\begin{figure*}[t]
  \centering
  \includegraphics[width=0.9\linewidth]{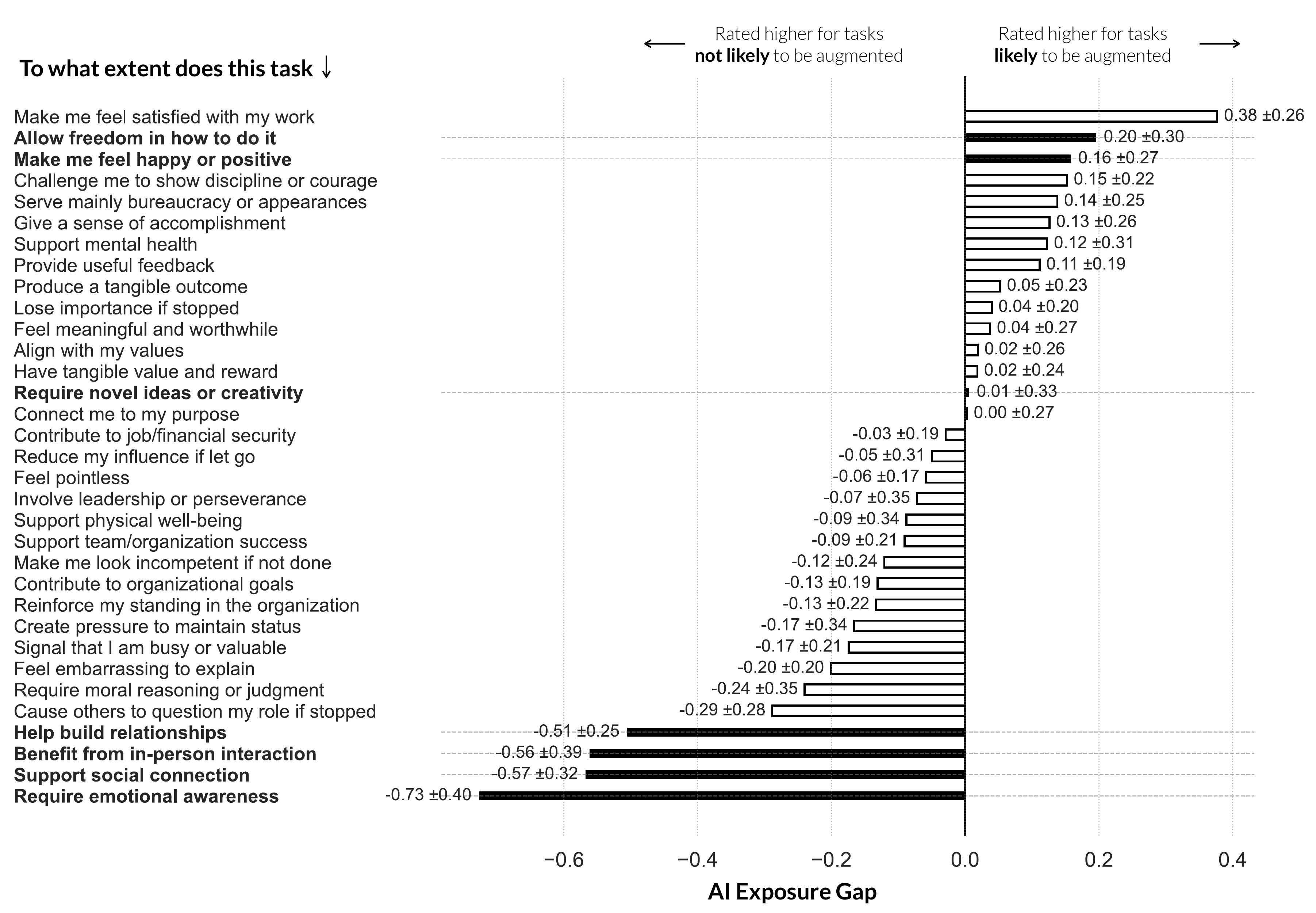}
  \caption{ \emph{RQ1:} AI Exposure Gap by dimensions of meaningful work (rows) based on small-scale human ratings for 171 tasks. A higher gap indicates that a dimension is more strongly associated with tasks likely to be augmented by AI. The gap is computed as the difference in the perceived importance of a dimension between two groups of tasks: those more likely and those less likely to be augmented. We estimate the gaps and 95\% confidence intervals by computing average differences for human ratings. Bold names and corresponding black bars indicate those statistically significant different dimensions of meaningful work identified with \textit{LM annotations} based model (Figure \ref{fig:rq1_bar}). Results based on human ratings are highly consistent with LM annotations: tasks rated as likely to be augmented by AI tend to involve satisfaction with work, happiness, and autonomy, whereas tasks rated as not likely to be augmented tend to involve emotional awareness, in-person interaction, relationship building, and social support. One major exception is ``Require novel ideas or creativity'' where human annotators tended to underestimate the novelty of tasks, whereas LMs captured more nuanced distinctions.}
    \Description{Horizontal bar chart showing AI exposure gaps across dimensions of meaningful work, based on human ratings. Bars extending right represent dimensions associated with tasks more likely to be augmented by AI, including satisfaction with work, happiness, autonomy, and novelty. Bars extending left represent dimensions associated with tasks not likely to be augmented, including emotional awareness, in-person interaction, relationships, and social support. Statistically significant dimensions, such as freedom in how to do a task and emotional awareness, are bolded with black bars. Values and error bars indicate effect sizes with 95\% confidence intervals.}

\label{fig:rq1_human_ratings}
\end{figure*}

\begin{table*}[t]
\centering
\small
\caption{Top sectors and exemplar tasks for each significant survey item in RQ1. 
For each item we select the three sectors with the highest normalized sector mean (z-score) within the relevant subset (likely vs.\ not likely to be automated). Within each sector we show one exemplar task from the 99th percentile of that item's ratings (fallback: sector max).}
\Description{Top sectors and exemplar tasks for each significant survey item in RQ1. 
For each item we select the three sectors with the highest normalized sector mean (z-score) within the relevant subset (likely vs.\ not likely to be automated). Within each sector we show one exemplar task from the 99th percentile of that item's ratings (fallback: sector max).}
\label{tab:rq1_top_sectors_examples}
\begin{tabular}{p{3.5cm} p{11cm}}
\toprule
\textbf{Task Characteristic}  & \textbf{Exemplar task (Occupation \;|\; Sector)} \\
\midrule
Novel ideas/creativity (+) & Formulate basic layout design or presentation approach and specify material details, such as style and size of type \ldots{} (Art Directors \;|\; Arts, Design, Entertainment, Sports, and Media) \\
 & Develop, present, or respond to proposals for specific customer requirements, including request for proposal responses \ldots{} (Sales Engineers \;|\; Sales and Related) \\
 & Prepare scale drawings or architectural designs, using computer-aided design or other tools. (Architects, Except Landscape and Naval \;|\; Architecture and Engineering) \\
\addlinespace
Happy/Positive (+) & Modify treatment plans to comply with changes in client status. (Substance Abuse and Behavioral Disorder Counselors \;|\; Community and Social Service) \\
 & Present lectures and conduct discussions to increase students' knowledge and competence using visual aids, such as \ldots{} (Career/Technical Education Teachers, Postsecondary \;|\; Educational Instruction and Library) \\
 & Create custom illustrations or other graphic elements. (Art Directors \;|\; Arts, Design, Entertainment, Sports, and Media) \\
\addlinespace
Freedom in how to do it (+) & Formulate basic layout design or presentation approach and specify material details, such as style and size of type \ldots{} (Art Directors \;|\; Arts, Design, Entertainment, Sports, and Media) \\
 & Develop or execute strategies to address issues such as energy use, resource conservation, recycling, pollution \ldots{} (Chief Sustainability Officers \;|\; Management) \\
 & Construct probability tables for events such as fires, natural disasters, and unemployment \ldots{} (Actuaries \;|\; Computer and Mathematical) \\
\addlinespace
Emotional awareness (-) & Counsel individuals or groups to help them understand and overcome personal, social, or behavioral problems affecting \ldots{} (Educational, Guidance, and Career Counselors and Advisors \;|\; Community and Social Service) \\
 & Perform surgery to prepare the mouth for dental implants and to aid in the regeneration of deficient bone and gum \ldots{} (Oral and Maxillofacial Surgeons \;|\; Healthcare Practitioners and Technical) \\
 & Provide assistance to the public, such as directions to court offices. (Bailiffs \;|\; Protective Service) \\
\addlinespace
In-person interaction (-) & Provide assistance to the public, such as directions to court offices. (Bailiffs \;|\; Protective Service) \\
 & Perform surgery to prepare the mouth for dental implants and to aid in the regeneration of deficient bone and gum \ldots{} (Oral and Maxillofacial Surgeons \;|\; Healthcare Practitioners and Technical) \\
 & Counsel individuals or groups to help them understand and overcome personal, social, or behavioral problems affecting \ldots{} (Educational, Guidance, and Career Counselors and Advisors \;|\; Community and Social Service) \\
\addlinespace
Build relationships (-) & Counsel individuals or groups to help them understand and overcome personal, social, or behavioral problems affecting \ldots{} (Educational, Guidance, and Career Counselors and Advisors \;|\; Community and Social Service) \\
 & Present purchase offers to sellers for consideration. (Real Estate Sales Agents \;|\; Sales and Related) \\
 & Mentor new faculty members. (Social Work Teachers, Postsecondary \;|\; Educational Instruction and Library) \\
\addlinespace
Socially supported (-) & Counsel individuals or groups to help them understand and overcome personal, social, or behavioral problems affecting \ldots{} (Educational, Guidance, and Career Counselors and Advisors \;|\; Community and Social Service) \\
 & Mentor new faculty members. (Social Work Teachers, Postsecondary \;|\; Educational Instruction and Library) \\
 & Plan, organize, and conduct occupational therapy programs in hospital, institutional, or community settings to help \ldots{} (Occupational Therapists \;|\; Healthcare Practitioners and Technical) \\
\bottomrule
\end{tabular}
\end{table*}

\begin{table*}[t]
\centering
\small
\caption{K-means clustering ($k=10$) of tasks in the 99th percentile for `novel ideas/creativity,' restricted to tasks judged likely to be augmented. Cluster labels were assigned using GPT-4o, and exemplar tasks come from sectors with the strongest heatmap $z$-scores.}
\Description{K-means clustering ($k=10$) of tasks in the 99th percentile for `novel ideas/creativity,' restricted to tasks judged likely to be augmented. Cluster labels were assigned using GPT-4o, and exemplar tasks come from sectors with the strongest heatmap $z$-scores.}
\label{tab:rq1_clusters_novel_ideas_creativity_likely}
\begin{tabular}{p{2.5cm} p{4.3cm} p{7.5cm}}
\toprule
\textbf{Cluster (size)} & \textbf{Label} & \textbf{Exemplar task (Occupation \textbar{} Sector)}\\
\midrule
2 (n=148) & Automated Task Management & Plan or coordinate investigation and resolution of customers' reports of technical problems with aircraft or aerospace vehicles. \textit{(Aerospace Engineers\,|\,Architecture and Engineering)} \\
6 (n=140) & Policy Development & Conduct educational programs that provide farmers or farm cooperative members with information that can help them improve agricultural prod… \textit{(Agricultural Engineers\,|\,Architecture and Engineering)} \\
5 (n=128) & Innovative Solutions Evaluation & Analyze project requests, proposals, or engineering data to determine feasibility, productibility, cost, or production time of aerospace or… \textit{(Aerospace Engineers\,|\,Architecture and Engineering)} \\
3 (n=124) & Sustainable Project Planning & Design environmentally sound structural upgrades to existing buildings, such as natural lighting systems, green roofs, or rainwater collect… \textit{(Architects, Except Landscape and Naval\,|\,Architecture and Engineering)} \\
8 (n=112) & Creative Engineering Tasks & Plan or conduct experimental, environmental, operational, or stress tests on models or prototypes of aircraft or aerospace systems or equip… \textit{(Aerospace Engineers\,|\,Architecture and Engineering)} \\
9 (n=107) & Design and Planning Tasks & Prepare scale drawings or architectural designs, using computer-aided design or other tools. \textit{(Architects, Except Landscape and Naval\,|\,Architecture and Engineering)} \\
0 (n=96) & Logistics and Market Analysis & Analyze new medical procedures to forecast likely outcomes. \textit{(Bioengineers and Biomedical Engineers\,|\,Architecture and Engineering)} \\
4 (n=95) & Biofuels Research and Development & Design sensing, measuring, and recording devices, and other instrumentation used to study plant or animal life. \textit{(Agricultural Engineers\,|\,Architecture and Engineering)} \\
7 (n=81) & Creative Visual Production & Create custom illustrations or other graphic elements. \textit{(Art Directors\,|\,Arts, Design, Entertainment, Sports, and Media)} \\
1 (n=79) & Marketing and Promotion & Confer with customers to assess customer needs or obtain feedback. \textit{(Craft Artists\,|\,Arts, Design, Entertainment, Sports, and Media)} \\
\bottomrule
\end{tabular}
\end{table*}

\begin{table*}[t]
\centering
\small
\caption{K-means clustering ($k=10$) of tasks in the 99th percentile for `happy/positive' restricted to tasks judged likely to be augmented. Cluster labels were assigned using GPT-4o, and exemplar tasks come from sectors with the strongest heatmap $z$-scores.}
\Description{K-means clustering ($k=10$) of tasks in the 99th percentile for `happy/positive' restricted to tasks judged likely to be augmented. Cluster labels were assigned using GPT-4o, and exemplar tasks come from sectors with the strongest heatmap $z$-scores.}
\label{tab:rq1_clusters_happy_positive_likely}
\begin{tabular}{p{2.5cm} p{4.3cm} p{7.5cm}}
\toprule
\textbf{Cluster (size)} & \textbf{Cluster label} & \textbf{Exemplar task (Occupation \textbar{} Sector)}\\
\midrule
6 (n=124) & Client Support and Training & Plan and promote career and employment-related programs and events, such as career planning presentations, work experience programs, job fa… \textit{(Educational, Guidance, and Career Counselors and Advisors\,|\,Community and Social Service)} \\
7 (n=79) & Positive Media Production & Use computers, audio-visual aids, and other equipment and materials to supplement presentations. \textit{(Kindergarten Teachers, Except Special Education\,|\,Educational Instruction and Library)} \\
9 (n=72) & Patient Eligibility Assessment & Arrange for medical, psychiatric, and other tests that may disclose causes of difficulties and indicate remedial measures. \textit{(Child, Family, and School Social Workers\,|\,Community and Social Service)} \\
5 (n=70) & Client Rehabilitation Plans & Modify treatment plans to comply with changes in client status. \textit{(Substance Abuse and Behavioral Disorder Counselors\,|\,Community and Social Service)} \\
1 (n=50) & Training and Development & Evaluate students' or individuals' abilities, interests, and personality characteristics, using tests, records, interviews, or professional… \textit{(Educational, Guidance, and Career Counselors and Advisors\,|\,Community and Social Service)} \\
4 (n=44) & Robotics and Mechatronics & Design advanced precision equipment for accurate or controlled applications. \textit{(Mechatronics Engineers\,|\,Architecture and Engineering)} \\
0 (n=41) & Emergency Response Tasks & Provide assistive devices, supportive technology, or assistance accessing facilities, such as restrooms. \textit{(Special Education Teachers, Preschool\,|\,Educational Instruction and Library)} \\
3 (n=37) & Creative Performance Tasks & Portray and interpret roles, using speech, gestures, and body movements, to entertain, inform, or instruct radio, film, television, or live… \textit{(Actors\,|\,Arts, Design, Entertainment, Sports, and Media)} \\
2 (n=35) & Biofuels Innovation & Propose new biofuels products, processes, technologies or applications based on findings from applied biofuels or biomass research projects. \textit{(Biofuels/Biodiesel Technology and Product Development Managers\,|\,Management)} \\
8 (n=28) & Speech and Language Therapy & Evaluate hearing or speech and language test results, barium swallow results, or medical or background information to diagnose and plan tre… \textit{(Speech-Language Pathologists\,|\,Healthcare Practitioners and Technical)} \\
\bottomrule
\end{tabular}
\end{table*}

\begin{table*}[t]
\centering
\small
\caption{K-means clustering ($k=10$) of tasks in the 99th percentile for `giving workers freedom and agency' restricted to tasks judged likely to be augmented. Cluster labels were assigned using GPT-4o, and exemplar tasks come from sectors with the strongest heatmap $z$-scores.}
\Description{K-means clustering ($k=10$) of tasks in the 99th percentile for `giving workers freedom and agency' restricted to tasks judged likely to be augmented. Cluster labels were assigned using GPT-4o, and exemplar tasks come from sectors with the strongest heatmap $z$-scores.}
\label{tab:rq1_clusters_freedom_in_how_to_do_it_likely}
\begin{tabular}{p{2.5cm} p{4.3cm} p{7.5cm}}
\toprule
\textbf{Cluster (size)} & \textbf{Cluster label} & \textbf{Exemplar task (Occupation \textbar{} Sector)}\\
\midrule
3 (n=188) & Automated Security Systems & Develop computer information resources, providing for data security and control, strategic computing, and disaster recovery. \textit{(Computer and Information Systems Managers\,|\,Management)} \\
4 (n=141) & Marketing and Strategy Evaluation & Identify, develop, or evaluate marketing strategy, based on knowledge of establishment objectives, market characteristics, and cost and mar… \textit{(Marketing Managers\,|\,Management)} \\
0 (n=136) & Sustainability Strategies & Develop or execute strategies to address issues such as energy use, resource conservation, recycling, pollution reduction, waste eliminatio… \textit{(Chief Sustainability Officers\,|\,Management)} \\
7 (n=124) & Engineering Design Tasks & Identify opportunities to improve plant electrical equipment, controls, or process control methodologies. \textit{(Geothermal Production Managers\,|\,Management)} \\
1 (n=123) & Health Program Management & Maintain awareness of advances in medicine, computerized diagnostic and treatment equipment, data processing technology, government regulat… \textit{(Medical and Health Services Managers\,|\,Management)} \\
6 (n=102) & Design and Layout Tasks & Plan store layouts or design displays. \textit{(General and Operations Managers\,|\,Management)} \\
5 (n=71) & Environmental Planning Tasks & Manage site assessments or environmental studies for wind fields. \textit{(Wind Energy Development Managers\,|\,Management)} \\
2 (n=71) & Marketing and Promotion Tasks & Develop or implement product-marketing strategies, including advertising campaigns or sales promotions. \textit{(General and Operations Managers\,|\,Management)} \\
9 (n=69) & Biofuels Research and Development & Design or conduct applied biodiesel or biofuels research projects on topics, such as transport, thermodynamics, mixing, filtration, distill… \textit{(Biofuels/Biodiesel Technology and Product Development Managers\,|\,Management)} \\
8 (n=44) & Training Program Development & Evaluate instructor performance and the effectiveness of training programs, providing recommendations for improvement. \textit{(Training and Development Managers\,|\,Management)} \\
\bottomrule
\end{tabular}
\end{table*}

\begin{table*}[t]
\centering
\small
\caption{K-means clustering ($k=10$) of tasks in the 99th percentile for `requires emotional awareness' restricted to tasks judged likely to be augmented. Cluster labels were assigned using GPT-4o, and exemplar tasks come from sectors with the strongest heatmap $z$-scores.}
\Description{K-means clustering ($k=10$) of tasks in the 99th percentile for `requires emotional awareness' restricted to tasks judged likely to be augmented. Cluster labels were assigned using GPT-4o, and exemplar tasks come from sectors with the strongest heatmap $z$-scores.}
\label{tab:rq1_clusters_emotion_awareness_not}
\begin{tabular}{p{2.5cm} p{4.3cm} p{7.5cm}}
\toprule
\textbf{Cluster (size)} & \textbf{Cluster label} & \textbf{Exemplar task (Occupation \textbar{} Sector)}\\
\midrule
2 (n=185) & Emotion Awareness in Leadership & Testify in depositions or trials as an expert witness. \textit{(Physicians, Pathologists\,|\,Healthcare Practitioners and Technical)} \\
7 (n=171) & Crisis Intervention and Support & Perform crisis interventions to help ensure the safety of the patients and others. \textit{(Mental Health Counselors\,|\,Community and Social Service)} \\
1 (n=165) & Emotion Awareness Tasks & Provide students with disabilities with assistive devices, supportive technology, and assistance accessing facilities, such as restrooms. \textit{(Educational, Guidance, and Career Counselors and Advisors\,|\,Community and Social Service)} \\
3 (n=140) & Emotion Awareness Tasks & Perform surgery to prepare the mouth for dental implants and to aid in the regeneration of deficient bone and gum tissues. \textit{(Oral and Maxillofacial Surgeons\,|\,Healthcare Practitioners and Technical)} \\
6 (n=129) & Personnel Management & Direct the operations of short stay or specialty units. \textit{(Hospitalists\,|\,Healthcare Practitioners and Technical)} \\
8 (n=115) & Emotion Awareness Tasks & Attend meetings, educational conferences, and training workshops, and serve on committees. \textit{(Educational, Guidance, and Career Counselors and Advisors\,|\,Community and Social Service)} \\
4 (n=105) & Networking and Representation & Deliver presentations to lay or professional audiences. \textit{(Preventive Medicine Physicians\,|\,Healthcare Practitioners and Technical)} \\
0 (n=104) & Emotion Awareness Counseling & Counsel individuals or groups to help them understand and overcome personal, social, or behavioral problems affecting their educational or … \textit{(Educational, Guidance, and Career Counselors and Advisors\,|\,Community and Social Service)} \\
5 (n=60) & Professional Development & Teach pharmacy students serving as interns in preparation for their graduation or licensure. \textit{(Pharmacists\,|\,Healthcare Practitioners and Technical)} \\
9 (n=40) & Student Behavior Management & Establish and enforce administration policies and rules governing student behavior. \textit{(Educational, Guidance, and Career Counselors and Advisors\,|\,Community and Social Service)} \\
\bottomrule
\end{tabular}
\end{table*}

\begin{table*}[t!]
\centering
\caption{\emph{RQ2:} Worker–developer misalignment by AI traits with human ratings on  171 tasks. Misalignment is defined as the average absolute difference between worker and developer ratings (Q34–Q45) of the traits they believe AI systems should possess when augmenting tasks. Differences ($\Delta$) are calculated as worker minus developer ratings, with the magnitude ($|\Delta|$) reflecting the size of the misalignment. Reported values aggregate over sectors and traits are grouped into high, mixed, or aligned categories based on percentile thresholds of average absolute misalignment. The bolded traits are those classified into the same categories based on human and LM ratings (Table \ref{tab:rq2_trait_lm_ratings}) on the same set of 171 tasks. As with LM rating–based results, the highest misalignments in human ratings occur for Explainable vs Fast/automatic, Straightforward vs. Polite, and Emotional vs. Apathetic. However, for the for the highly misaligned traits, discrepancies emerge: Handle complex vs. Routine work are highly misaligned in human ratings but fall only into mixed misalignment categories in LM ratings. After manual inspection, we found that this is partly due to bias in the small-scale human ratings data, which are more concentrated in certain sectors/occupations.}
\Description{RQ2: Worker–developer misalignment by AI traits with human ratings. Misalignment is defined as the average absolute difference between worker and developer ratings (Q34–Q45) of the traits they believe AI systems should possess when augmenting tasks. Differences (Delta) are calculated as worker minus developer ratings, with the magnitude (absolute value of Delta) reflecting the size of the misalignment. Reported values aggregate over sectors and traits are grouped into high, mixed, or aligned categories based on percentile thresholds of average absolute misalignment based on human ratings. The bolded traits are those classified into the same categories based on human and LM ratings. As with LM rating–based results, the highest misalignments in human ratings occur for Explainable vs Fast/automatic, Straightforward vs. Polite, and Emotional vs. Apathetic. However, for the for the highly misaligned traits, discrepancies emerge: Handle complex vs. Routine work are highly misaligned in human ratings but fall only into mixed misalignment categories in LM ratings. After manual inspection, we found that this is partly due to bias in the small-scale human ratings data, which are more concentrated in certain sectors/occupations.}
\label{tab:rq2_trait_human_ratings}
\begin{tabular}{p{9cm}r}
\toprule
\textbf{Trait} & \(\mathbf{\mu|\Delta|}\) \\
\midrule
\multicolumn{2}{l}{\textbf{High misalignment}} \\
\midrule
\emph{\textbf{(Q36) Explainable vs. Fast/automatic}}    & 1.367 \\
\emph{\textbf{(Q40) Straightforward vs. Polite}}        & 1.211 \\
(Q34) Handle complex vs. Routine work   & 1.165 \\
\emph{\textbf{(Q35) Address emotions vs. Apathetic}}     & 1.165 \\
\midrule
\multicolumn{2}{l}{\textbf{Mixed misalignment}} \\
\midrule
\emph{\textbf{(Q43) Flexible vs. Determined}}          & 1.0917 \\
\emph{\textbf{(Q39) Business-like vs. Warm/caring}}       & 1.073 \\
\emph{\textbf{(Q41) Tolerant/Open-minded vs. Strict}}    & 1.064 \\
(Q38) Generalized vs. Personalized    & 0.936 \\
\midrule
\multicolumn{2}{l}{\textbf{Aligned}} \\
\midrule
(Q42) Precise vs. Simple                 & 0.936 \\
\emph{\textbf{(Q45) Practical vs. Imaginative}}       & 0.872 \\
\emph{\textbf{(Q37) Definitive vs. Open to challenge}}  & 0.853 \\
\emph{\textbf{(Q44) Simple vs. Comprehensive}}       & 0.817 \\
\bottomrule
\end{tabular}
\end{table*}


\begin{table*}[t!]
\centering
\caption{\emph{RQ2:}  Worker–developer misalignment by AI traits with LM-simulated ratings on 171 tasks. The bolded traits are those classified into the same categories based on human and LM ratings (Table \ref{tab:rq2_trait_human_ratings}) on the same set of 171 tasks.}
\Description{RQ2: Worker–developer misalignment by AI traits with LM-simulated ratings. The bolded traits are those classified into the same categories based on human and LM ratings on the same set of 171 tasks.}
\label{tab:rq2_trait_lm_ratings}
\begin{tabular}{p{9cm}r}
\toprule
\textbf{Trait} & \(\mathbf{\mu|\Delta|}\) \\
\midrule
\multicolumn{2}{l}{\textbf{High misalignment}} \\
\midrule
\emph{\textbf{(Q40) Straightforward vs. Polite}} & 1.936 \\
\emph{\textbf{(Q36) Explainable vs. Fast/automatic}}    & 0.780 \\
\emph{\textbf{(Q35) Address emotions vs. Apathetic}}     & 0.706 \\
(Q42) Precise vs. Simple                 & 0.670 \\
\midrule
\multicolumn{2}{l}{\textbf{Mixed misalignment}} \\
\midrule
\emph{\textbf{(Q41) Tolerant/Open-minded vs. Strict}}    & 0.6514 \\
\emph{\textbf{(Q39) Business-like vs. Warm/caring}}       & 0.523 \\
(Q34) Handle complex vs. Routine work   & 0.450 \\
\emph{\textbf{(Q43) Flexible vs. Determined}}          & 0.450 \\
\midrule
\multicolumn{2}{l}{\textbf{Aligned}} \\
\midrule
\emph{\textbf{(Q45) Practical vs. Imaginative}}       & 0.404 \\
\emph{\textbf{(Q37) Definitive vs. Open to challenge}}  & 0.330 \\
(Q38) Generalized vs. Personalized    & 0.275 \\
\emph{\textbf{(Q44) Simple vs. Comprehensive}}       & 0.110 \\
\bottomrule
\end{tabular}
\end{table*}

\begin{table*}[t]
\centering
\small
\caption{Clusters of tasks with the lowest worker–developer misalignment on the trait definitive vs. open to challenge, identified using MPNet embeddings and K-means clustering. Cluster labels were generated using GPT-4o.}
\Description{Clusters of tasks with the lowest worker–developer misalignment on the trait definitive vs. open to challenge, identified using MPNet embeddings and K-means clustering. Cluster labels were generated using GPT-4o.}
\label{tab:rq2_q37_least}
\begin{tabular}{p{2.5cm} p{4.3cm} p{7.5cm}}
\toprule
\textbf{Cluster (size)} & \textbf{Cluster label} & \textbf{Exemplar task (Occupation \textbar{} Sector)}\\
\midrule
0 (n=263) & Strategic Decision Making & Investigate traffic problems and recommend methods to improve traffic flow or safety. \textit{(Transportation Engineers \textbar{} Architecture and Engineering)} \\
3 (n=254) & Strategic Evaluation and Analysis & Analyze data on conditions such as site location, drainage, or structure location for environmental reports or landscaping plans. \textit{(Landscape Architects \textbar{} Architecture and Engineering)} \\
5 (n=217) & Flexible Decision-Making & Analyze new medical procedures to forecast likely outcomes. \textit{(Bioengineers and Biomedical Engineers \textbar{} Architecture and Engineering)} \\
1 (n=212) & Flexible Decision-Making & Develop or assist in the development of transportation-related computer software or computer processes. \textit{(Transportation Engineers \textbar{} Architecture and Engineering)} \\
2 (n=187) & Research and Development Planning & Prepare scale drawings or architectural designs, using computer-aided design or other tools. \textit{(Architects, Except Landscape and Naval \textbar{} Architecture and Engineering)} \\
7 (n=168) & Task Coordination and Improvement & Develop processes to separate components of liquids or gases or generate electrical currents, using controlled chemical processes. \textit{(Chemical Engineers \textbar{} Architecture and Engineering)} \\
9 (n=158) & Flexible Decision Making & Document equipment or process details of radio frequency identification device (RFID) technology. \textit{(Radio Frequency Identification Device Specialists \textbar{} Architecture and Engineering)} \\
6 (n=132) & Quality Control Analysis & Plan or conduct experimental, environmental, operational, or stress tests on models or prototypes of aircraft or aerospace systems or equip… \textit{(Aerospace Engineers \textbar{} Architecture and Engineering)} \\
8 (n=110) & Data Analysis and Review & Design sensing, measuring, and recording devices, and other instrumentation used to study plant or animal life. \textit{(Agricultural Engineers \textbar{} Architecture and Engineering)} \\
4 (n=89) & Task Evaluation and Development & Train users in task techniques or ergonomic principles. \textit{(Human Factors Engineers and Ergonomists \textbar{} Architecture and Engineering)} \\
\bottomrule
\end{tabular}
\end{table*}

\begin{table*}[t]
\centering
\small
\caption{Clusters of tasks with the lowest worker–developer misalignment on the trait generalization vs. personalized, identified using MPNet embeddings and K-means clustering. Cluster labels were generated using GPT-4o.}
\Description{Clusters of tasks with the lowest worker–developer misalignment on the trait generalization vs. personalized, identified using MPNet embeddings and K-means clustering. Cluster labels were generated using GPT-4o.
}
\label{tab:rq2_q38_least}
\begin{tabular}{p{2.5cm} p{4.3cm} p{7.5cm}}
\toprule
\textbf{Cluster (size)} & \textbf{Cluster label} & \textbf{Exemplar task (Occupation \textbar{} Sector)}\\
\midrule
9 (n=278) & Financial Coordination & Prepare responses to customer requests for information, such as product data, written regulatory affairs statements, surveys, or questionna… \textit{(Regulatory Affairs Specialists \textbar{} Business and Financial Operations)} \\
6 (n=255) & Quality and Compliance Monitoring & Examine damaged vehicle to determine extent of structural, body, mechanical, electrical, or interior damage. \textit{(Insurance Appraisers, Auto Damage \textbar{} Business and Financial Operations)} \\
0 (n=240) & Technical Project Management & Develop and implement technical project management tools, such as plans, schedules, and responsibility and compliance matrices. \textit{(Logisticians \textbar{} Business and Financial Operations)} \\
5 (n=231) & Medical Task Execution & Perform medicolegal examinations and autopsies, conducting preliminary examinations of the body to identify victims, locate signs of trauma… \textit{(Coroners \textbar{} Business and Financial Operations)} \\
3 (n=211) & Health Program Management & Prepare reports of findings, illustrating data graphically and translating complex findings into written text. \textit{(Market Research Analysts and Marketing Specialists \textbar{} Business and Financial Operations)} \\
1 (n=205) & Marketing Strategy Development & Verify and analyze data used in settling claims to ensure that claims are valid and that settlements are made according to company practice… \textit{(Claims Adjusters, Examiners, and Investigators \textbar{} Business and Financial Operations)} \\
7 (n=171) & Design and Media Tasks & Compose images of products, using video or still cameras, lighting equipment, props, or photo or video editing software. \textit{(Online Merchants \textbar{} Business and Financial Operations)} \\
2 (n=165) & Energy and Resource Management & Evaluate the use of technologies, such as global positioning systems (GPS), radio-frequency identification (RFID), route navigation softwar… \textit{(Logistics Engineers \textbar{} Business and Financial Operations)} \\
8 (n=139) & Communication Systems Management & Investigate, evaluate, and settle claims, applying technical knowledge and human relations skills to effect fair and prompt disposal of cas… \textit{(Claims Adjusters, Examiners, and Investigators \textbar{} Business and Financial Operations)} \\
4 (n=96) & Training Program Evaluation & Obtain, organize, or develop training procedure manuals, guides, or course materials, such as handouts or visual materials. \textit{(Training and Development Specialists \textbar{} Business and Financial Operations)} \\
\bottomrule
\end{tabular}
\end{table*}
\begin{table*}[t]
\centering
\small
\caption{Clusters of tasks with the lowest worker–developer misalignment on the trait simple vs. deeply insightful/comprehensive, identified using MPNet embeddings and K-means clustering. Cluster labels were generated using GPT-4o.}
\Description{Clusters of tasks with the lowest worker–developer misalignment on the trait simple vs. deeply insightful/comprehensive, identified using MPNet embeddings and K-means clustering. Cluster labels were generated using GPT-4o.}
\label{tab:rq2_q44_least}
\begin{tabular}{p{2.5cm} p{4.3cm} p{7.5cm}}
\toprule
\textbf{Cluster (size)} & \textbf{Cluster label} & \textbf{Exemplar task (Occupation \textbar{} Sector)}\\
\midrule
2 (n=410) & Health Program Management & Monitor patients' performance in therapy activities, providing encouragement. \textit{(Occupational Therapy Assistants \textbar{} Healthcare Support)} \\
0 (n=394) & Strategic Operations Management & Work under the direction of occupational therapists to plan, implement, or administer educational, vocational, or recreational programs tha… \textit{(Occupational Therapy Assistants \textbar{} Healthcare Support)} \\
9 (n=360) & Task Management and Support & Prepare, maintain, and record records of inventories, receipts, purchases, or deliveries, using a variety of computer screen formats. \textit{(Pharmacy Aides \textbar{} Healthcare Support)} \\
4 (n=339) & Sales Promotion Analysis & Collect and compile data to document clients' performance or assess program quality. \textit{(Speech-Language Pathology Assistants \textbar{} Healthcare Support)} \\
8 (n=284) & Biofuels Research Tasks & Fabricate and fit orthodontic appliances and materials for patients, such as retainers, wires, or bands. \textit{(Dental Assistants \textbar{} Healthcare Support)} \\
3 (n=275) & Quality Control Tasks & Consult with managers or other personnel to resolve problems in areas such as equipment performance, output quality, or work schedules. \textit{(First-Line Supervisors of Office and Administrative Support Workers \textbar{} Office and Administrative Support)} \\
6 (n=259) & Equipment Maintenance Coordination & Design, fabricate, or repair assistive devices or make adaptive changes to equipment or environments. \textit{(Occupational Therapy Assistants \textbar{} Healthcare Support)} \\
5 (n=229) & Evaluate Training Effectiveness & Select or prepare speech-language instructional materials. \textit{(Speech-Language Pathology Assistants \textbar{} Healthcare Support)} \\
1 (n=165) & Environmental Monitoring & Read and effectively interpret small-scale maps and information from a computer screen to determine locations and provide directions. \textit{(Public Safety Telecommunicators \textbar{} Office and Administrative Support)} \\
7 (n=150) & Biofuels Data Analysis & Expose dental diagnostic x-rays. \textit{(Dental Assistants \textbar{} Healthcare Support)} \\
\bottomrule
\end{tabular}
\end{table*}

\end{document}